\documentclass[11pt,preprint]{aastex}

\usepackage{epsf}
\usepackage{lscape}

\shorttitle{Luminosity and Color Dependence}
\shortauthors{}

\def\eg{{e.g.}}

\newcommand{\kms}{\,{\rm km}\;{\rm s}^{-1}}
\newcommand{\hubunits}{\,\kms\;{\rm Mpc}^{-1}}
\newcommand{\hmpc}{\,h^{-1}\;{\rm Mpc}}

\newcommand{\xir}{{\xi(r)}}
\newcommand{\xrp}{{\xi(r_p,\pi)}}
\newcommand{\xsirpi}{{\xi(r_p,\pi)}}
\newcommand{\wrp}{{w_p(r_p)}}
\newcommand{\gr}{{g-r}}
\newcommand{\Mmin}{M_{\rm min}}

\newcommand{\Mr}{M_r}
\newcommand{\rp}{r_p}
\newcommand{\zmax}{z_{\rm max}}
\newcommand{\zmin}{z_{\rm min}}

\def\ngavg{\bar{n}_g}
\def\nravg{\bar{n}_r}
\def\nbavg{\bar{n}_b}
\def\Navg{\langle N\rangle_M}
\def\Mmin{M_{\rm min}}
\def\NNm1{\langle N(N-1) \rangle}
\def\xis{\xi_{\rm 1h}}
\def\xid{\xi_{\rm 2h}}
\def\Rvir{R_{\rm vir}}
\def\intdn{\int_0^\infty dM\frac{dn}{dM}}

\def\hMpc{h^{-1}{\rm Mpc}}
\def\Ncen{N_{\rm cen}}
\def\Nsat{N_{\rm sat}}
\def\Ncenavg{\langle N_{\rm cen}\rangle_M}
\def\Nsatavg{\langle N_{\rm sat}\rangle_M}
\def\Nrcen{N_{r,{\rm cen}}}
\def\Nbcen{N_{b,{\rm cen}}}
\def\Nrsat{N_{r,{\rm sat}}}
\def\Nbsat{N_{b,{\rm sat}}}

\def\Lthres{L_{\rm thres}}
\def\M\sun{M_\odot}
\def\hMsun{h^{-1}M_\odot}

\begin{document}
\title{The Luminosity and Color Dependence of the Galaxy 
Correlation Function}
\author{
Idit Zehavi\altaffilmark{1}, Zheng Zheng\altaffilmark{2,3,4},
David H.\ Weinberg\altaffilmark{2}, Joshua A.\ Frieman\altaffilmark{5,6},
Andreas A.\ Berlind\altaffilmark{7}, Michael R.\ Blanton\altaffilmark{7},
Roman Scoccimarro\altaffilmark{7}, Ravi K.\ Sheth\altaffilmark{8},
Michael A.\ Strauss\altaffilmark{9},
Issha Kayo\altaffilmark{10,11}, Yasushi Suto\altaffilmark{10}, 
Masataka Fukugita\altaffilmark{12}, Osamu Nakamura\altaffilmark{13}, 
Neta A.\ Bahcall\altaffilmark{9}, Jon Brinkmann\altaffilmark{14}, 
James E.\ Gunn\altaffilmark{9}, Greg S.\ Hennessy\altaffilmark{15},
\v{Z}eljko Ivezi\'{c}\altaffilmark{16},
Gillian R.\ Knapp\altaffilmark{9}, Jon Loveday\altaffilmark{17}, 
Avery Meiksin\altaffilmark{18}, David J.\ Schlegel\altaffilmark{9}, 
Donald P.\ Schneider\altaffilmark{19}, Istvan Szapudi\altaffilmark{20}, 
Max Tegmark\altaffilmark{21}, Michael S.\ Vogeley\altaffilmark{22}, 
and Donald G.\ York\altaffilmark{5} (for the SDSS Collaboration)}

\altaffiltext{1}{Steward Observatory, University of Arizona, 933 N. Cherry
Avenue, Tucson, AZ 85721}
\altaffiltext{2}{Department of Astronomy, Ohio State University,
Columbus, OH 43210}
\altaffiltext{3}{Institute for Advanced Study, School of Natural Sciences, 
Einstein Drive, Princeton, NJ 08540}
\altaffiltext{4}{Hubble Fellow}
\altaffiltext{5}{Astronomy and Astrophysics Department, University of
Chicago, Chicago, IL 60637}
\altaffiltext{6}{Fermi National Accelerator Laboratory, P.O.\ Box 500, Batavia,
IL 60510}
\altaffiltext{7}{Department of Physics, New York University, 4 Washington
Place, New York, NY 10003}
\altaffiltext{8}{University of Pittsburgh, Department of Physics and
Astronomy, 3941 O'Hara Street, Pittsburgh, PA 15260}
\altaffiltext{9}{Department of Astrophysical Sciences, Princeton University, 
Peyton Hall, Princeton, NJ 08544}
\altaffiltext{10}{Department of Physics, The University of Tokyo,  Tokyo 
113-0033, Japan}
\altaffiltext{11}{Department of Physics and Astrophysics, Nagoya University, 
Nagoya 464-8602, Japan}
\altaffiltext{12}{Institute for Cosmic Ray Research, University of Tokyo,
Kashiwa 277-8582, Japan}
\altaffiltext{13}{School of Physics and Astronomy, University of Nottingham,
Nottingham NG7 2RD, UK}
\altaffiltext{14}{Apache Point Observatory, P.O.\ Box 59, Sunspot, NM 88349}
\altaffiltext{15}{US Naval Observatory, 3450 Massachusetts Avenue NW, 
Washington DC 20392}
\altaffiltext{16}{Department of Astronomy, University of Washington, Box
351580, Seattle, WA 98195}
\altaffiltext{17}{Sussex Astronomy Centre, University of Sussex, Falmer, 
Brighton, BN1 9QJ, UK}
\altaffiltext{18}{Institute for Astronomy, University of Edinburgh, 
Blackford Hill, Edinburgh EH9 3HJ, UK}
\altaffiltext{19}{Department of Astronomy and Astrophysics, Pennsylvania
State University, 525 Davey Laboratory, University Park, PA 16802}
\altaffiltext{20}{Institute for Astronomy, University of Hawaii, 2680
Woodlawn Drive, Honolulu, HI 96822}
\altaffiltext{21}{Department of Physics, University of Pennsylvania, 
Philadelphia, PA 19104}
\altaffiltext{22}{Department of Physics, Drexel University, Philadelphia,
PA 19104}

\begin{abstract}
We study the luminosity and color dependence of the galaxy
two-point correlation function in the Sloan Digital Sky Survey (SDSS),
starting from a sample of $\sim 200,000$ galaxies over 2500 deg$^2$.
We concentrate our analysis on volume-limited subsamples of specified
luminosity ranges, for which we measure the projected correlation 
function $\wrp$, which is directly related to the real space correlation
function $\xi(r)$.  The amplitude of $\wrp$ rises continuously with 
luminosity from $\Mr \approx -17.5$ to $\Mr \approx -22.5$, with the
most rapid increase occurring above the characteristic luminosity
$L_*$ ($\Mr \approx -20.5$).  Over the scales $0.1\hmpc < \rp < 10\hmpc$,
the measurements for samples with $\Mr>-22$ can be approximated, imperfectly,
by power-law three-dimensional correlation functions 
$\xi(r)=(r/r_0)^{-\gamma}$ with
$\gamma \approx 1.8$ and $r_0(L_*) \approx 5.0\hmpc$.  The brightest
subsample, $-23 < \Mr < -22$, has a significantly steeper $\xi(r)$.
When we divide samples by color, redder galaxies exhibit a higher amplitude
and steeper correlation function at all luminosities.  The correlation
amplitude of blue galaxies increases continuously with luminosity, but
the luminosity dependence for red galaxies is less regular, with bright
red galaxies exhibiting the strongest clustering at large scales and 
faint red galaxies exhibiting the strongest clustering at small scales.
We interpret these results using halo occupation distribution (HOD) models
assuming concordance cosmological parameters.  For most samples, an HOD
model with two adjustable parameters fits the $\wrp$ data better than 
a power-law, explaining inflections at $\rp \sim 1-3\hmpc$ as the transition
between the one-halo and two-halo regimes of $\xi(r)$.  The implied minimum
mass for a halo hosting a {\it central} galaxy more luminous than $L$
grows steadily, with $\Mmin \propto L$ at low luminosities and a steeper
dependence above $L_*$.  The mass at which a halo has, on average, one
{\it satellite} galaxy brighter than $L$ is $M_1 \approx 23\Mmin(L)$, at
all luminosities.  These results imply a conditional luminosity function
(at fixed halo mass) in which central galaxies lie far above a Schechter
function extrapolation of the satellite population. The HOD model fits nicely
explain the color dependence of $\wrp$ and the cross correlation between
red and blue galaxies.  For galaxies with $\Mr<-21$, halos slightly above
$\Mmin$ have blue central galaxies, while more massive halos have red central
galaxies and predominantly red satellite populations.  The fraction of blue
central galaxies increases steadily with decreasing luminosity and host 
halo mass.  The strong clustering of faint red galaxies follows from the
fact that nearly all of them are satellite systems in high mass halos.
The HOD fitting results are in good qualitative agreement with the predictions
of numerical and semi-analytic models of galaxy formation.
\end{abstract}



\section{Introduction}
\label{sec:intro}

Over the course of many decades, studies of large scale structure have
established a dependence of galaxy clustering on morphological type
(e.g., 
\citealt{hubble36,zwicky68,davis76,dressler80,postman84,einasto91,guzzo97,
willmer98,zehavi02,goto03}), 
luminosity (e.g., \citealt{davis88,hamilton88,white88,einasto91,park94,
loveday95,guzzo97,benoist96,norberg01,zehavi02}),
color (e.g., \citealt{willmer98,brown00,zehavi02}), 
and spectral type (e.g., \citealt{norberg02,budavari03,madgwick03}).
Galaxies with bulge-dominated morphologies, red colors, and spectral
types indicating old stellar populations preferentially reside in
dense regions, and they exhibit stronger clustering even on the largest
scales because these dense regions are themselves biased tracers
of the underlying matter distribution \citep{kaiser84}.
Luminous galaxies cluster more strongly than faint galaxies, with
the difference becoming marked above the characteristic luminosity
$L_*$ of the \cite{schechter76} luminosity function, but the detailed
luminosity dependence has been difficult to establish because of the
limited dynamic range even of large galaxy redshift surveys.
The dependence of clustering on galaxy properties is a fundamental
constraint on theories of galaxy formation, providing clues to the
role of initial conditions and environmental influences in determining
these properties.  A detailed understanding of this dependence
is also crucial to any attempt to constrain cosmological models
with galaxy redshift surveys, since different types of galaxies trace
the underlying large scale structure of the dark matter distribution
in different ways.

Achieving this understanding is one of the central design goals of the
Sloan Digital Sky Survey (SDSS; \citealt{york00}), which provides high
quality photometric information and redshifts for hundreds of thousands
of galaxies, sufficient to allow high precision clustering measurements
for many distinct classes of galaxies.  In this paper we analyze the
luminosity and color dependence of galaxy clustering in a sample of
$\sim 200,000$ galaxies drawn from the SDSS, roughly corresponding
to the main galaxy sample \citep{strauss02} of the Second Data Release
(DR2; \citealt{abazajian04}).  Our methods are similar to those used
in our study of clustering in an early sample of $\sim 30,000$ SDSS
galaxies (\citealt{zehavi02}, hereafter Z02) and in studies of the
luminosity, spectral type, and color dependence of clustering in the
Two-Degree Field Galaxy Redshift Survey (2dFGRS; \citealt{colless01})
by \cite{norberg01,norberg02}, \cite{madgwick03}, and \cite{hawkins03}.  
Specifically, we concentrate on the projected correlation function $\wrp$, 
where integration of the redshift space correlation function $\xsirpi$ 
over the redshift dimension $\pi$ yields a quantity that depends only on 
the real space correlation function $\xi(r)$ \citep{davis83}.

Examining traditional large scale structure statistics for different
classes of galaxies complements studies of the correlation between
galaxy properties and the local environment
\citep{dressler80,postman84,einasto87,whitmore93,lewis02}.
Studies using the SDSS have allowed precise quantification
of many of the trends recognized in earlier galaxy surveys,
and the size and detail of the SDSS data set have allowed
some qualitatively new results to emerge.
\cite{hogg03}
show that the local density
increases sharply with luminosity at the bright end of the luminosity
function and depends mainly on color for lower luminosity systems,
with faint red galaxies in particular occupying high density regions.
\cite{blanton03e} demonstrate that the dependence 
of local density
on galaxy morphology and surface brightness can be largely understood
as a consequence of the luminosity and color dependence, since these
quantities themselves correlate strongly with luminosity and color.
\cite{goto03} further examine the morphology-density
relation in the SDSS and show that the transition from late to 
intermediate type populations occurs at moderate overdensity and
the transition from intermediate to early types occurs at high overdensity.
Studies using the SDSS spectroscopic properties
reveal that star formation rates decrease sharply in high density
environments \citep{gomez03,kauffmann04} and that luminous AGN arise
in systems with large bulges but relatively low density environments
\citep{miller03,kauffmann04}.  Our focus on the projected
two-point correlation function
at scales of $r \la 30\hmpc$ also complements Tegmark et al.'s
(\citeyear{tegmark04a}) examination of the luminosity 
dependence of the large scale galaxy power spectrum,
and Kayo et al.'s (\citeyear{kayo04}) study of the luminosity, color, and
morphology dependence of the two-point and three-point correlation
functions in redshift space.

Section~\ref{sec:xi} takes a traditional, empirical approach to our
task.  We measure $\wrp$ for the full, flux-limited data sample and for
volume-limited subsets with different luminosity and color cuts, and
we fit the measurements with power-laws, which generally provide a good
but not perfect description at $r_p \la 20\hmpc$
(\citealt{zehavi04}, hereafter Z04; see also 
\citealt{hawkins03,gaztanaga01}).
Figures~\ref{fig:bias_vs_L} and~\ref{fig:wrp_lumcol} 
below summarize our empirical results
on the luminosity and color dependence of the projected correlation function.

In section~\ref{sec:hod}, we interpret these measurements in the framework
of the Halo Occupation Distribution (HOD; see, e.g., 
\citealt{ma00,peacock00,seljak00,scoccimarro01,berlind02}).
The HOD formalism describes the ``bias'' relation between galaxies
and mass in terms of the probability distribution $P(N|M)$ that 
a halo of virial mass $M$ contains $N$ galaxies of a given type,
together with prescriptions for the relative bias of galaxies and
dark matter within virialized halos.  
(Throughout
this paper, we use the term ``halo'' to refer to a structure of overdensity
$\rho/\bar{\rho} \sim 200$ in approximate dynamical equilibrium, which
may contain a single galaxy or many galaxies.)
This description is complete,
in the sense that any galaxy clustering statistic on any scale can
be predicted given an HOD and a cosmological model, using numerical
simulations or analytic methods.\footnote{We implicitly assume that halos
of the same mass in different environments have, on average, the
same galaxy populations, as expected on the basis of fairly general
theoretical arguments \citep{kauffmann99a,berlind03,sheth04}.
The correlation of galaxy properties with environment emerges naturally
from the environmental dependence of the halo mass function \citep{berlind04}.}
The HOD approach to modeling $\wrp$
comes with strong theoretical priors: we assume a $\Lambda$CDM cosmological
model (inflationary cold dark matter with a cosmological constant)
with parameters motivated by independent measurements, and we adopt
parameterized forms of the HOD loosely motivated by contemporary theories of
galaxy formation 
\citep{kauffmann97,kauffmann99,benson00,berlind03,kravtsov04,zheng04b}.
HOD modeling transforms $\wrp$ data on galaxy pair
counts into a physical relation between galaxies and dark matter halos,
and it sets the stage for detailed tests of galaxy formation
models and sharpened cosmological parameter constraints that draw
simultaneously on a range of galaxy clustering statistics.
\citet{jing98a} pioneered HOD modeling of correlation function data in their
study of the Las Campanas Redshift Survey, using an $N$-body approach. 
HOD modeling (or the closely 
related ``conditional luminosity function" method) has since been applied
to interpret clustering data from a number of surveys at low and high 
redshift (e.g., \citealt{jing98b,jing02,bullock02,moustakas02,bosch03a,
magliocchetti03,yan03,zheng04,porciani04}). In Z04, we used 
this approach to show that the observed deviations of the correlation
function of $\Mr<-21$ SDSS galaxies from a power-law form have a 
natural explanation in terms of the transition between galaxy pairs
within a single virialized halo and galaxy pairs in separate halos.

Section~\ref{sec:observations} describes our data samples and methods.
Section~\ref{sec:xi} presents the empirical results for the galaxy correlation
function. Section~\ref{sec:hod} describes the details and results of the
HOD modeling. Section~\ref{sec:disc}
presents a summary of our results, comparison to previous work,
and directions for future investigation.

\section{Observations and Analysis}
\label{sec:observations}

\subsection{Data}
\label{subsec:data}

The Sloan Digital Sky Survey (SDSS; \citealt{york00}) is an ongoing
project that aims to map nearly a quarter of the sky in the northern
Galactic cap, and a small  portion of the southern
Galactic cap, using a dedicated 2.5 meter telescope located at
Apache Point Observatory in New Mexico. A drift-scanning mosaic CCD camera 
\citep{gunn98} is used to image the sky in five photometric bandpasses 
\citep{fukugita96,smith02} to a limiting magnitude of $r \sim 22.5$. The 
imaging data are processed through a series of 
pipelines that perform astrometric calibration \citep{pier03}, photometric
reduction \citep{lupton01}, and photometric calibration (\citealt{hogg01}),
and objects are then selected for spectroscopic
followup using specific algorithms for the main galaxy sample 
\citep{strauss02}, luminous red galaxy sample \citep{eisenstein01},
and quasars \citep{richards02}. To a good approximation, the main 
galaxy sample consists of all galaxies with Petrosian magnitude $r<17.77$. 
The targets are assigned to spectroscopic plates (tiles) using an adaptive 
tiling algorithm \citep{blanton03a} and observed with a pair of fiber-fed
spectrographs. Spectroscopic data reduction and redshift determination are 
performed by automated pipelines.  Redshifts are 
measured with a success rate greater than $99\%$ and with estimated accuracy
of $30\kms$.  A summary description of the hardware, pipelines, and data 
outputs can be found in \citet{stoughton02}. 

Considerable effort has been invested in preparing the SDSS redshift data
for large-scale structure studies (see, \eg, \citealt{blanton05};
\citealt{tegmark04a}, Appendix A).  
The radial selection function is derived from the sample
selection criteria using the K-corrections of \cite{blanton03b} and 
a modified version of 
the evolving luminosity function model of \cite{blanton03c}.
All magnitudes are corrected for Galactic extinction \citep{schlegel98}.
We K-correct and evolve the luminosities to rest-frame magnitudes
at $z=0.1$, near the median redshift of the sample.
When we create volume-limited samples below, we include a galaxy
if its evolved, redshifted spectral energy distribution would put
it within the main galaxy sample's apparent magnitude and surface
brightness limits at the limiting redshift of the sample.
The angular completeness is characterized carefully for each sector
(a unique region of overlapping spectroscopic plates) on the sky. 
An operational constraint
of using the fibers to obtain spectra is that no two fibers on the same
plate can be closer than $55''$. This fiber collision constraint
is partly alleviated by having roughly
a third of the sky covered by overlapping plates, but it still results in 
$\sim 7\%$ of targeted galaxies not having a measured redshift. These 
galaxies are assigned the redshift of their nearest neighbor.
We show below that this treatment is adequate for our purposes. 

\begin{figure}[tb]
\plotone{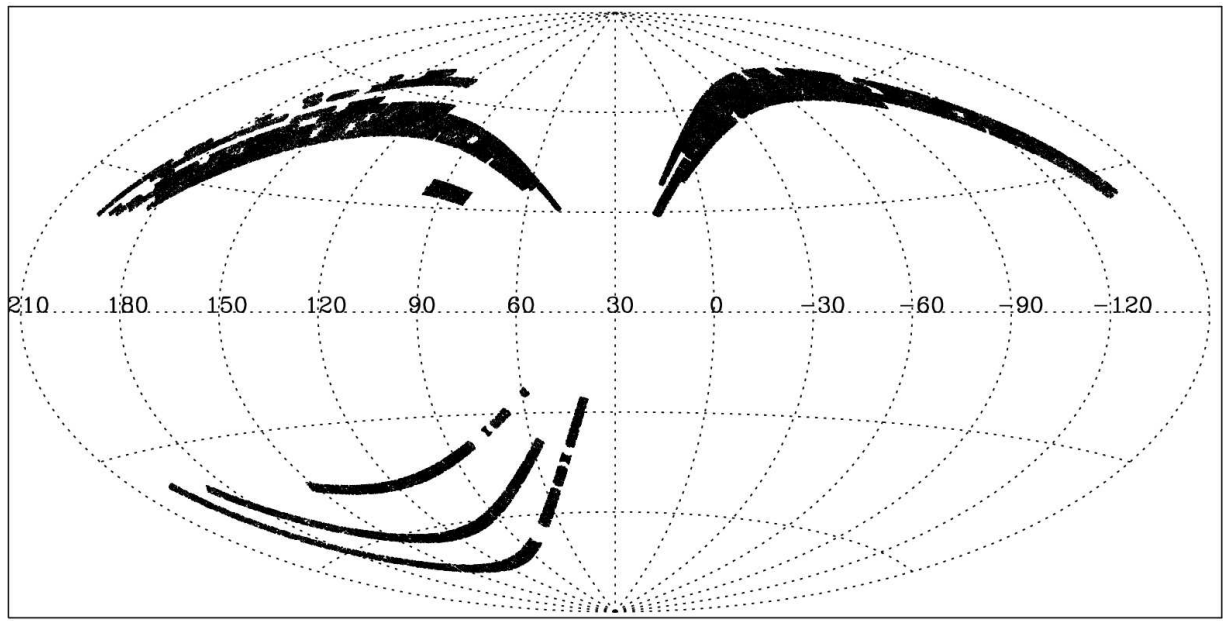}
\caption[]{\label{fig:aitoff}
An Aitoff projection of our galaxy sample in Galactic coordinates.
}
\end{figure}

The clustering  measurements in this paper are based on SDSS Large
Scale Structure {\tt sample12}, based on data taken as of July 2002
(essentially equivalent to the Second Data Release, DR2, 
\citealt{abazajian04}).  It
includes 204,584 galaxies over 2497 deg$^2$ of the sky.
The angular coverage of this sample can be seen in Figure~\ref{fig:aitoff}.
This can be compared to the much smaller sky coverage of the sample
analyzed in Z02 (its figure 1, $\sim 28\%$ of current area).  
The details of the construction and illuminating plots of the LSS sample are 
described by Tegmark et al.\ (\citeyear{tegmark04a}; \S2 and 
Appendix A).\footnote{The sample described there is actually {\tt sample11}.
Our {\tt sample 12} has almost exactly the same set of galaxies,
but it has slightly different
weights and selection function, incorporating an improved technical
treatment of fiber collisions and an improved 
luminosity evolution model that includes dependence on absolute magnitude and 
is valid for a larger redshift range.  These changes make minimal 
difference to our results.} 
Throughout the paper, when measuring distances we refer to 
comoving separations, and for all distance calculations and absolute 
magnitude definitions we adopt a flat $\Lambda$CDM model with $\Omega_m=0.3$.
We quote distances in $\hmpc$ (where $h\equiv H_0/100\hubunits$), and 
we use $h=1$ to compute absolute magnitudes; one should add $5\log h$ to
obtain magnitudes for other values of $H_0$.

We carry out some analyses of a full, flux-limited sample,
with $14.5 \le r \la 17.77$, with the bright limit imposed to avoid
small incompleteness associated with galaxy deblending.
The survey's faint-end apparent magnitude limit
varies slightly over the area of the sample, 
as the target selection criteria changed during the early phases of the survey.
The radial completeness is computed independently for each of these regions
and taken into account appropriately in our analysis. We have verified that 
our results do not change substantively if we cut the sample to a uniform 
flux limit of 17.5 (as done for simplicity in previous 
SDSS large-scale structure 
analyses), but we choose to incorporate the most expansive limits and gain 
in statistical accuracy. 
We also impose an absolute magnitude cut of $-22 < M_r < -19$ (for $h=1$),
thus limiting our analysis to a broad but well-defined range of absolute 
magnitudes around $M_*$ ($-20.44$; \citealt{blanton03c}), and reducing the 
effects of luminosity dependent bias within the sample.   
This cut maintains the majority of the galaxies in the sample, extending
roughly from $(1/4)L_*$ to $4L_*$. We use galaxies in the redshift range 
$0.02<z<0.167$, resulting in a total of $154,014$ galaxies.  

In addition to the flux-limited sample, we analyze a set of 
volume-limited subsamples that span a wider absolute magnitude range. 
For a given
luminosity bin we discard the galaxies that are too faint to be included
at the far redshift limit or too bright to be included at the near limit, 
so that the clustering measurement describes a well defined class of
galaxies observed throughout the sample volume.  
We further cut these samples by color, using the K-corrected $\gr$ color 
as a separator into blue and red populations.\footnote{Cuts using
the $u-r$ color give similar results.  While $u-r$ is a more sensitive
diagnostic of star formation histories, the SDSS $g$-band photometry
is more precise and more uniformly calibrated than the $u$-band
photometry, so we adopt $g-r$ to define our color-selected samples.}
In addition to luminosity-bin samples, we utilize
a set of luminosity-threshold samples, which are volume-limited
samples of all galaxies brighter than a given threshold. 
This set is particularly useful for the HOD modeling in \S~\ref{sec:hod}.
For these samples we relax the bright flux limit to $r>10.5$;
otherwise the sample volumes become too small as the lower redshift
limit for the most luminous objects approaches the upper redshift
limit of the faintest galaxies.
While there are occasional problems with galaxy deblending
or saturation at $r<14.5$, the affected galaxies are a small fraction of the
total samples, and we expect the impact on clustering measurements
to be negligible.

\subsection{Clustering Measures}
\label{subsec:estimates}

We calculate the galaxy correlation function on a two-dimensional
grid of pair separations parallel ($\pi$) and perpendicular ($r_p$) to
the line of sight. To estimate the background counts expected for 
unclustered objects while accounting for the
complex survey geometry, we generate random catalogs with the
detailed radial and angular selection functions of the samples. We
estimate $\xi(r_p,\pi)$ using the \citet{landy93} estimator
\begin{equation}
\xi(r_p,\pi)=\frac{DD-2DR+RR}{RR} ,
\label{eq:LS}
\end{equation}
where DD, DR and RR are the suitably normalized numbers of weighted
data-data, data-random and random-random pairs in each separation
bin. For the flux-limited sample we weight pairs using the minimum variance
scheme of \citet{hamilton93}.
 
To learn about the real-space correlation function, we follow standard practice
and compute the projected correlation function
\begin{equation}
w_p(r_p) = 2 \int_0^{\infty} d\pi \, \xi(r_p,\pi). 
\label{eq:wp}
\end{equation}
In practice we integrate up to $\pi=40\hmpc$, which is large enough
to include most correlated pairs and gives a stable result
by suppressing noise from distant, uncorrelated pairs.
The projected correlation function can in turn be related to the
real-space correlation function, $\xir$,
\begin{equation}
w_p(r_p) = 2 \int_0^{\infty} dy \, \xi\left[(r_p^2+y^2)^{1/2}\right]
= 2 \int_{r_p}^{\infty} r\, dr\, \xir  (r^2-{r_p}^2)^{-1/2}
\label{eq:wp2}
\end{equation}
\citep{davis83}.  In particular, for a power-law 
$\xi(r) = (r/r_0)^{-\gamma}$,  one obtains 
\begin{equation}
w_p(r_p) = r_p \left(\frac{r_p}{r_0}\right)^{-\gamma} 
           \Gamma\left(\frac{1}{2}\right) 
           \Gamma\left(\frac{\gamma-1}{2}\right)\,
          \Bigr/ \,\Gamma\left(\frac{\gamma}{2}\right),
\label{eq:wp3}
\end{equation}
allowing us to infer the best-fit power-law for $\xir$ from $w_p$.
The above measurement methods are those used in Z02, to which we
refer the reader for more details. 

Alternatively, one can directly invert $w_p$ to get $\xir$ independent of 
the power-law assumption.  Equation~(\ref{eq:wp2}) can be recast as
\begin{equation}
\label{eq:xi}
\xir = - \frac{1}{\pi} \int_r^{\infty} {w_p}'(r_p)({r_p}^2-r^2)^{-1/2}dr_p
\end{equation}
(\eg, \citealt{davis83}).  We calculate the integral analytically
by linearly interpolating between the binned $\wrp$ values, following
\citet{saunders92}. As this is still a somewhat approximate treatment,
we focus our quantitative modeling on $\wrp$.

We estimate statistical errors on our different measurements using 
jackknife resampling. We define 104 spatially contiguous subsamples of 
the full data set, each covering approximately 24 deg$^2$ on the sky, and 
our jackknife samples are  then created by omitting each of these subsamples 
in turn. The covariance error matrix is estimated from the total dispersion
among the jackknife samples, 
\begin{equation}
\label{eq:jk}
{\rm Covar}(\xi_i,\xi_j) = \frac{N-1}{N} \sum_{l=1}^{N} 
({\xi_i}^l - {\bar{\xi}_i})({\xi_j}^l - {\bar{\xi}_j}),
\end{equation}
where $N=104$ in our case, and $\bar{\xi}_i$ is the mean value of
the statistic ${\xi}_i$ measured in the samples
($\xi$ denotes here the statistic at hand, whether it is $\xi$ or $w_p$).
In Z02 we used $N=10$ for a much smaller sample, while here the larger 
number, 104, enables us to estimate the full covariance matrix and still
allows each excluded subvolume to be sufficiently large.

\begin{figure}[bp]
\plotone{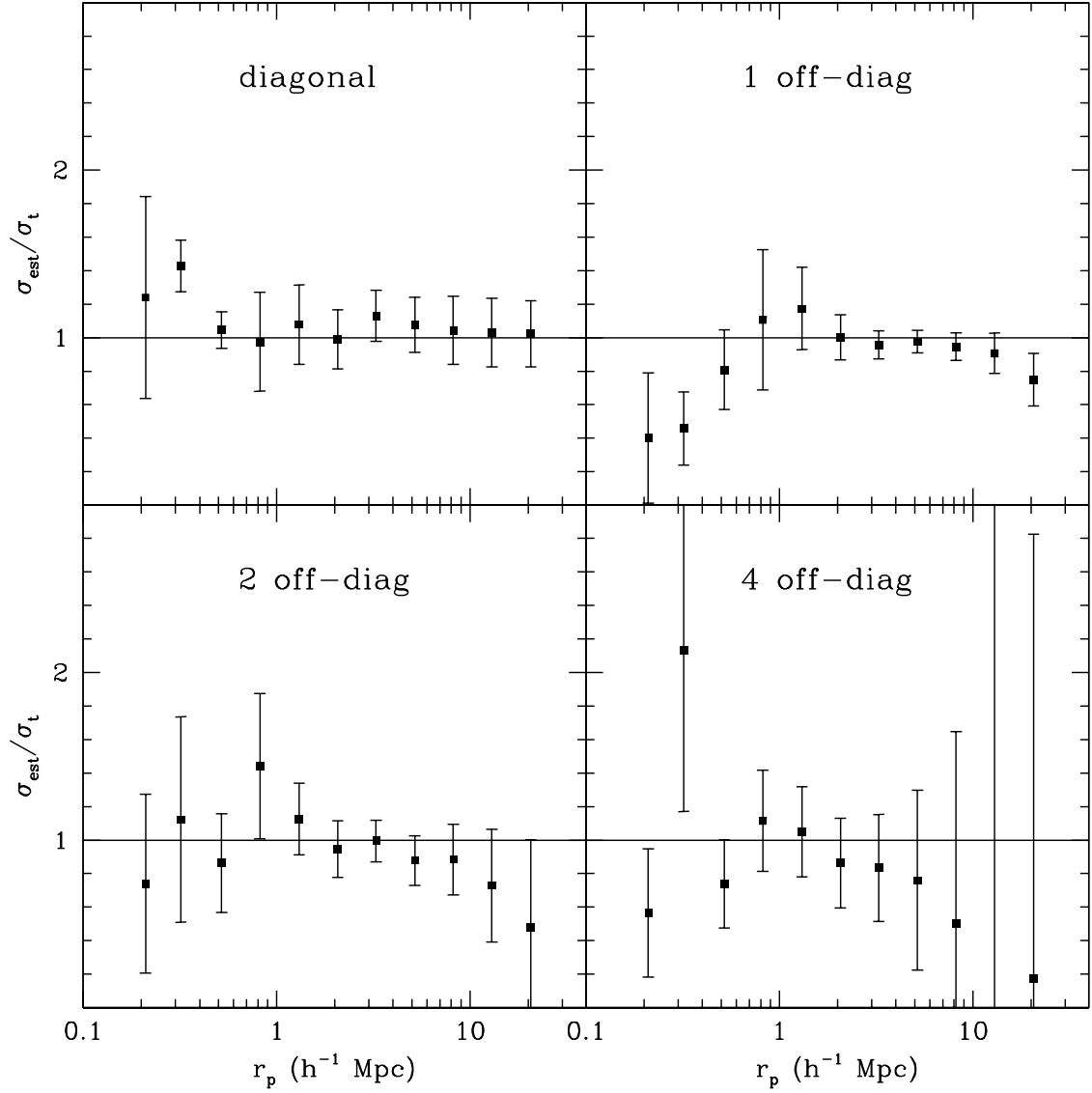}
\caption[]{\label{fig:jackknife}
Accuracy of jackknife error estimates on $\wrp$, tested using the PTHalos mock 
catalogs. Points and errorbars show the mean and 1$\sigma$ scatter of 
the error estimates derived by applying our jackknife procedure
to 100 mock catalogs, divided by the ``true'' errors defined by the
scatter in $\wrp$ among the 100 fully independent 
catalogs. The upper left panel shows diagonal terms in the covariance matrix,
and other panels show terms one, two, and four elements from the diagonal
as indicated.
}
\end{figure}

Following Z02, we repeat and extend the tests with mock catalogs to check 
the reliability of the jackknife error estimates. We use 100 mock
catalogs with the same geometry and angular completeness as the SDSS
sample and similar clustering properties, created using the PTHalos method 
of \citet{scoccimarro02}.
(The mocks correspond to the sample analyzed in Z04, a slightly 
earlier version of our current sample, but we expect the results to be
the same).  For each mock catalog we calculate the projected correlation
function $\wrp$ and compute jackknife error estimates with the same
procedure that we use for the SDSS data.
Figure~\ref{fig:jackknife} compares these error estimates to the ``true'' 
errors, defined as the dispersion among the 100 $\wrp$ estimates
from the fully independent mock catalogs.
The jackknife 
estimates recover the true errors reasonably well for most separations
(with $\sim 20\%\ 1 \sigma$ scatter for the diagonal elements), without gross 
systematics.  The jackknife errors seem to fare well also for the
off-diagonal elements of the covariance error matrix, but with larger
deviations.   Since fits can be sensitive to off-diagonal elements
when errors are strongly correlated, we will present some fits below
using both the full jackknife covariance matrix and the diagonal elements
alone.
For a specified clustering model, the mock catalog approach is probably
the best way to assess agreement of the model with the data. 
However, the jackknife approach
is much more practical when analyzing multiple samples that
have different sizes and clustering properties, as
it automatically accounts for these differences without requiring
a new clustering model in each case.
The tests presented here indicate that parameter errors derived using
the jackknife error estimates should
be representative of the true statistical errors.  We have verified this
expectation using the SDSS data sample of Z04, finding that 
the jackknife covariance matrix produces fits and $\chi^2$ values
similar to those obtained with a mock catalog covariance matrix,
for either power-law or HOD model fitting. 

\begin{figure}[bp]
\plotone{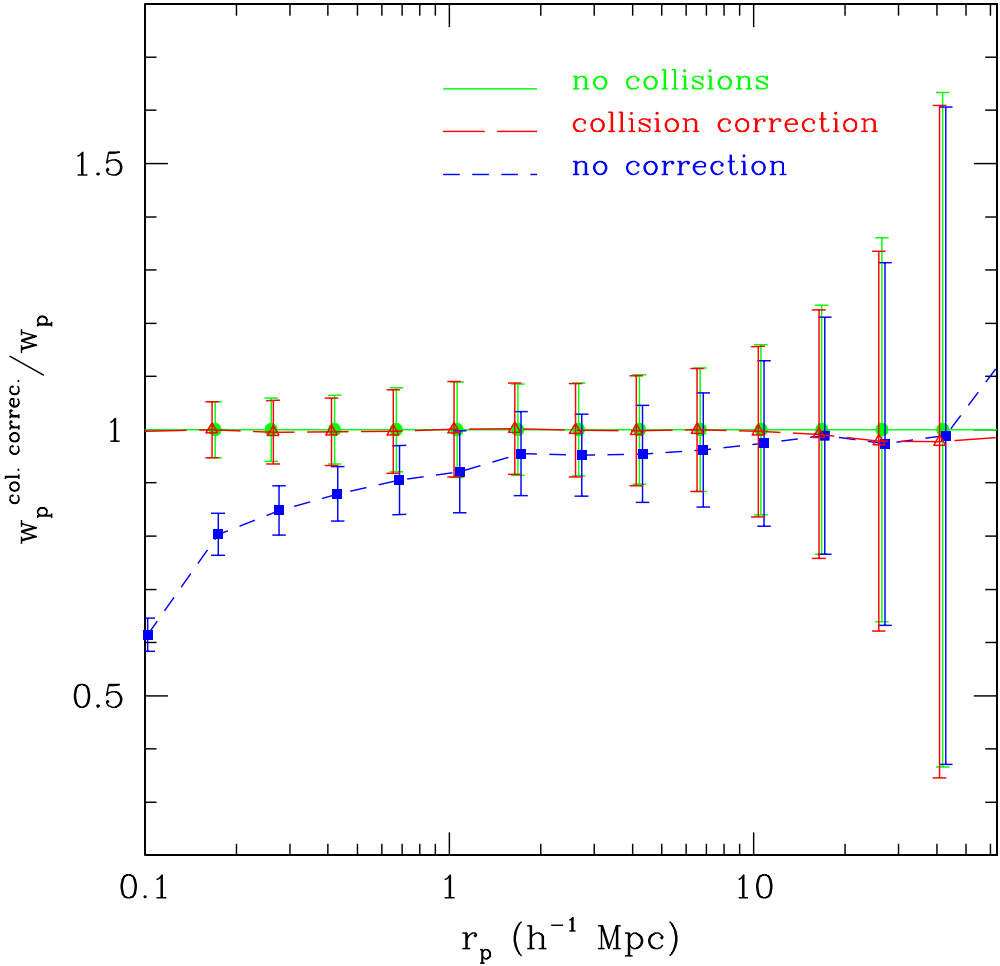}
\caption[]{\label{fig:fiberc}
Test of the accuracy of our correction for fiber collisions, using
a mock catalog drawn from a high resolution $N$-body simulation.
The (green) solid  line and errorbars denote the exemplary ideal case
obtained for the full mock catalog, with no galaxies eliminated by 
fiber collisions. Errorbars throughout are obtained by jackknife 
resampling. The (red) long-dashed line and errorbars show the $\wrp$ 
estimate from a mock catalog that includes fiber collisions, corrected 
using the same procedure we apply to the data, divided by $\wrp$ for the
full catalog.  No systematic effect is present for scales above our 
minimum separation. 
The (blue) short-dashed line and square symbols with errorbars show
the case where we apply no correction for
collisions but simply drop the collided galaxies from the catalog.
}
\end{figure}

Another issue we re-visited with simulations is the effect of
fiber collisions. As mentioned above, we are unable to obtain
redshifts of approximately $7\%$ of the galaxies because of 
the finite fiber size constraints; when two galaxies lie within
$55''$ of each other, one is selected at random for spectroscopic
observations. 
At $cz=50,000 \kms$, the outer edge of our flux-limited sample, $55''$
corresponds to a comoving transverse separation of $0.13 \hmpc$, and we
thus restrict our measurements to separations larger than that. 
We assign to each ``collided'' (unobserved) galaxy 
the redshift of its nearest neighboring galaxy in angle.
This approach is roughly equivalent to double weighting the galaxies
for which we {\it do} obtain redshifts, but using the angular
position of the unobserved galaxy better preserves the small scale pair
distribution (see further discussion in Z02 and \citealt{strauss02}).  
In Z02 we tested this procedure using the tile overlap regions, where 
redshifts of collided galaxies are obtained when the area of sky is 
reobserved, and found this to be an adequate treatment:
residual systematics for the redshift space correlation function were
considerably smaller than the statistical errors, 
and this was even more true for $\wrp$. 
We have since carried out improved tests using mock catalogs created from 
the \citet{white02} $\Lambda$CDM N-body simulation, which 
was run using a TreePM code in a periodic box of $300 \hmpc$ on a side.
We impose on it the SDSS {\tt sample12} mask and populate galaxies 
in dark matter halos (as in \citealt{berlind02}) using a realistic HOD model 
derived from the $\Mr<-20$ volume-limited SDSS sample (see \S\ref{sec:hod}). 
We identify galaxies that would have been collided according to
the $55''$ fiber separation criterion.
Figure~\ref{fig:fiberc} shows the $\wrp$ estimate for this mock catalog
corrected using our standard treatment, divided by the ``true'' $\wrp$
calculated from the mock catalog with no galaxies eliminated by fiber
collisions.  The correction procedure works spectacularly well, with any
residual bias being much smaller than the statistical errors on scales
above our adopted minimum separation.
The correction is important, however, as simply discarding the collided
galaxies causes $\wrp$ to be underestimated at all scales, especially
$\rp \la 1\hmpc$.  Our correction works particularly well for $\wrp$
because it is integrated over the line-of sight direction. 
The post-correction biases can be larger for other statistics --- 
\eg, the redshift-space correlation function shows a small systematic
deviation at $s \la 1\hmpc$, though this is still well within the
statistical uncertainty.

\section{The Galaxy Correlation Function}
\label{sec:xi}

\subsection{Clustering Results for the Flux-Limited Sample}
\label{subsec:xi_flux}

\begin{figure}[bp]
\plotone{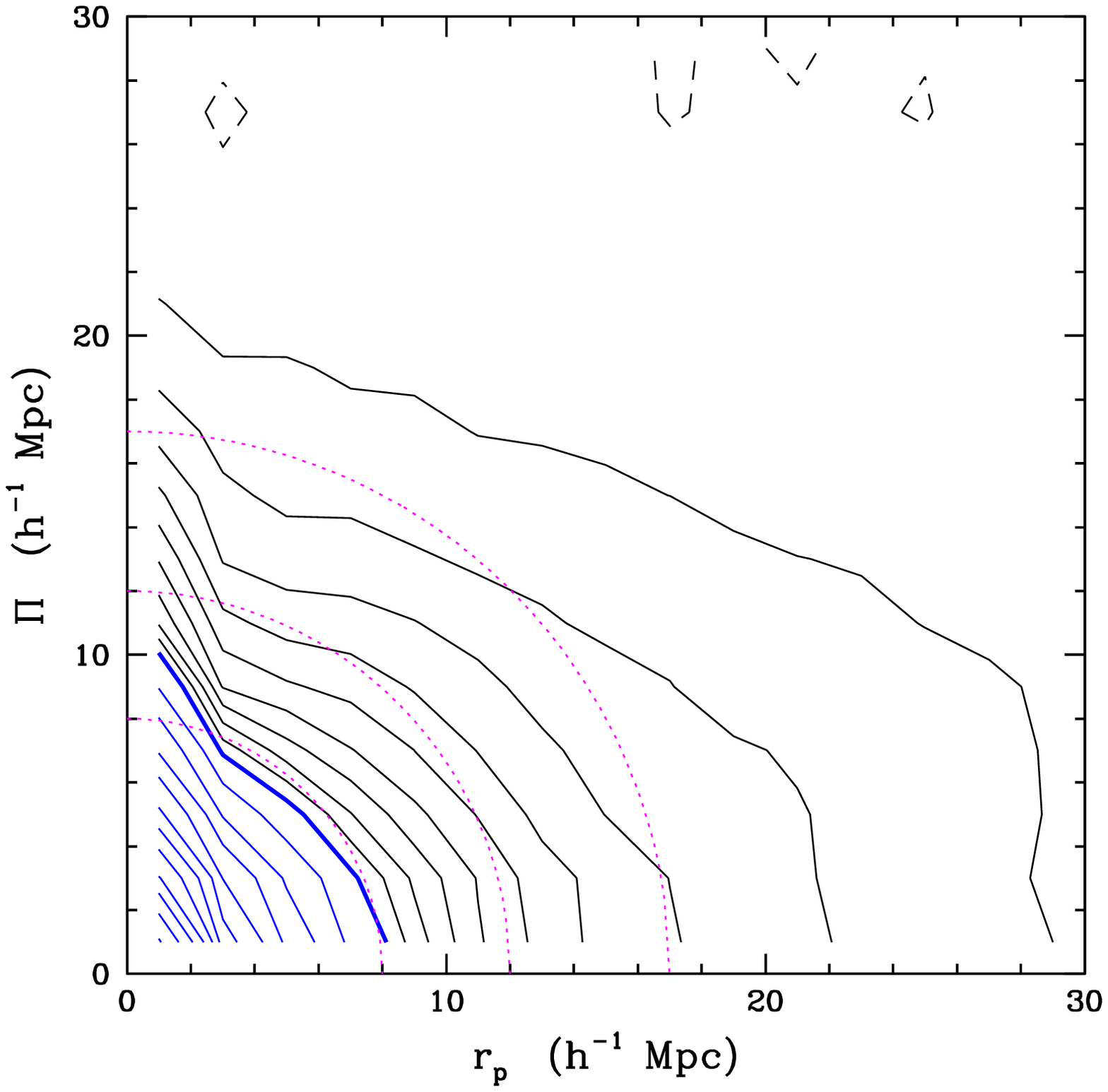}
\caption[]{\label{fig:xsirpi}
Contours of the galaxy correlation function as a function of tangential
separation $r_p$ and line-of-sight separation $\pi$, evaluated for the
full flux-limited sample in $2\hmpc$ bins.
The heavy contour marks $\xsirpi=1$;
inner contours are spaced by 0.1 in $\log \xi$ and outer contours by
0.1 in $\xi$.  The dashed contour marks $\xsirpi=0$.
Dotted lines show the isotropic behavior expected in
the absence of redshift-space distortions.  Contours show the compression
at large scales caused by coherent peculiar velocities and the elongation
at small $r_p$ caused by ``finger-of-God'' distortions in collapsed
structures.  
}
\end{figure}

Figure~\ref{fig:xsirpi} shows contours of the correlation function as a 
function of projected $(r_p)$ and line-of-sight ($\pi$) separation
for our full flux-limited sample, where we bin $r_p$ and
$\pi$ in linear bins of $2\hmpc$. 
One can clearly see the effects of redshift distortions in $\xrp$.
At small projected separations the contours are elongated along the line of 
sight due to small-scale virial motions in clusters,
the so-called ``finger-of-God'' effect.  At large projected separations
$\xrp$ shows compression in the $\pi$ direction caused by coherent 
large-scale streaming \citep{sargent77,kaiser87,hamilton92}.

\begin{figure}[bp]
\plotone{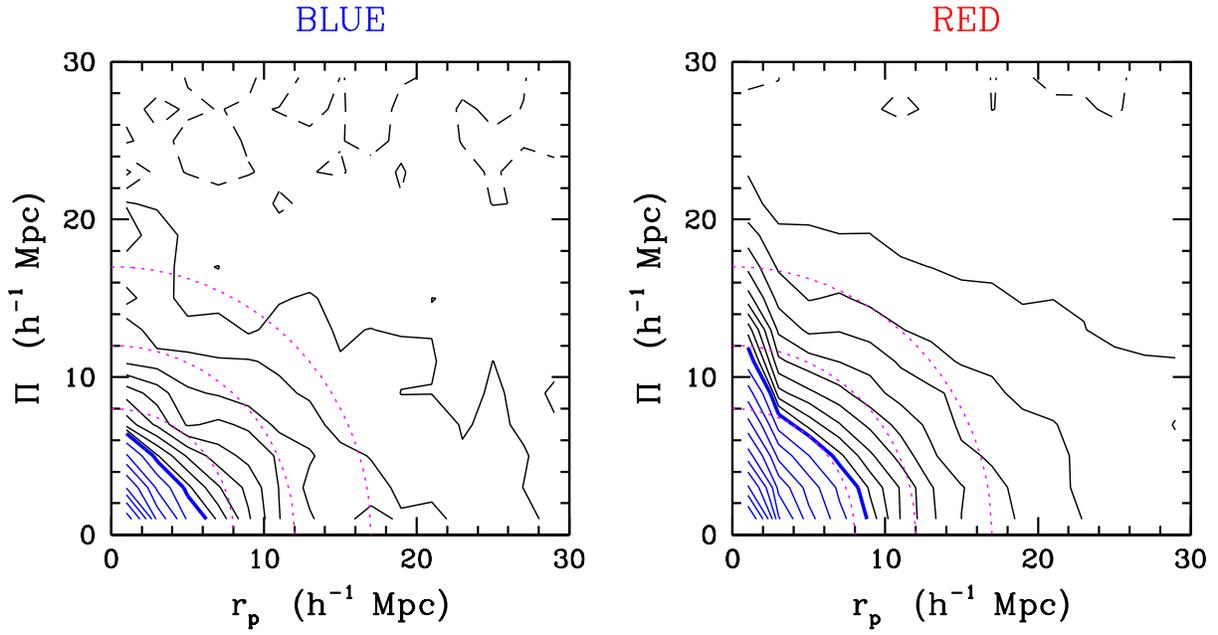}
\caption[]{\label{fig:xsirpi_br}
Correlation function contours for galaxies with 
$\gr < 0.7$ (left) and $\gr>0.7$ (right) in the flux-limited sample.
Contour specifications are as in Fig.~\ref{fig:xsirpi}.
Red galaxies have a higher amplitude correlation at a given separation,
and they show stronger finger-of-God distortions because of their
preferential location in dense regions.  Both classes of galaxies show
large scale compression, though the results for blue galaxies are 
noisier because of the lower $\xsirpi$ amplitude and smaller sample. 
}
\end{figure}

Figure~\ref{fig:xsirpi_br} shows $\xrp$ separately for red and blue galaxies.
The $\gr$ galaxy color distribution is bimodal, similar to $u-r$ 
\citep{strateva01}, so we divide the sample at a rest-frame $\gr=0.7$,
which naturally separates the two populations.
The red sample contains roughly twice as many galaxies as the blue one.
As expected, the red galaxies exhibit a larger clustering 
amplitude than do the blue galaxies. The difference in the 
anisotropy is striking, with the red galaxies exhibiting much stronger 
finger-of-God distortions on small-scales.  
Both samples show clear signatures of large scale distortion.
We examine the dependence
of real space clustering on galaxy color in \S~\ref{subsec:xi_color}.

\begin{figure}[bp]
\plotone{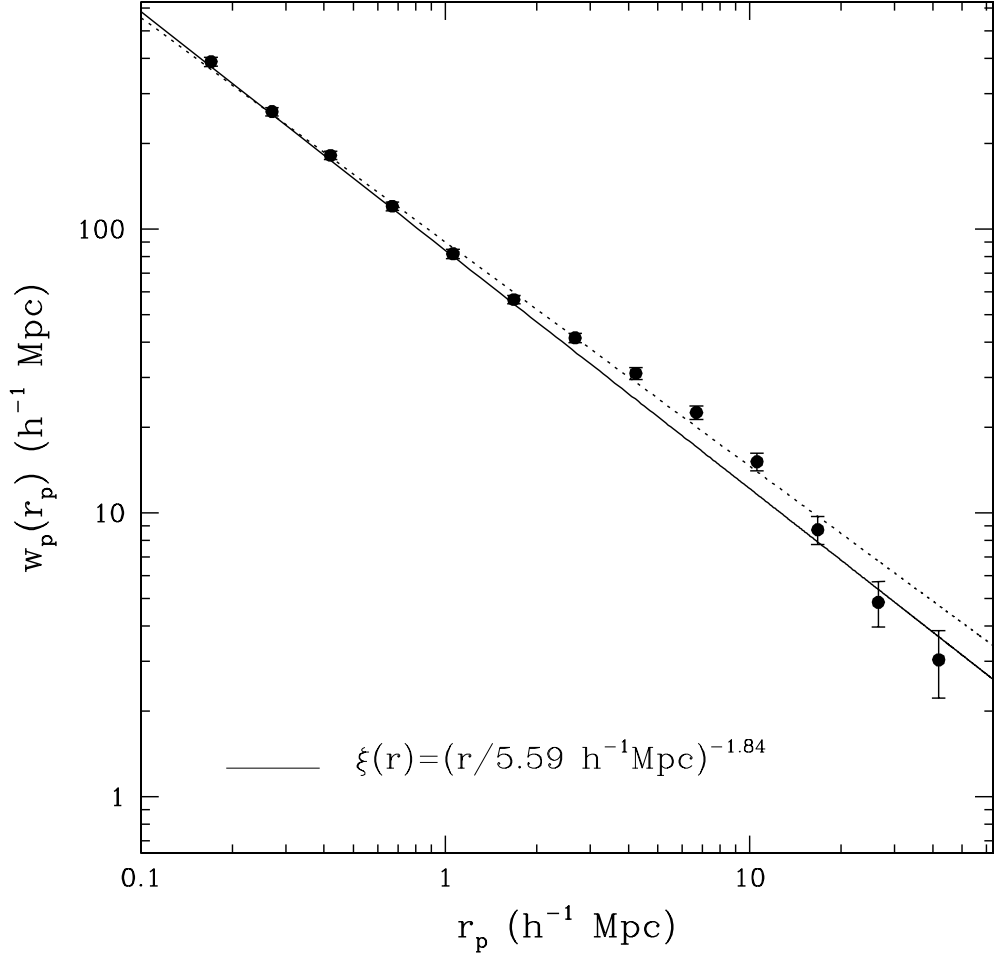}
\caption[]{\label{fig:wp}
Projected galaxy correlation function $\wrp$ for the flux-limited galaxy
sample.  The solid line shows a power-law fit to the data points,
using the full covariance matrix, which corresponds to a real-space 
correlation function $\xi(r)=(r/5.59\hmpc)^{-1.84}$. The dotted line
shows the fit when using only the diagonal error elements, corresponding
to $\xi(r)=(r/5.94\hmpc)^{-1.79}$. The fits are performed for $r_p<20\hmpc$.
}
\end{figure}

We disentangle the effects of redshift distortions from real space
correlations by estimating the projected correlation
function $\wrp$ via equation~(\ref{eq:wp}), now using logarithmic
bins of 0.2 in $r_p$.  The resulting $w_p$ for 
the full flux-limited sample is shown in Figure~\ref{fig:wp}
together with fits of the data points to a power-law. The fits are
done using the measured data points in the range $0.13\hmpc < r_p < 20\hmpc$,
as this is the range where the measurements are robust.
As discussed in \S\ref{subsec:estimates}, the power-law fits can be 
directly related to the real-space correlation function (eq.~\ref{eq:wp3}).
The inferred real space correlation function is 
$\xir=(r/r_0)^{-\gamma}$ with $r_0=5.59 \pm 0.11 \hmpc$ and
$\gamma = 1.84 \pm 0.01$,
when the fit is done using the full covariance matrix (solid line).
When using only the diagonal elements (dotted line), i.e., ignoring the 
correlation of errors between bins, one gets
a slightly higher and shallower power-law
as the strongly correlated points at large separation are effectively
given higher weight when they are treated as independent.
The parameters of this diagonal fit are 
$r_0=5.94 \pm 0.05 \hmpc$ and $\gamma= 1.79\pm 0.01$, but the
errorbars are not meaningful in this case.
The power-law provides an approximate description of the
projected correlation function, but, as emphasized by Z04,
there are notable and systematic deviations from it.
The $\chi^2/{\rm d.o.f.}$ for the power-law fit 
when using the jackknife covariance matrix is $\sim 5$.
The deviations from a power-law can be naturally explained in the HOD
framework as discussed by Z04 and in \S\ref{sec:hod} below.
Power-law fits are nonetheless useful as approximate characterizations
of the data and for facilitating the comparison to other measurements.

\begin{figure}[bp]
\plotone{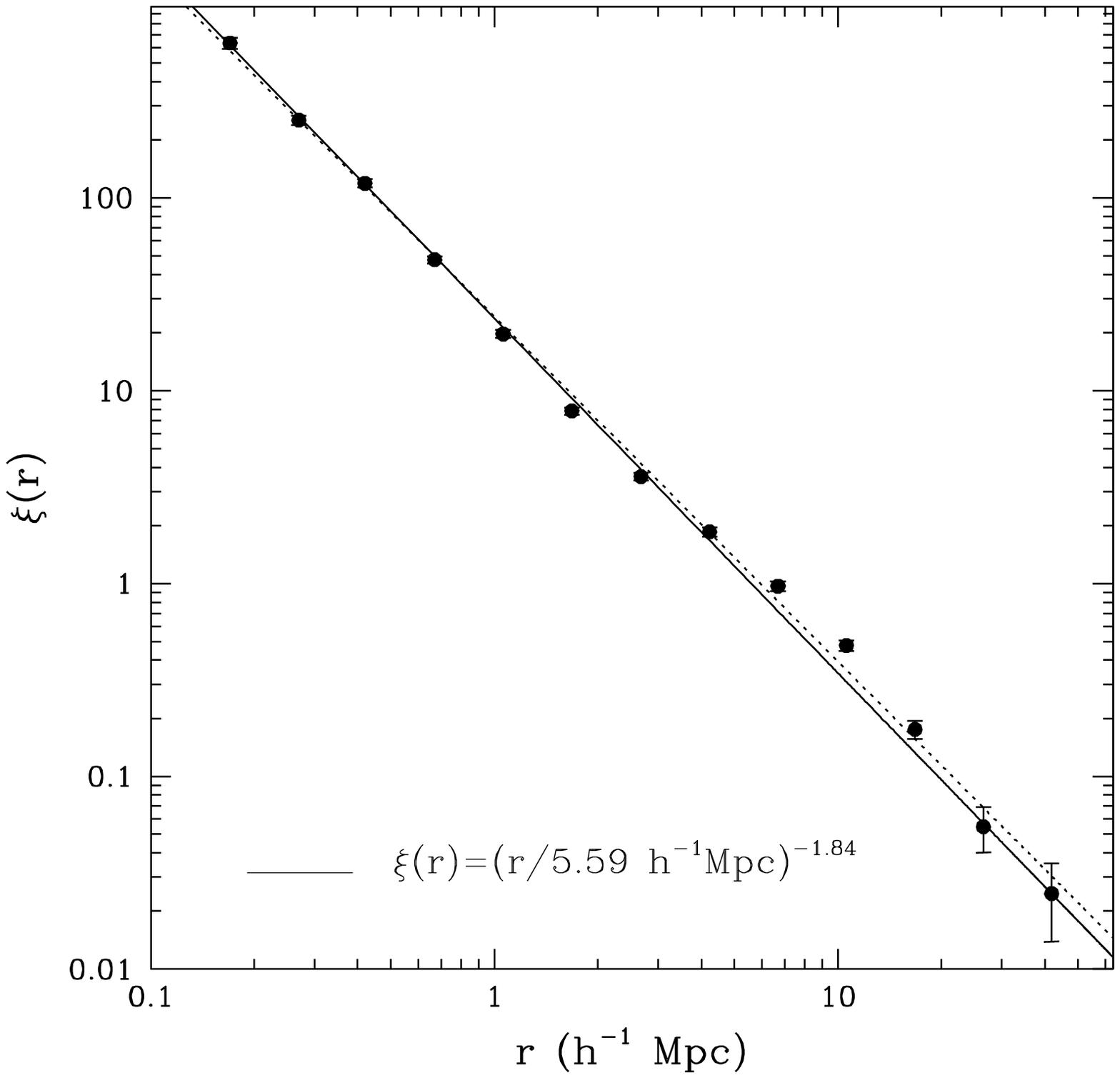}
\caption{\label{fig:xi}
The real-space correlation function $\xir$ for the flux-limited galaxy
sample, obtained from $\wrp$ as discussed in the text. The solid and
dotted lines show the corresponding power-law fits obtained by fitting 
$\wrp$ using the full covariance matrix or just the diagonal elements,
respectively.
}
\end{figure}

Figure~\ref{fig:xi} shows the real-space correlation function,
$\xir$, obtained by inverting $\wrp$ for the flux-limited sample
using equation~(\ref{eq:xi}),
independent of the power-law assumption. The lines plotted are 
the same corresponding power-law fits obtained by fitting $\wrp$.
We can see that the characteristic deviation from a power-law is 
also apparent in $\xir$,  but we still choose to do all model 
fitting to $\wrp$ because it is more accurately measured and has
better understood errors.

\subsection{Luminosity Dependence}
\label{subsec:xi_lum}

\begin{figure}[bp]
\plotone{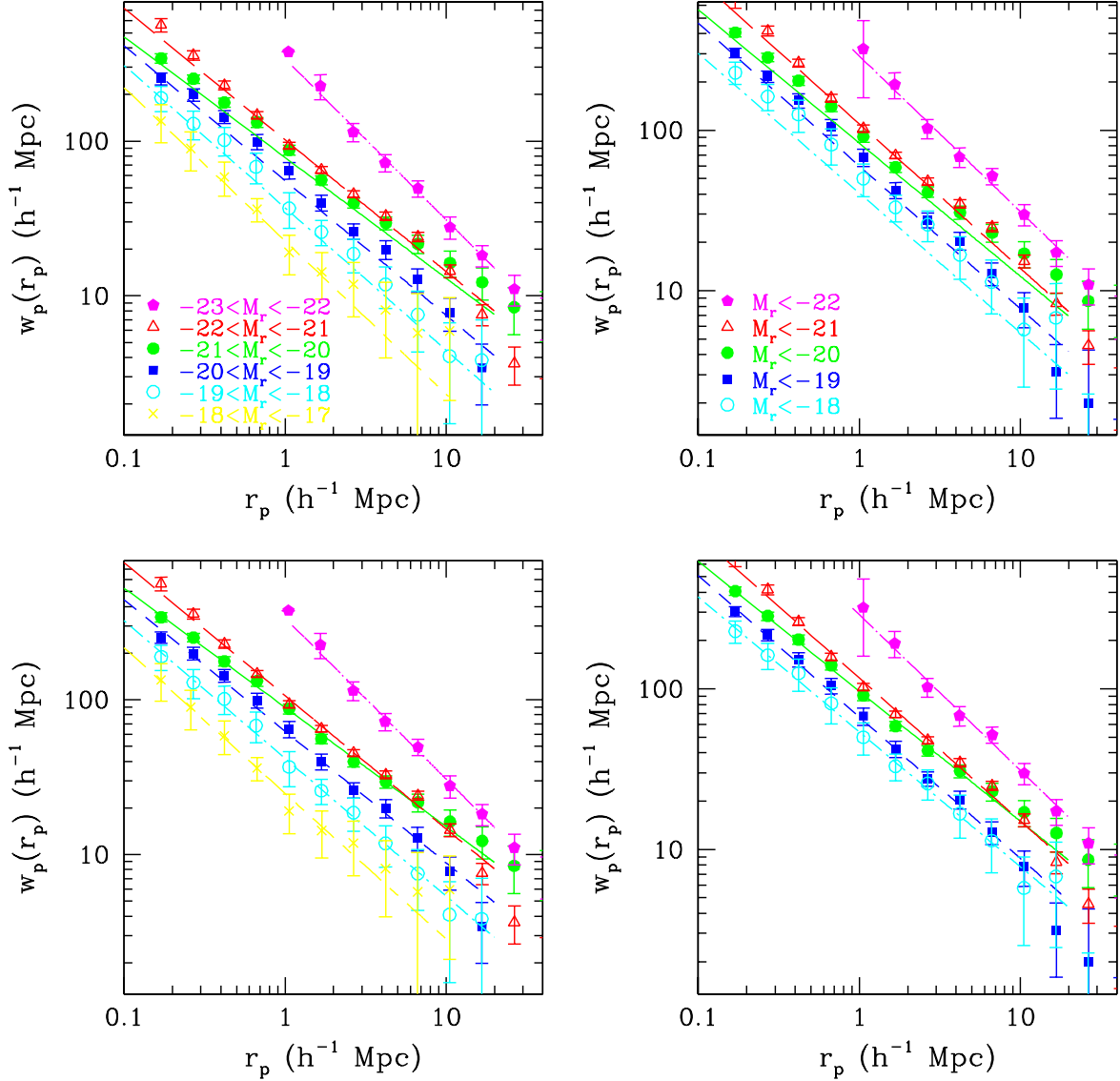}
\caption[]{\label{fig:wp_vl}
Top left: Projected galaxy correlation functions $\wrp$ for 
volume-limited samples with the indicated absolute magnitude and redshift 
ranges. Lines show power-law fits to each set of data points, using the
full covariance matrix.  Top right: Same as top left, but now the samples 
contain all galaxies brighter than the indicated absolute magnitude, i.e., 
they are defined by luminosity thresholds rather than luminosity ranges. 
Bottom panels: Same as the top panels, but now with power-law fits that use
only the diagonal elements of the covariance matrix.
}
\end{figure}

We examine the clustering dependence on luminosity using sets of 
volume-limited samples constructed from the full sample, corresponding to 
different absolute magnitude bins and thresholds. The details of the 
individual samples are given in Table~\ref{tab:vl} and Table~\ref{tab:gtmag}. 
Figure~\ref{fig:wp_vl} shows the projected correlation functions
obtained for the different volume-limited samples corresponding to 
galaxies in specified absolute magnitude bins (top-left panel) and 
to galaxies brighter than the indicated absolute magnitude (top right
panel).  For clarity, we omit some of the latter subsamples
from the plot, but we list their properties in Table~\ref{tab:gtmag},
and we will use them in \S~\ref{sec:hod}. 
The dependence of clustering on luminosity is clearly evident in
Figure~\ref{fig:wp_vl}, with the
more luminous galaxies exhibiting higher clustering amplitude.
This steady trend holds throughout the luminosity range, confirming
the early results of Z02 (but see further discussion below).
The slopes of power-law fits to $\wrp$ for the different samples 
are $\gamma \sim 1.8 - 2.0$, with a notable steepening
for the most luminous bin.  These trends are in agreement with 
an analogous study of the galaxy-mass correlation function from
weak lensing measurements in the SDSS \citep{sheldon04}.

\begin{deluxetable}{ccccccccc}
\tablewidth{0pt}
\tablecolumns{9}
\tablecaption{\label{tab:vl} Volume-limited Correlation
Function Samples Corresponding to Magnitude Ranges}
\tablehead{
$M_r$ & $z$ & $N_{\mathrm{gal}}$ & ${\bar n}$ & $r_0$ & $\gamma$ & 
$\frac{\chi^2}{d.o.f.}$ 
& $r_0^d$ & $\gamma^d$ 
}
\startdata
-23 $-$ -22 & $0.10 - 0.23$ & 3,499 & 0.005 & 
10.04 (0.37) & 2.04 (0.08) & 0.4 
& 10.00 (0.29) & 2.04 (0.08) 
\cr
-22 $-$ -21 & $0.07 - 0.16$ & 23,930 & 0.114 & 
6.16 (0.17) & 1.85 (0.03) & 3.4 
& 6.27 (0.07) & 1.86 (0.02) 
\cr
-21 $-$ -20 & $0.04 - 0.10$ & 31,053 & 0.516 & 
5.52 (0.19) & 1.78 (0.03) & 2.2 
& 5.97 (0.11) & 1.77 (0.02) 
\cr
-21 $-$ -20 & $0.04 - 0.07^*$ & 5,670 & 0.482 & 
5.02 (0.30) & 1.80 (0.05) & 1.1 
& 4.96 (0.16) & 1.86 (0.03) 
\cr
-20 $-$ -19 & $0.03 - 0.07$ & 14,223 & 0.850 & 
4.41 (0.23) & 1.87 (0.04) & 1.9 
& 4.74 (0.11) & 1.85 (0.03) 
\cr
-19 $-$ -18 & $0.02 - 0.04$ & 4,545 & 1.014 & 
3.51 (0.32) & 1.92 (0.05) & 0.9 
& 3.77 (0.17) & 1.89 (0.06) 
\cr
-18 $-$ -17 & $0.01 - 0.03$ & 1,950 & 1.209 & 
2.68 (0.39) & 1.99 (0.09) & 0.3 
& 2.83 (0.19) & 1.94 (0.11) 
\cr
\enddata
\tablecomments{All samples use $14.5 < r \la 17.77$.
${\bar n}$ is measured in units of $10^{-2}$ $h^{3}$ Mpc$^{-3}$.
$r_0$ and $\gamma$ are obtained from a fit for $w_p(r_p)$ using
the full error covariance matrix. ${r_0}^d$ and ${\gamma}^d$ are
the corresponding values when using just the diagonal elements.
Values in brackets are the fitting error.
The clipped $-21 < \Mr < -20$ sample, indicated with a $^*$, 
is confined to a limiting redshift $\zmax=0.07$ to avoid the
effects of the large supercluster at $z=0.08$ (see text).
}
\end{deluxetable}

\begin{deluxetable}{ccccccccc}
\tablewidth{0pt}
\tablecolumns{9}
\tablecaption{\label{tab:gtmag} Volume-limited Correlation
Function Samples Corresponding to Magnitude Thresholds}
\tablehead{
${M_r}^{\mathrm{max}}$ & $z^{\mathrm{max}}$ & $N_{\mathrm{gal}}$ 
& ${\bar n}$ & $r_0$ & $\gamma$ & $\frac{\chi^2}{d.o.f.}$ 
& $r_0^d$ & $\gamma^d$ 
}
\startdata
-22.0 & 0.22 &  3,626 & 0.006 & 9.81 (0.39) & 1.97 (0.08) & 0.8 
& 9.81 (0.30) & 1.97 (0.08) 
\cr
-21.5 & 0.19 & 11,712 & 0.031 & 7.70 (0.22) & 1.88 (0.03) & 1.7 
& 7.77 (0.12) & 1.88 (0.02) 
\cr
-21.0 & 0.15 & 26,015 & 0.117 & 6.24 (0.16) & 1.90 (0.02) & 4.0 
& 6.49 (0.08) & 1.89 (0.02) 
\cr
-20.5 & 0.13 & 36,870 & 0.308 & 5.81 (0.15) & 1.88 (0.02) & 1.6 
& 5.98 (0.07) & 1.86 (0.02) 
\cr
-20.0 & 0.10 & 40,660 & 0.611 & 5.58 (0.20) & 1.83 (0.03) & 2.8 
& 6.12 (0.11) & 1.81 (0.02) 
\cr 
-20.0 & 0.06$^*$ & 9,161 & 0.574 & 5.02 (0.24) & 1.88 (0.04) & 0.8 
& 5.09 (0.13) & 1.90 (0.03) 
\cr
-19.5 & 0.08 & 35,854 & 1.015 & 4.86 (0.17) & 1.85 (0.02) & 2.0 
& 5.19 (0.10) & 1.85 (0.02) 
\cr
-19.0 & 0.06 & 23,560 & 1.507 & 4.56 (0.23) & 1.89 (0.03) & 1.7 
& 4.85 (0.11) & 1.88 (0.03) 
\cr
-18.5 & 0.05 & 14,244 & 2.060 & 3.91 (0.27) & 1.90 (0.05) & 1.0
& 4.37 (0.15) & 1.92 (0.04) 
\cr
-18.0 & 0.04 &  8,730 & 2.692 & 3.72 (0.30) & 1.87 (0.05) & 1.9 
& 4.39 (0.20) & 1.84 (0.06) 
\cr
\enddata
\tablecomments{All samples use $10.0 < r <17.5$.
$z^{\mathrm{min}}$ for the samples is $0.02$.
${\bar n}$ is measured in units of $10^{-2}$ $h^{3}$ Mpc$^{-3}$.
$r_0$ and $\gamma$ are obtained from a fit for $w_p(r_p)$ using
the full error covariance matrix. ${r_0}^d$ and ${\gamma}^d$ are
the corresponding values when using just the diagonal elements.
Values in brackets are the fitting errors.
The clipped $\Mr < -20$ sample, indicated with a $^*$, 
is confined to a limiting redshift $\zmax=0.06$ to avoid the
effects of the large supercluster at $z=0.08$ (see text).
}
\end{deluxetable}

The lines plotted in the top panels are power-law fits to the 
measurements obtained using the full error covariance matrix. The fitted 
values of $r_0$ and $\gamma$ are specified in Tables~\ref{tab:vl} 
and~\ref{tab:gtmag}. 
Note that some of the fits, particularly for the smaller volume
subsamples, lie systematically below the data points.  While initially
counter-intuitive, these fits do indeed have lower $\chi^2$ than 
higher amplitude power-laws that pass closer to the points.
This kind of behavior is not uncommon when fitting strongly correlated
data points, and in tests with the PTHalo mock catalogs 
on some small volume samples we find similar
results using mock catalog covariance matrices in place of jackknife
covariance matrices.  We thus have no reason to think that these are
not the ``best'' values of $r_0$ and $\gamma$, in a statistical sense.
However, the covariance matrices do have noise because they are estimated
from the finite data samples themselves.  For good measure, the
bottom panels in Figure~\ref{fig:wp_vl} show the same measurements,
but now with fits using only the diagonal components of the jackknife
covariance matrix.  These fits pass through the points in agreement
with the ``chi-by-eye'' expectation. The best-fitting values of 
$r_0$ and $\gamma$ for these cases are quoted in the tables, but the 
$\chi^2$ values are no longer meaningful as goodness-of-fit estimates.

The $-21<\Mr<-20$ luminosity-bin sample and the $\Mr<-20$ 
luminosity-threshold sample in Figure~\ref{fig:wp_vl} exhibit
anomalously high $\wrp$ at large separations, with a flat slope
at $\rp \ga 3\hmpc$ that is clearly out of line with other samples.
We believe that this anomalous
behavior is a ``cosmic variance'' effect caused by an enormous
supercluster at $z\sim 0.08$, slightly inside the limiting
redshift $z_{\rm max}=0.10$ of these two samples.
This ``Sloan Great Wall'' at $\alpha \sim 200^\circ$, 
$\delta \sim 0^\circ$ is the largest structure detected in the SDSS
to date, or, indeed, in any galaxy redshift survey 
(see \citealt{gott03}).  It has an important effect on the 
$\zmax=0.1$ samples, but no effect on the fainter samples,
which have $\zmax < 0.08$, and little effect on brighter samples,
which cover a substantially larger volume.  If we repeat our
analysis excluding the supercluster region in an {\it ad hoc}
fashion, then the large scale $\wrp$ amplitude drops for these
two $\zmax=0.10$ samples but changes negligibly for other samples.
Because our jackknife subsamples are smaller than the supercluster,
the jackknife errorbars do not properly capture the large variance
introduced by this structure.
We note that while this super-cluster is certainly a striking feature in 
the data, its existence is in no contradiction to concordance cosmology: 
preliminary tests with the PTHalo mock catalogs reveal similar structures 
in more than $10\%$ of the cases. 

\begin{figure}[bp]
\plotone{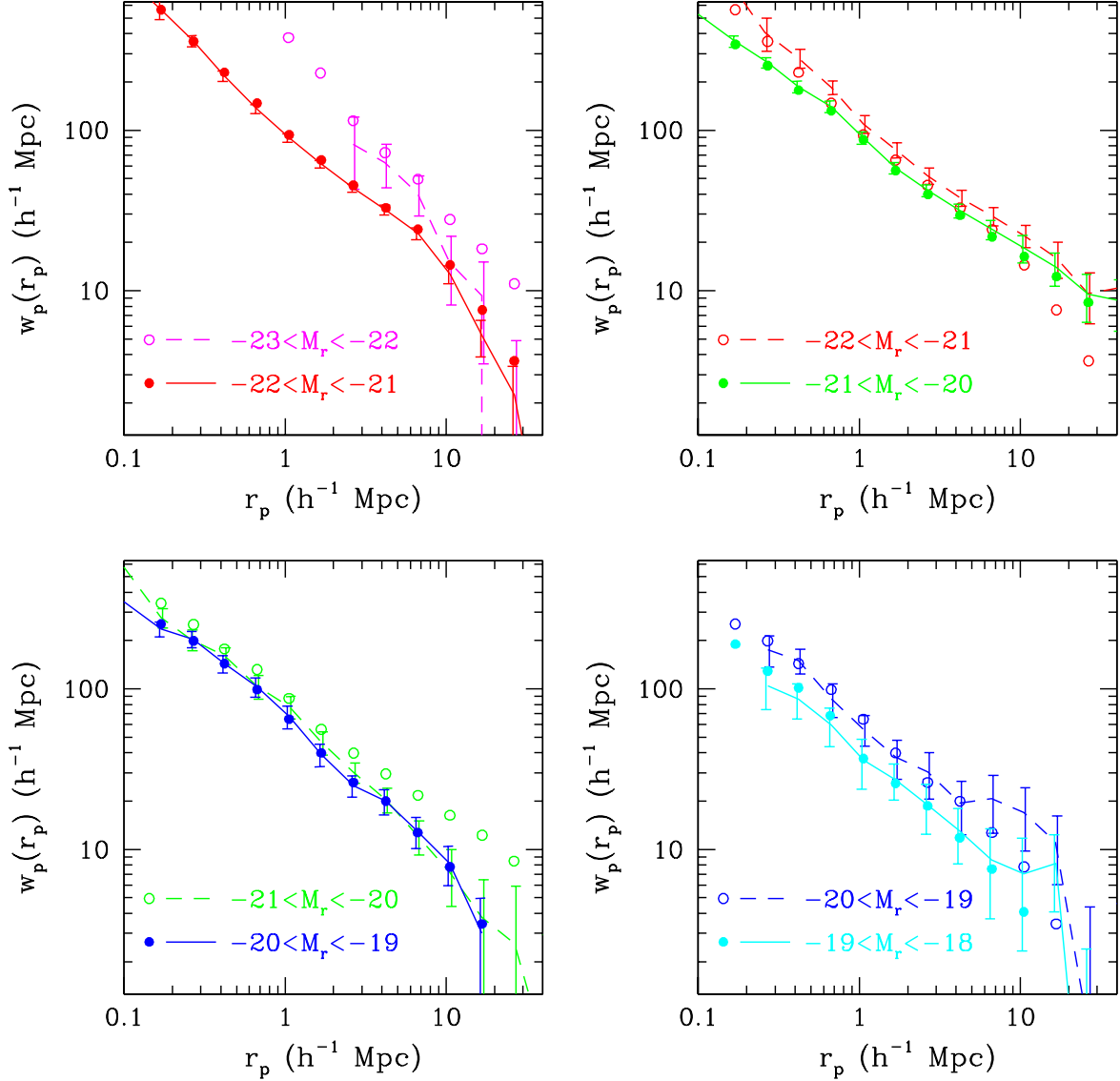}
\caption[]{\label{fig:wp_vl_z}
The impact of finite volume fluctuations on the 
measured luminosity dependence of $\wrp$.
Each panel compares the projected correlation function of two neighboring
absolute-magnitude bins.  The filled and open symbols show the results using
the maximum accessible redshift range for each bin, as in
Fig.~\ref{fig:wp_vl}.  Lines show the measurements of 
$\wrp$ restricted to the overlap redshift range in which both sets of 
galaxies are observable.  
Error bars, attached to the lines (not the points), represent
$1\sigma$ errors for these smaller, overlap samples.
}
\end{figure}

A more general cosmic variance problem is that we measure the clustering
of each galaxy luminosity subset over a different volume, and variations
in the true underlying structure could masquerade as luminosity dependence
of galaxy bias.  To test for this problem, we compare the $\wrp$ 
measurements for each pair of adjacent luminosity bins to the values
measured when we restrict the two samples to the volume where they 
overlap (and thus trace identical underlying structure; 
see the corresponding redshift ranges in Table~\ref{tab:vl}). 
In each
panel of Figure~\ref{fig:wp_vl_z}, the points show
our standard $\wrp$ measurement for the maximum volume accessible
to each luminosity bin; brighter and fainter luminosity bins are
represented by open and filled circles, respectively.  The dashed and
solid curves and errorbars show the corresponding measurements when 
the samples
are restricted to the volume where they overlap.  In the absence of
any cosmic variance, the dashed curve should pass through the open
points and the solid curve through the filled points.  This is essentially
what we see for the two faintest bins, shown in the lower right,
except for a low significance fluctuation at large scales in the overlap 
measurement for $-20 < \Mr < -19$.

Moving to the next comparison in the lower left, we see the dramatic
effect of the $z\sim 0.08$ supercluster.  When the $-21 < \Mr < -20$
sample is restricted to the overlap volume, reducing $\zmax$ from 0.10
to 0.07, its projected correlation function drops and steepens, coming
into good agreement with that of the $-20 < \Mr < -19$ sample.
Conversely, when the $-22 < \Mr < -21$ sample is restricted to $\zmax=0.1$
(dashed curve, upper right), it acquires an anomalous large separation
tail like that of the (full) $-21 < \Mr < -20$ sample.  Increasing 
the minimum redshift of this sample to $\zmin=0.10$ 
(solid curve, upper left), on the
other hand, has minimal impact, suggesting that the influence of the
supercluster is small for the full $-22 < \Mr < -21$ sample, which
extends from $\zmin=0.07$ to $\zmax=0.16$.  The large scale amplitude of
the $-23 < \Mr < -22$ sample drops when it is restricted to $\zmax=0.16$,
but this drop again has low significance because of the limited overlap
volume, which contains only about 1000 galaxies in this luminosity range.
In similar fashion, the overlap between the $-19 < \Mr < -18$ and
$-18 < \Mr < -17$ volumes is too small to allow a useful cosmic variance
test for our faintest sample.  We have carried out the volume overlap
test for the luminosity-threshold samples in Table~\ref{tab:gtmag}, and
we reach a similar conclusion to that for the luminosity bins: the
$\Mr<-20$ sample, with $\zmax=0.10$, is severely affected by the
$z\sim 0.08$ supercluster, but other samples appear robust to changes
in sample volume.  

Given these results, we have chosen to use the measurements from the
$-21 < \Mr < -20$ sample limited to $\zmax=0.07$ (the same limiting
redshift as for $-20 < \Mr < -19$) and the $\Mr<-20$ sample limited to 
$\zmax=0.06$ (same as $\Mr < -19$) in our subsequent analyses.
We list properties of these reduced samples in Tables~\ref{tab:vl} 
and~\ref{tab:gtmag}. 
This kind of data editing should become unnecessary as the SDSS grows
in size, and even structures as large as the Sloan Great Wall are 
represented with their statistically expected frequency.
As an additional test of
cosmic variance effects, we have measured $\wrp$ separately in each
of the three main angular regions of the survey 
(see Figure~\ref{fig:aitoff}), and despite significant fluctuations
from region to region, we find the same continuous trend of clustering
strength with luminosity in each case.

\begin{figure}[bp] 
\plotone{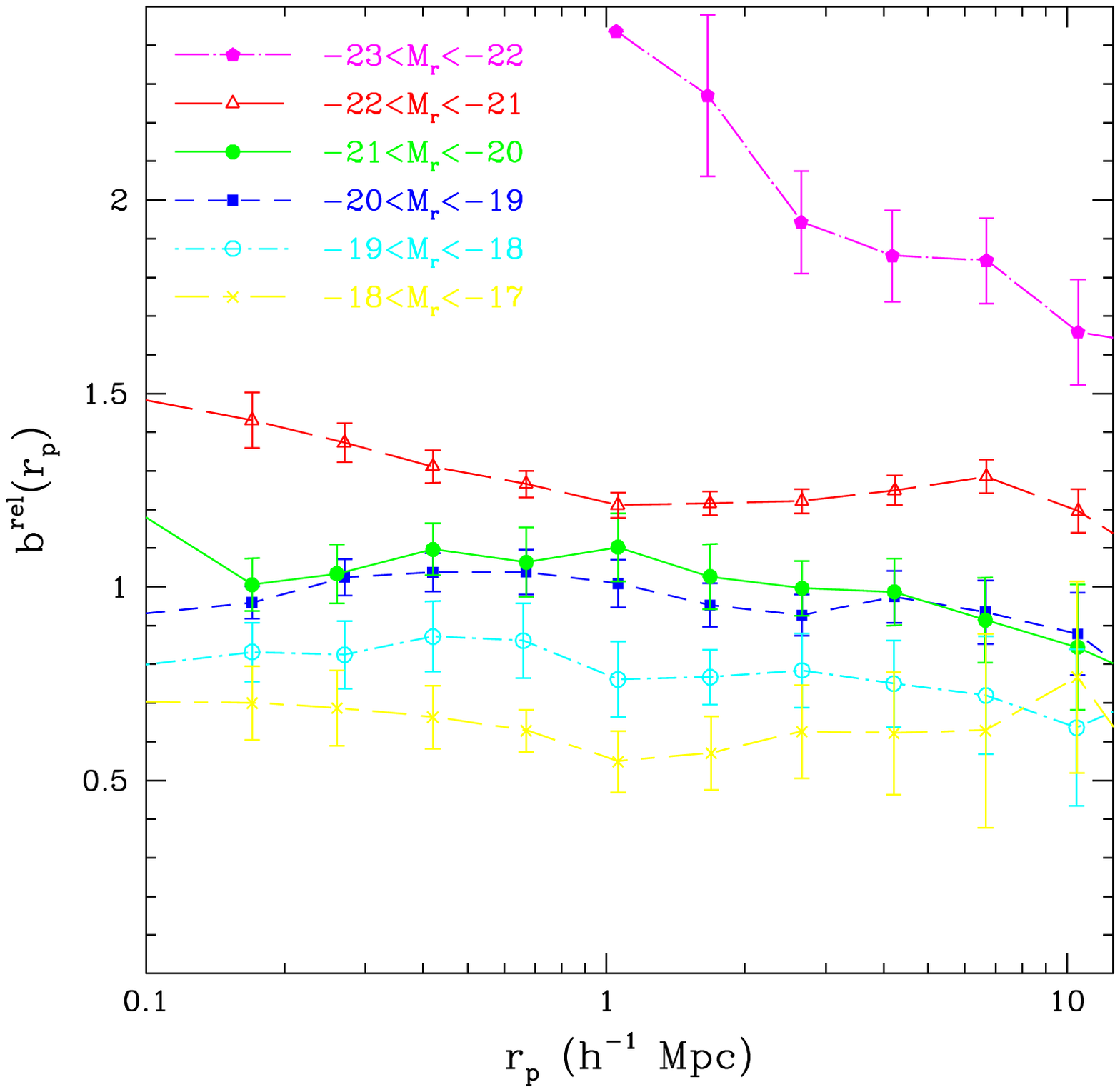}
\caption[]{\label{fig:bias_vs_R}
Relative bias factors as a function of separation $r_p$ for samples
defined by luminosity ranges. Bias factors are defined by 
$b_{\rm rel}(\rp)\equiv [\wrp/w_{p,{\rm fid}}(\rp)]^{1/2}$
relative to a fiducial
power-law corresponding to $\xir=(r/5\hmpc)^{-1.8}$.
}
\end{figure}

Figure~\ref{fig:bias_vs_R} presents the luminosity dependence in the
form of relative bias functions, 
$b_{\rm rel}(\rp)\equiv [\wrp/w_{p,{\rm fid}}(\rp)]^{1/2},$
where $\wrp$ is the measured result for a luminosity bin and 
$w_{p,{\rm fid}}(\rp)$ is the projected correlation function corresponding
to $\xi(r)=(r/5.0\hmpc)^{-1.8}$.  We take a power-law rather than
a given sample as our fiducial so that measurement noise does
not propagate into the definition of bias functions.
The bias factors increase steadily with luminosity, and they
are roughly scale independent, with $\sim 10-20\%$ fluctuations,
for all samples except the brightest one.  The $-23<\Mr<-22$
galaxies have a $\wrp$ slope steeper than $-1.8$, so their bias
relative to fainter galaxies increases with decreasing $\rp$.

\begin{figure}[bp]
\plotone{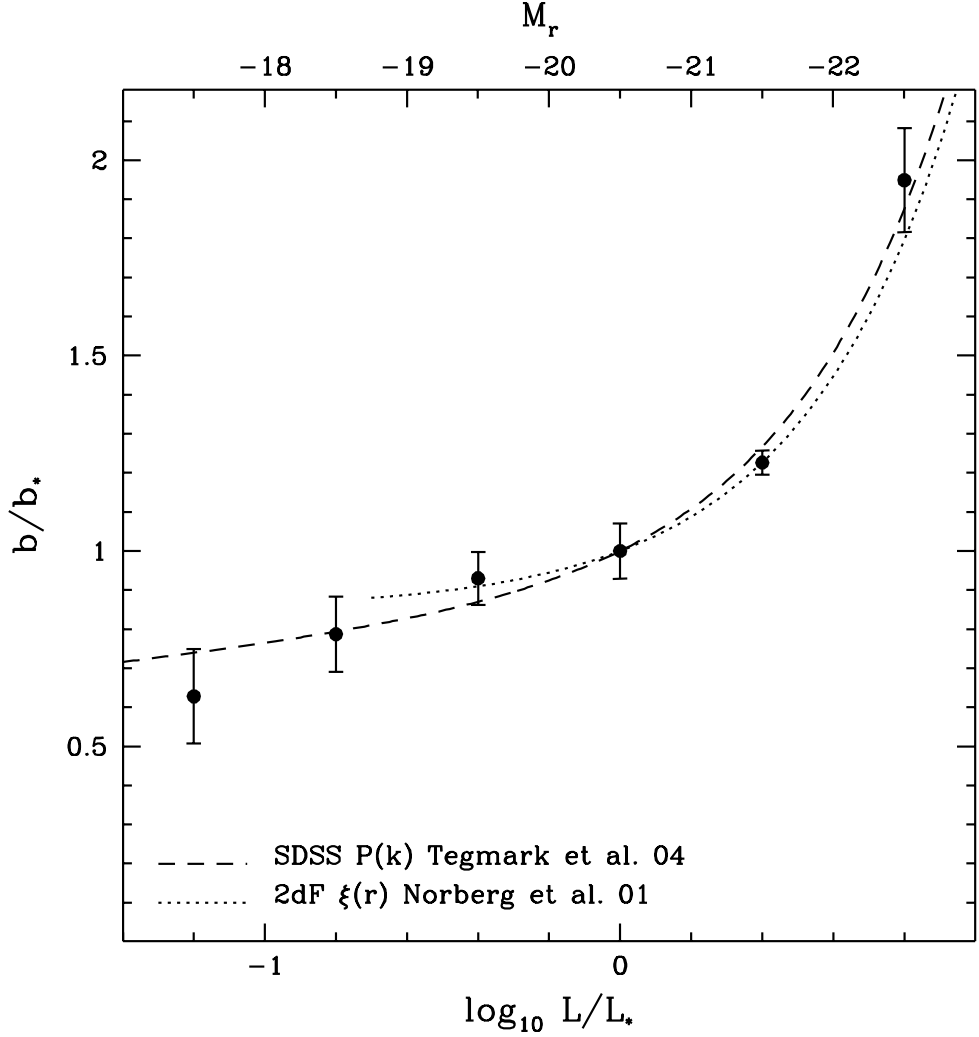}
\caption[]{\label{fig:bias_vs_L}
Relative bias factors for samples defined by luminosity ranges.
Bias factors are defined by the relative amplitude of the  $\wrp$ estimates 
at a fixed separation of $r_p=2.7 \hmpc$ and are normalized by the
$-21 < M_r < -20$ sample ($L\approx L_*$). The dashed curve is a
fit obtained from measurements of the SDSS power spectrum, 
$b/b_*=0.85+0.15L/L_*-0.04(M-M_*)$ \citep{tegmark04a}, and the dotted
curve is a fit to similar $\wrp$ measurements in the 2dF survey, 
$b/b_*=0.85+0.15L/L_*$ \citep{norberg01}.
}
\end{figure}

To summarize our results and compare to previous work, we take
the bias factors at $\rp=2.7\hmpc$ and divide them by the bias
factor $b_*$ of the $-21<\Mr<-20$ sample, which has luminosity
$L\approx L_*$.  We choose $2.7\hmpc$ because it is out of the
extremely nonlinear regime and all samples are well measured there;
one can see from Figure~\ref{fig:bias_vs_R} that other choices
would give similar but not identical results.  
Figure~\ref{fig:bias_vs_L} plots $b/b_*$ vs. $\log L/L_*$, where
the solid points show the results from our $\wrp$ measurements.
The dashed curve is the fit to SDSS results by \cite{tegmark04a}, 
where bias factors are derived from the galaxy power spectrum 
at wavelengths $2\pi/k \sim 100\hmpc$, in the linear
(or at least near-linear) regime.
The $\wrp$ and $P(k)$ results agree remarkably well, despite being
measured at very different scales.
The dotted curve in Figure~\ref{fig:bias_vs_L} shows the
fit of \cite{norberg01}, based on $\wrp$ measurements of
galaxies with $\log L/L_* > -0.7$ in the 2dFGRS.
Agreement is again very good, over the range of the \cite{norberg01}
measurements, with all three relative bias measurements (from two independent 
data sets) showing that the bias factor increases sharply for $L > L_*$,
as originally argued by \cite{hamilton88}.
At luminosities $L \la 0.2 L_*$, the \cite{tegmark04a} formula
provides a better fit to our data than the extrapolation of the
\cite{norberg01} formula.

\subsection{Color Dependence}
\label{subsec:xi_color}

In addition to luminosity, the clustering of galaxies is known to depend
on color, spectral type, morphology, and surface brightness.
These quantities are strongly correlated with each other, 
and in Z02 we found that dividing galaxy samples based on any of these
properties produces similar changes to $\wrp$.  This result holds
true for the much larger sample investigated here.
For this paper, we have elected to focus on color, since it is
more precisely measured by the SDSS data than the other quantities.
In addition, \cite{blanton03e} find that luminosity and color are
the two properties most predictive of local density, and that any
residual dependence on morphology or surface brightness at
fixed luminosity and color is weak.

\begin{figure}[bp] 
\plotone{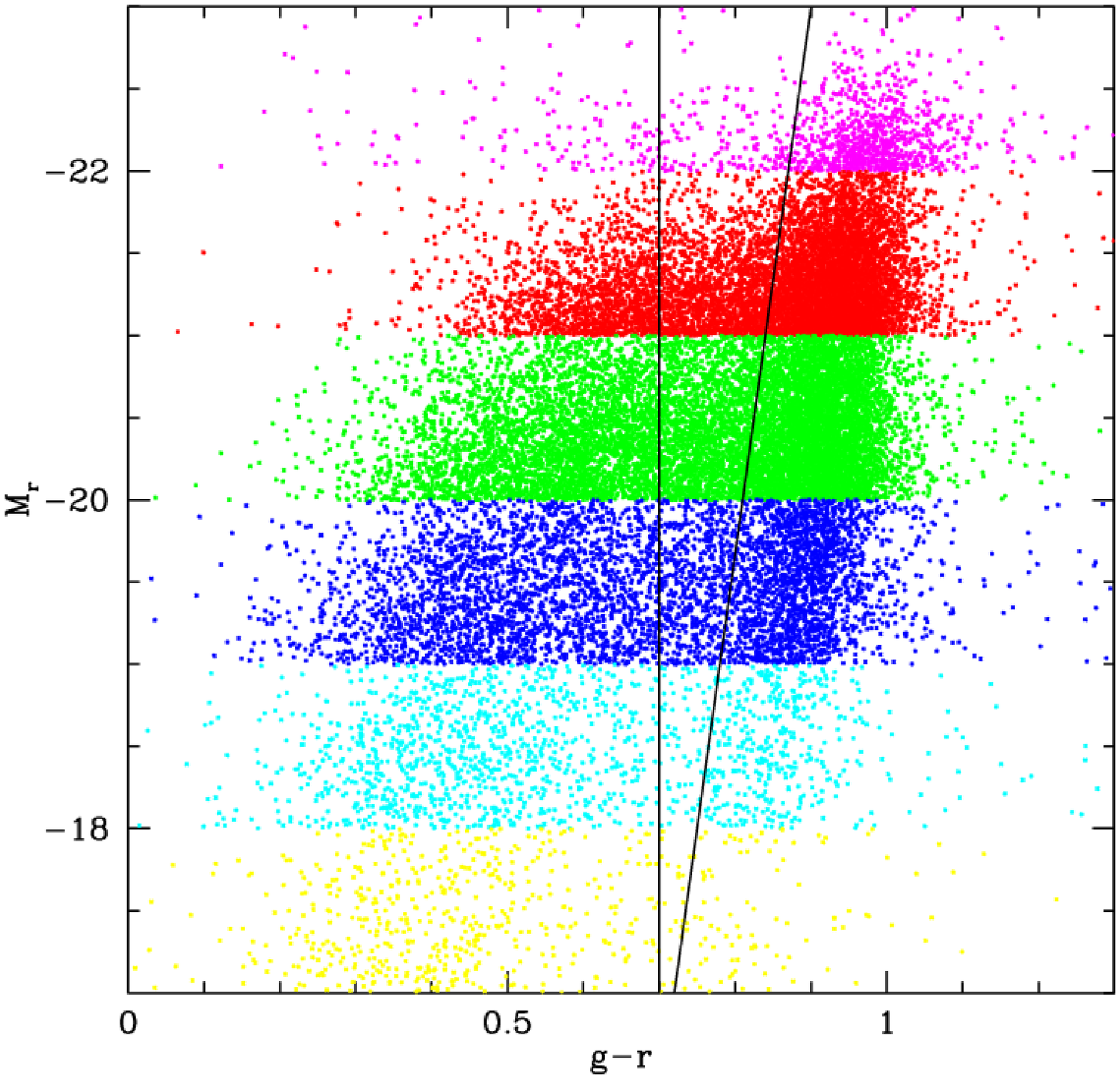}
\caption[]{\label{fig:colormag}
K-corrected $\gr$ color vs. absolute magnitude for all galaxies comprising
our volume-limited luminosity bins samples. A clear color-magnitude trend 
is evident. The vertical line demarcates a simple cut at $\gr=0.7$, while
the tilted line indicates the luminosity-dependent color cut
that we adopt for the analyses in \S\S\ref{subsec:xi_color} 
and~\ref{subsec:hod_color}.
}
\end{figure}

Figure~\ref{fig:colormag} shows a 
color-magnitude diagram constructed from a random subsampling of
the volume-limited samples used in our analysis.  The gradient along 
each magnitude bin reflects the fact that in each volume faint galaxies 
are more common than bright ones, while the offset from bin to bin reflects 
the larger volume sampled by the brighter bins. 
While we used $g-r=0.7$ for the color division of the flux-limited
sample (Fig.~\ref{fig:xsirpi_br}), in this section we adopt the
tilted color cut shown in Figure~\ref{fig:colormag}, which better
separates the E/S0 ridgeline from the rest of the population.
It has the further
advantage of keeping the red:blue ratio closer to unity in our
different luminosity bins, though it remains the case that 
red galaxies predominate in bright bins and blue galaxies in faint ones
(with roughly equal numbers for the $L_*$ bin).
The dependence of the color separation on luminosity has been
investigated more quantitatively by \cite{baldry04}.

\begin{figure}[bp] 
\plotone{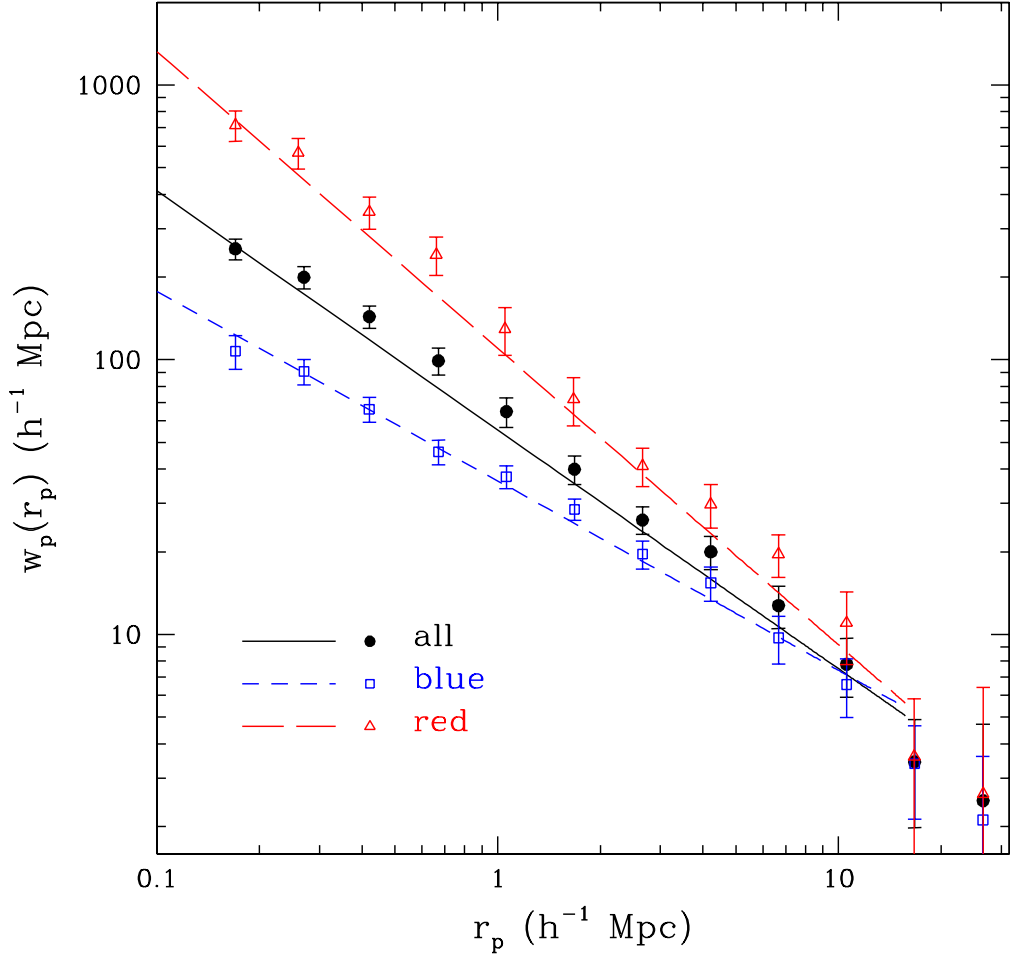}
\caption[]{\label{fig:wrp_color}
Projected correlation function of the full volume-limited sample
of all galaxies with $-20 < M_r < -19$ and of the blue and red galaxies in this
sample, with the color cut indicated by the tilted line in 
Fig.~\ref{fig:colormag}.  Lines show the best-fit power-laws.
}
\end{figure}

Figure~\ref{fig:wrp_color} shows, as a representative case, the projected 
correlation function obtained with the tilted color division for the 
$-20<M_r<-19$ volume-limited sample. 
The red galaxy $\wrp$ has a steeper slope and a higher amplitude
at all $r_p \la 10\hmpc$; at $r_p > 10\hmpc$ the two correlation
functions are consistent within the (large) statistical errors.
Power-law fits for these samples using the full covariance 
matrix give $r_0=5.7\hmpc$ and $\gamma=2.1$ for the red sample, and
$r_0=3.6\hmpc$ and $\gamma=1.7$ for the blue sample. 
The change in slope contrasts with the results for the luminosity dependence,
where (with small variations) the slope remains
fairly constant and only the clustering amplitude changes.
The results for the color dependence in the other luminosity bins, 
and in luminosity-threshold samples and the flux-limited sample,
are qualitatively similar (see Figures~\ref{fig:color_hod_m21}
and~\ref{fig:color_lumbin} below). The behavior in Figure~\ref{fig:wrp_color}
is strikingly similar to that found by 
\citeauthor{madgwick03} (\citeyear{madgwick03}, Fig. 2) for 
flux-limited samples of active and passive galaxies in the 2dFGRS,
where spectroscopic properties are used to distinguish galaxies
with ongoing star formation from those without.

\begin{figure}[bp] 
\plotone{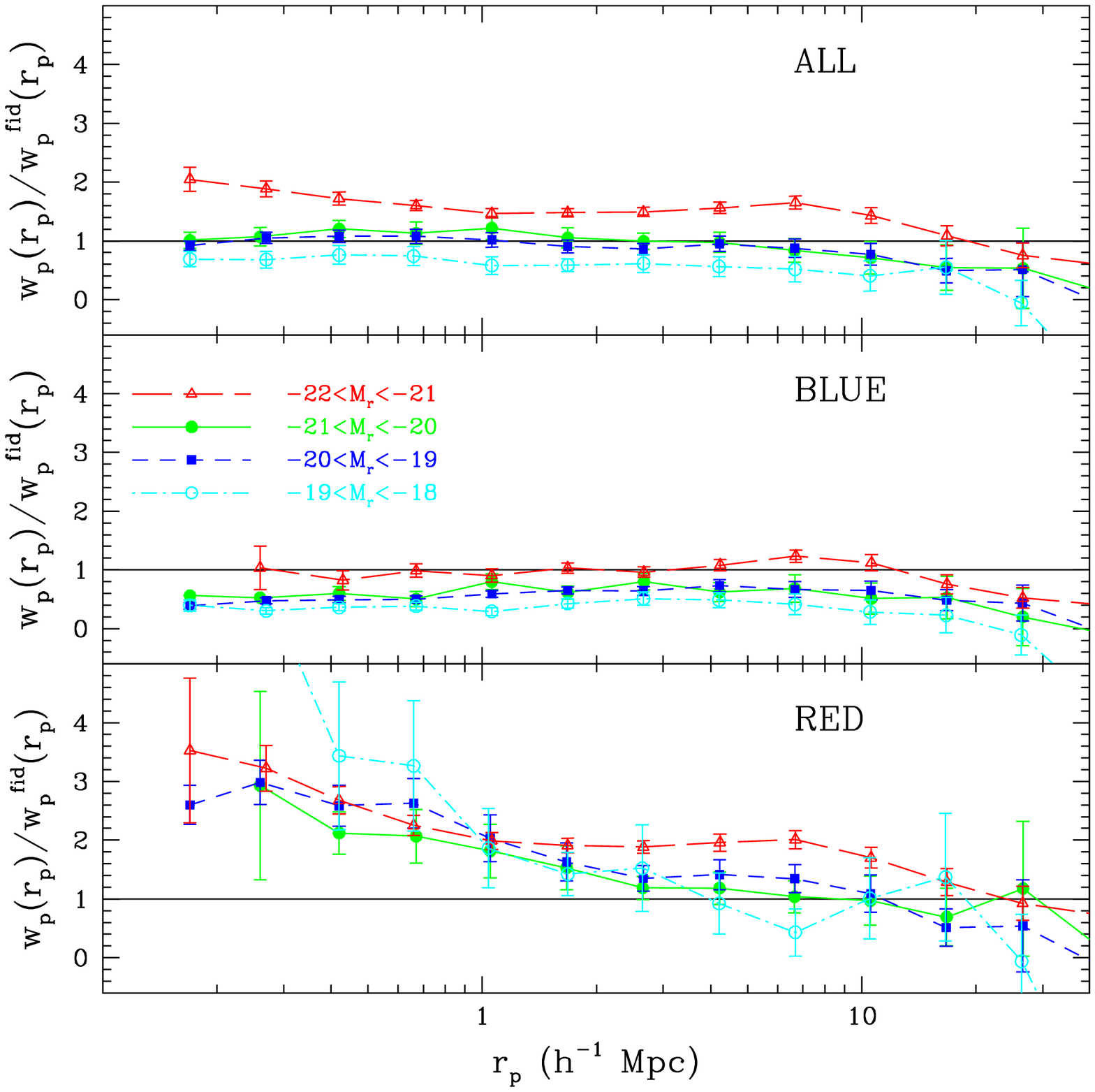}
\caption[]{\label{fig:wrp_lumcol}
Luminosity and color dependence of the 
galaxy correlation function.  Upper, middle, and lower panels
show projected correlation functions of all galaxies, blue galaxies,
and red galaxies, respectively, in the indicated absolute-magnitude 
ranges.  All projected correlation functions are divided by a
fiducial power-law corresponding to $\xi(r)=(r/5\hmpc)^{-1.8}$.
}
\end{figure}

Figure~\ref{fig:wrp_lumcol} shows the luminosity dependence of $\wrp$
separately for blue galaxies (middle panel) and red galaxies 
(bottom panel).  We divide $\wrp$ by a fiducial power-law corresponding
to $\xi(r)=(r/5.0\hmpc)^{-1.8}$, and we show the luminosity dependence
for the full (red and blue) samples again in the top panel
(repeating Fig.~\ref{fig:bias_vs_R}, but here showing $b^2$
instead of $b$).  We focus on the four central luminosity bins,
since the $-18<\Mr<-17$ sample is too small once it is divided
by color, and the $-23<\Mr<-22$ sample consists mainly of red
galaxies alone.  Blue galaxies exhibit a roughly scale-independent
luminosity dependence reminiscent of the full sample, but even
the most luminous blue galaxy bin only has $r_0 \approx 5\hmpc$.
The red galaxies are always more clustered than the fiducial
power-law on small scales, regardless of luminosity, and
the luminosity dependence for red galaxies is more complex.
At large scales, the luminous red galaxies are the most strongly
clustered, but at small scales faint red galaxies have the highest
and steepest correlation function, with $\wrp$ of all samples
intersecting at $\rp \sim 1\hmpc$.
All of these trends would also hold if we adopted a fixed color
cut at $g-r=0.7$ instead of the tilted color cut used in this section.

Figure~\ref{fig:wrp_lumcol} demonstrates that the luminosity and color
dependence of the galaxy correlation function is not trivially separable,
nor is the luminosity dependence a simple consequence of the color
dependence (or vice versa).  The effect of color is in some sense
stronger, since red galaxies of any luminosity are more clustered than
blue galaxies of any luminosity (at least for $r_p \la 3\hmpc$).  However,
luminosity dependence of clustering
remains evident within the red and blue populations separately.
The overall appearance of Figure~\ref{fig:wrp_lumcol} is roughly
like that of Figure~7 of \cite{norberg02}, who divide their samples into
early and late spectral types, but we find systematically different
slopes for red and blue galaxies, a steadier luminosity trend
for blue galaxies, and a much more noticeable scale dependence of
relative bias for red galaxies.  These differences could reflect the 
difference in the sample definitions (color versus spectral type) and 
overall selection ($r$-band versus $b_J$-band).
The strong small scale clustering
of faint red galaxies in our sample agrees with the results for
faint early type galaxies in \cite{norberg02}, and with the
results of \cite{hogg03}, who find that these galaxies reside in
denser environments than red galaxies of intermediate luminosity.
\cite{kayo04} find qualitatively similar trends for the dependence
of the {\it redshift}-space two-point correlation amplitude on
luminosity, color, and morphological type.

\section{HOD Modeling of the SDSS Galaxy Clustering}
\label{sec:hod}

\subsection{HOD Framework and Formalism}
\label{subsec:hod}

We now turn to physical interpretation of these results using the 
Halo Occupation Distribution (HOD) framework, which describes
the bias between galaxies and mass in terms of
the probability distribution  $P(N|M)$ that a halo of virial mass $M$ 
contains $N$ galaxies of a given type, together with prescriptions for
the relative bias of
galaxies and dark matter within virialized halos.
Because the real-space 
correlation function describes a limited (but important) subset of the 
information encoded in galaxy clustering, we will have to fit restricted 
HOD models with a small number of free parameters, and we will assume that
the underlying cosmological model is known {\it a priori}. However, relative 
to power-law fits,
HOD modeling fits the data more accurately (in most cases) and
in a way that we consider more physically informative.
In the longer run, constraints from multiple galaxy clustering
statistics can be combined to test the HOD predictions of galaxy
formation models
(e.g., \citealt{kauffmann97, kauffmann99,benson00,somerville01,yoshikawa01,
white01,berlind03,kravtsov04}), and to obtain simultaneous constraints
on cosmological parameters (see discussions by 
\citealt{berlind02,zheng02,weinberg02,bosch03b}; Z.\ Zheng \& D.\ Weinberg
2005, in preparation; and an initial application to SDSS data by 
\citealt{abazajian04b}).

We assume a spatially flat $\Lambda$CDM cosmological model with matter 
density parameter $\Omega_m=0.3$. For the matter fluctuation power spectrum, 
we adopt the parameterization of \citet{efstathiou92} 
and assume that the spectral 
index of the inflationary power spectrum is $n_s=1$, the rms matter 
fluctuation (linearly evolved to $z=0$) at a scale of 8$\hMpc$ is 
$\sigma_8=0.9$, and the shape parameter is $\Gamma=0.21$.
These parameters are in good agreement with joint analyses
of CMB anisotropies and the 2dFGRS or SDSS galaxy power spectrum
\citep{percival03,spergel03,tegmark04b} or with a more recent
analysis that incorporates constraints from the SDSS Ly$\alpha$
forest and galaxy-galaxy lensing \citep{seljak04a}.
We have verified that our results do not change significantly
if we use CMBFAST \citep{seljak96} to compute the linear theory 
power spectrum instead of the \cite{efstathiou92} form.

We focus first on luminosity-threshold samples, mainly because the
theoretical predictions for HODs have been studied more extensively
for samples defined by mass or luminosity thresholds
(e.g., \citealt{seljak00,white01,yoshikawa01,berlind03,kravtsov04,zheng04}).
The larger galaxy numbers in luminosity-threshold samples also allow
higher precision $\wrp$ measurements.  Our adopted
HOD parameterization is motivated by 
Kravtsov et al.'s (2004) recent work on substructures in high-resolution 
dissipationless simulations. They find that when the HOD is divided into 
contributions of central and satellite objects, it assumes a simple form. 
For a subhalo sample above a threshold in maximum circular velocity 
(known empirically to
correlate with luminosity), the mean occupation 
number for central substructures can be modeled as a step function, i.e., 
$\langle \Ncen\rangle=1$ for halos with mass $M\geq\Mmin$ and 
$\langle \Ncen\rangle=0$ for 
$M<\Mmin$, while the distribution of satellite substructures can be well 
approximated by a Poisson distribution with the mean following a power-law, 
$\langle \Nsat\rangle=(M/M_1)^\alpha$, with $\alpha\approx 1$. This way of 
separating central and satellite substructures naturally explains both the
general shape of the mean occupation function $\Navg$ and, more
importantly, the transition 
from sub-Poisson fluctuations at low occupation number to Poisson 
fluctuations at high occupation number found in semi-analytic and numerical
galaxy formation models (e.g., \citealt{benson00,berlind03}).  
\cite{zheng04b} show that the \cite{kravtsov04} formulation also provides
a good description of results from the semi-analytic models and 
hydrodynamic simulations.

As implemented here, this HOD formulation
has three free parameters: $\Mmin$, the minimum halo mass for galaxies above
the luminosity threshold, $M_1$, the mass of a halo that on average hosts 
one {\it satellite} galaxy above the threshold, and $\alpha$, the power-law
slope of the satellite mean occupation function. One of these, which we take
to be $\Mmin$, is fixed by matching the observed space density of the sample,
leaving $M_1$ and $\alpha$ as free parameters to fit $w_p(r_p)$. This 
parameterization thus has the same number of adjustable degrees of freedom 
as an ($r_0$, $\gamma$) power-law, allowing a fair comparison of goodness of 
fit. However, this parameterization is not a unique choice (we discuss some 
variations below), and achieving a fully accurate fit to the predictions of 
galaxy formation models requires additional parameters to describe the shapes 
of the low mass cutoff for central and satellite galaxies. 
The HOD parameterization adopted in Z04, 
with the mean occupation function changing from a plateau of $\Navg = 1$ to a 
power-law above a given halo mass, can be regarded as a simplified version of 
the one used in this paper.

In halo-based calculations, the two-point correlation function $\xi(r)$ is
decomposed into two components (see, e.g., \citealt{zheng04}),
\begin{equation}
\label{eqn:1h2h}
\xi(r)=[1+\xis(r)]+\xid(r), 
\end{equation}
where the one-halo term $\xis(r)$ (dominant at small scales) and the two-halo 
term $\xid(r)$ (dominant at large scales) represent contributions by galaxy
pairs from the same halos and from different halos, respectively.  
The ``1+'' in equation~(\ref{eqn:1h2h}) 
arises because the total number of pairs
(proportional to $1+\xi$) is the sum of the number of one-halo and two-halo
pairs (proportional to $1+\xis$ and $1+\xid$).
Our computations of 
these two terms follow those in Z04 and \citet{zheng04}, as briefly reviewed 
below. 

We calculate the one-halo term in real space through (e.g., 
\citealt{berlind02})
\begin{equation}
\label{eqn:1halo}
1+\xis(r)=\frac{1}{2\pi r^2\ngavg^2}
              \intdn\frac{\langle N(N-1)\rangle_M}{2}
              \frac{1}{2\Rvir(M)} F^\prime\left(\frac{r}{2\Rvir}\right),
\end{equation}
where $\ngavg$ is the mean number density of galaxies of the given sample, 
$dn/dM$ is the halo mass function (\citealt{sheth99,jenkins01}), 
$\langle N(N-1)\rangle_M/2$ is the average number of galaxy pairs in a halo 
of mass $M$, and $F(r/2\Rvir)$ is the cumulative radial distribution of 
galaxy pairs. For luminosity-threshold samples, one galaxy is always 
assumed to reside at the center of a halo. With the separation of central 
and satellite galaxies, $F^\prime(x)$ is then the pair-number weighted average
of the central-satellite pair distribution $F^\prime_{\rm cs}(x)$ and the
satellite-satellite pair distribution $F^\prime_{\rm ss}(x)$ (see, e.g.,
\citealt{berlind02,yang03}), 
\begin{equation}
\label{eqn:dFdx}
        \frac{\langle N(N-1)\rangle_M}{2}F^\prime(x) =
        \langle \Ncen\Nsat \rangle_M F^\prime_{\rm cs}(x) 
      + \frac{\langle \Nsat(\Nsat-1)\rangle_M}{2}F^\prime_{\rm ss}(x).
\end{equation}
For our parameterization, the occupation number of satellite galaxies follows
a Poisson distribution, which implies that 
$\langle \Nsat (\Nsat -1)\rangle=\langle \Nsat\rangle^2$. 
In cases where we allow a smooth cutoff in $\Ncen$, we further assume that 
$\langle \Ncen\Nsat \rangle_M = \langle \Ncen\rangle_M\langle \Nsat 
\rangle_M$, but $\Nsat \ll 1$ when $\Ncen$ is significantly below one in any 
case. The central-satellite galaxy pair distribution, $F^\prime_{\rm cs}(x)$,
is just the normalized radial distribution of galaxies. In this paper, we 
assume that the satellite galaxy distribution follows the dark matter 
distribution within the halo, which we describe by a spherically
symmetric NFW profile 
(\citealt{NFW95,NFW96,NFW97}) truncated at the virial radius
(defined to enclose a mean overdensity of 200).  
The satellite-satellite galaxy pair distribution $F^\prime_{\rm ss}(x)$ is 
then the convolution of the NFW profile with itself (see \citealt{sheth01a}).
For the dependence of NFW halo concentration on halo mass, we use the 
relation given by \citet{bullock01}, after modifying it to be consistent 
with our slightly different definition of the halo virial radius.

On large scales, the two-halo term is a weighted
average of halo correlation functions, where the weight is
proportional to the halo number density times the mean galaxy occupation.
On intermediate scales, one must also convolve with the finite
halo size, and it is easier to do the calculation in Fourier space 
and transform it to obtain the real space 
correlation function. To achieve the accuracy needed to model 
the SDSS data, we improve upon the original calculations of
\cite{seljak00} and \cite{scoccimarro01} by taking into account the 
nonlinear evolution of matter clustering \citep{smith03}, halo exclusion,
and the scale-dependence of the halo bias factor 
(see Z04 and \citealt{zheng04} for details).  

We project $\xi(r)$ to obtain $w_p(r_p)$ using the first part of 
equation~(\ref{eq:wp2}) and setting $r_{\rm max}=40\hMpc$. Under the 
plane-parallel approximation, for an ideal case where 
$r_{\rm max}\rightarrow \infty$, the projected correlation 
function is not affected at all by redshift space distortion. However,
since the measured $w_p(r_p)$ is derived from a finite projection 
out to $r_{\rm max}$, redshift distortion cannot be completely 
eliminated, especially at large $r_p$. We have verified that increasing
both $\pi_{\rm max}$ in the measurement and $r_{\rm max}$ in the HOD 
modeling to $80\hMpc$ has almost no effect on the inferred HOD. 
Still, we choose to only fit data points with $r_p<20 \hMpc$  
to avoid any possible contamination by redshift space distortion.
When fitting and evaluating $\chi^2$, we use the full jackknife
covariance matrix.

\subsection{Modeling the Luminosity Dependence}
\label{subsec:hod_lum} 

\begin{figure}[bp]
\plotone{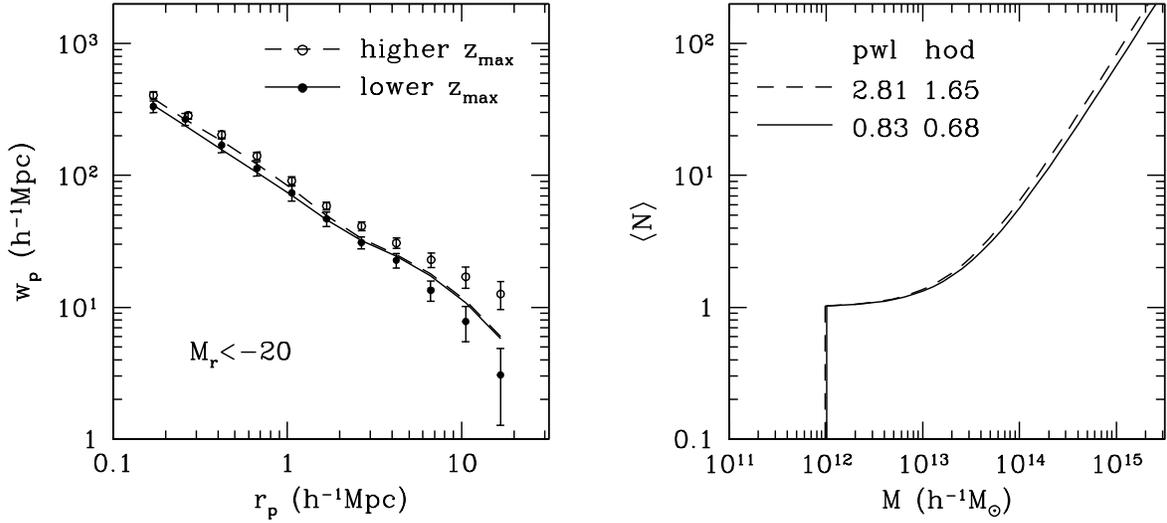}   
\caption[]{\label{fig:m20_zmax}
HOD fits to the projected correlation function of the $M_r<-20$ sample, 
which has the greatest sensitivity to limiting redshift 
(Fig.~\ref{fig:wp_vl_z}). In the
left panel, open circles show the measured $w_p(r_p)$ for $z_{\rm max}=0.10$,
and the dashed curve shows the predicted $w_p(r_p)$ for the best-fit HOD
model, whose mean occupation function is shown by the dashed curve in the
right panel. Filled circles show the measured $w_p(r_p)$ for 
$z_{\rm max}=0.06$, the same cutoff used for the $M_r<-19$ sample. Solid 
curves show the mean occupation function (right panel) and predicted
$w_p(r_p)$ (left panel) from fitting these data points. Values of 
$\chi^2/{\rm d.o.f.}$ for the best-fit power-laws and HOD models are listed in 
the right panel. Reducing $z_{\rm max}$ eliminates the anomalous high
amplitude tail of $w_p(r_p)$ at large $r_p$, thereby greatly improving
the statistical quality of the HOD (and power-law) fit, but it has little
effect on the values of the best-fit HOD parameters. 
}
\end{figure}

As discussed in 
\S~\ref{subsec:xi_lum}, the Sloan Great Wall produces an anomalous 
high amplitude tail at large $r_p$ for the $\Mr<-20$ sample with
$z_{\rm max}=0.10$, and we therefore use $\zmax=0.06$ to get a more
reliable estimate of $\wrp$ for this luminosity threshold.
Figure~\ref{fig:m20_zmax} shows fits to $w_p(r_p)$ for 
$M_r<-20$ samples with $z_{\rm max}=0.10$ and 0.06, with the mean 
occupation function $\Navg$ shown in the right-hand panel and the 
predicted and observed $w_p(r_p)$ in the left-hand panel. The quality 
of fit is much better for the shallower sample ($\chi^2/{\rm d.o.f.}=0.68$ 
vs. $1.65$), but the fit parameters are nearly identical. 
Given the underlying matter correlation function of the adopted 
cosmology and the requirement of matching the observed number density, 
there is simply not much freedom to increase the large-scale values of 
$\wrp$ while remaining consistent with the data at $r_p\la 3\hMpc$, where 
the one-halo contribution is important. Our derived HOD parameters (though 
not the $\chi^2$ values) are thus relatively insensitive to statistical 
fluctuations or systematic uncertainties in $w_p(r_p)$ at $r_p\gtrsim 5 
\hMpc$, where the difference in $w_p(r_p)$ is largest.
The HOD parameters are similarly insensitive to the choice of halo
bias factors.  We generally adopt the formula of \cite{sheth01}
for halo bias as a function of mass, since we have tuned our treatment
of halo exclusion and the scale dependence of halo bias assuming
these results.  If we instead use the formula of \cite{seljak04b}
(with the same treatment of exclusion and scale dependence), then
we find negligible change in the best-fit HODs, but the predicted
amplitude of $\wrp$ at large scales is generally lower, increasing
$\chi^2$ for some samples and decreasing it for others.

\begin{figure}[bp]
\plotone{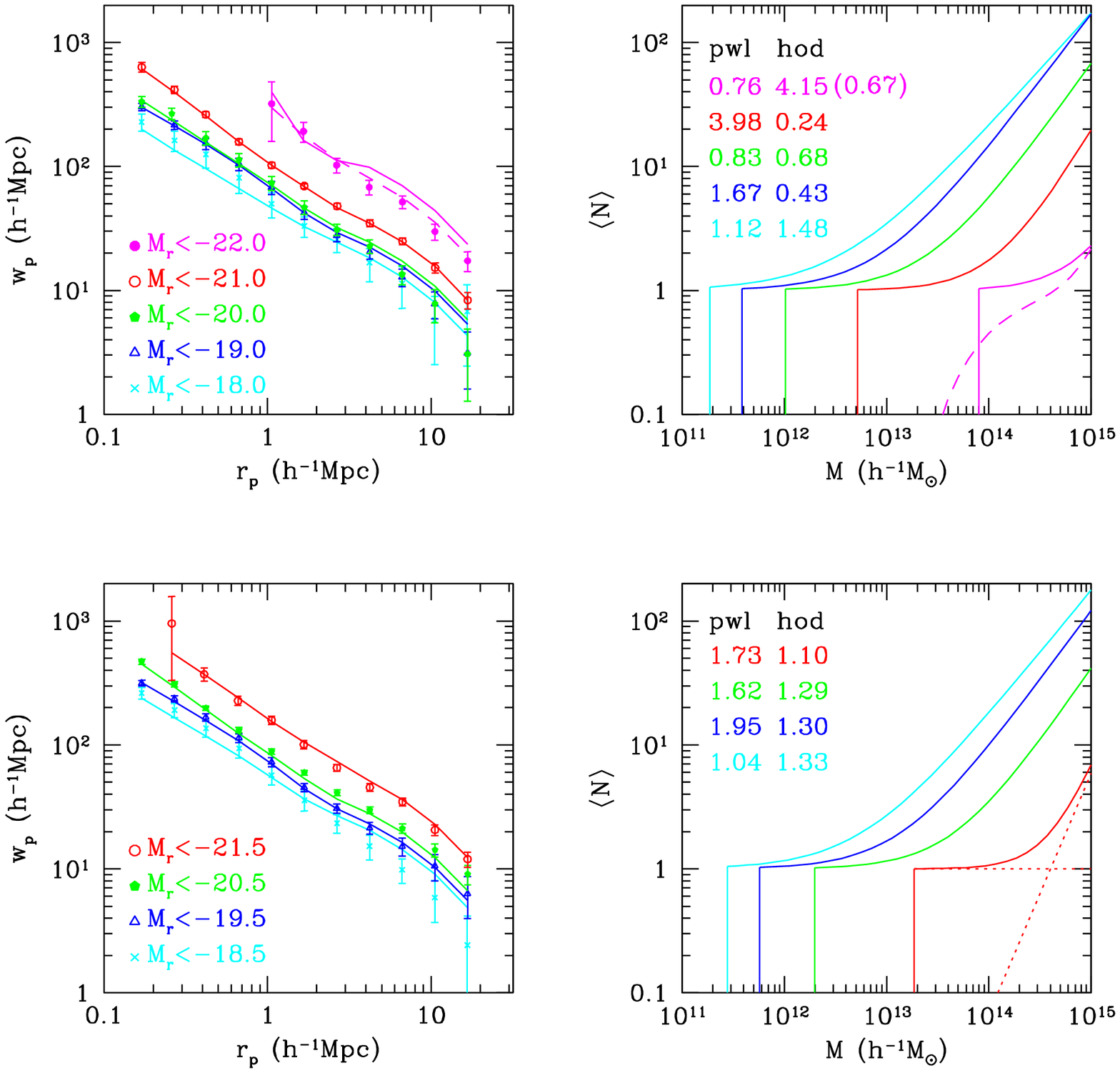} 
\caption[]{\label{fig:L_hod}
Luminosity dependence of the HOD. Left panels show the measured 
$w_p(r_p)$ and the best HOD fit for each luminosity-threshold
sample. Right panels show the corresponding $\Navg$ curves (which shift
to the right as the luminosity threshold increases).
Also labeled are values of reduced $\chi^2$ for the best power-law fit 
and HOD fit (ordered from top to bottom by
decreasing luminosity threshold).
The separation of central and satellite galaxies is illustrated for the 
right-most $\Navg$ curve in the lower-right panel (dotted lines).
For the brightest sample ($M_r<-22$), we show the effect of adding an 
exponential cutoff profile in $\Navg$ (dashed curve in the
upper-right panel), which leads to an improved HOD fit (dashed curve in 
the upper-left panel). 
}
\end{figure}

\begin{figure}[bp]
\plotone{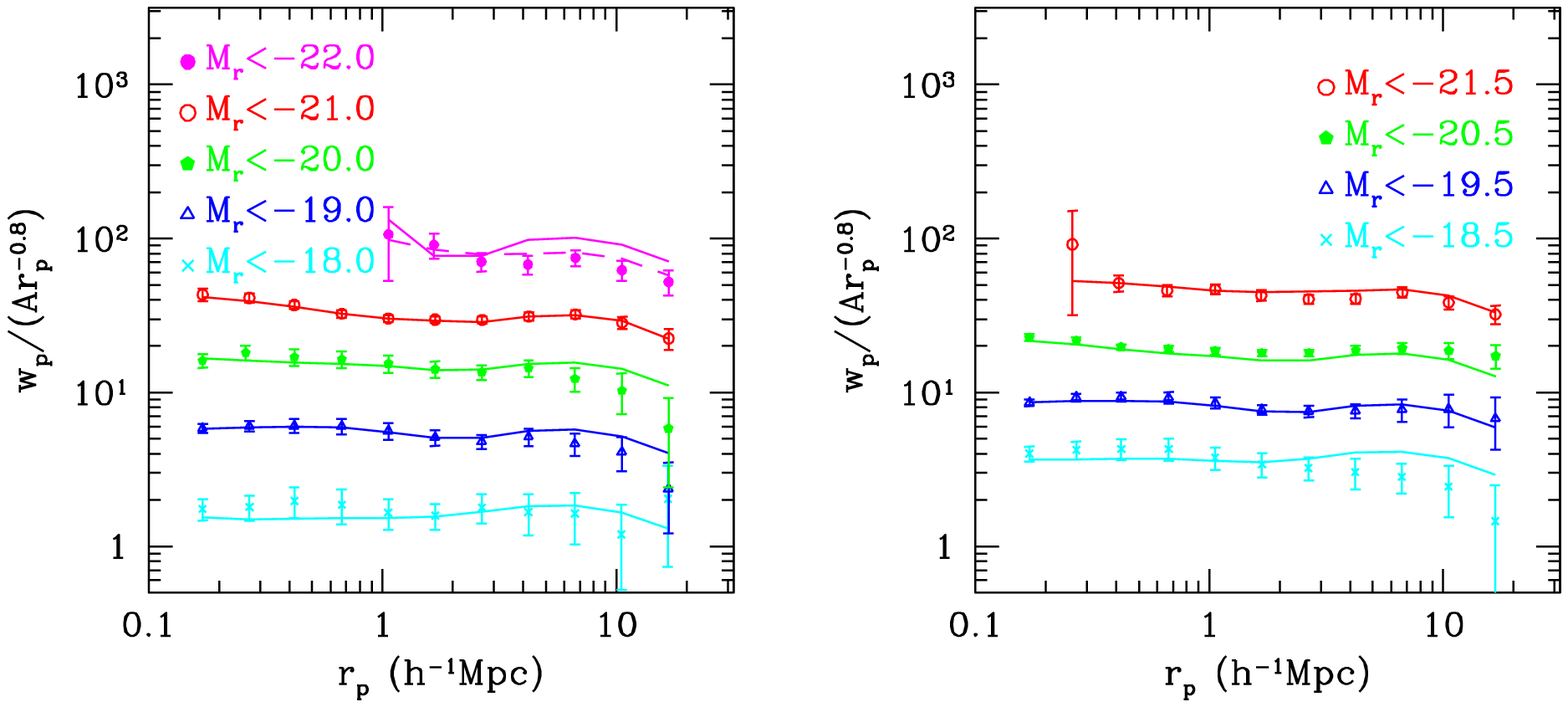}
\caption[]{\label{fig:pwl_div}
Same as left panels of Fig.~\ref{fig:L_hod}, but each of the measured and  
best-fit projected correlation functions is divided by a power-law 
$\propto r_p^{-0.8}$. An arbitrary vertical displacement is applied for 
each sample. This plot allows a close inspection of departures from a
power-law in $w_p$. 
}
\end{figure}

Figure~\ref{fig:L_hod} shows HOD fits to the projected correlation functions
of samples of different luminosity thresholds. Table~\ref{tab:hodfit} lists
the HOD parameters and $\chi^2$ values for HOD and power-law fits to these 
samples. With the same number of degrees of freedom, HOD modeling generally
yields a better fit than a power-law correlation function. The brightest 
sample is a strong exception, which we will discuss below. The two faintest
samples are mild exceptions, but the small volume 
probed by these samples makes their overall normalization somewhat uncertain,
perhaps by an amount that exceeds the internal jackknife error estimates.
The $M_r < -21$ correlation function has a marked inflection at 
$r_p\sim 2\hMpc$.  Z04 showed that this feature is naturally
explained in the HOD framework by the transition 
near the virial diameter of large halos from the steeply falling
one-halo term dominant at smaller scales to the flatter two-halo term dominant
at larger scales (see their Figs.~2 and 3). 
Figure~\ref{fig:pwl_div} plots
the $w_p(r_p)$ data and the HOD fits divided by an $r_p^{-0.8}$ power-law.
While the power-law departures are not as striking for less luminous samples,
nearly all of them show some change in slope at $r_p\sim 2\hMpc$, as the 
HOD fits generally predict, lending further support to the results of Z04.
 
\begin{deluxetable}{ccccccc}
\tablewidth{0pt}
\tablecolumns{7}
\tablecaption{\label{tab:hodfit}
Best-fit HOD Parameters for Luminosity-Threshold Samples\tablenotemark{a}}
\tablehead{${M_r}^{\mathrm{max}}$  & $\log_{10}\Mmin$\tablenotemark{b}& $\log_{10}M_1$\tablenotemark{c} & $\alpha$ & N\tablenotemark{d} & $\chi^2_{HOD}$ & $\chi^2_{power-law}$}
\startdata
-22.0 & 13.91 & 14.92 & 1.43 &  7 & 20.74 (3.33)\tablenotemark{e} &  
3.81 \cr   
-21.5 & 13.27 & 14.60 & 1.94 & 10 &  8.81        & 13.81 \cr  
-21.0 & 12.72 & 14.09 & 1.39 & 11 &  2.18        & 35.78 \cr  
-20.5 & 12.30 & 13.67 & 1.21 & 11 & 11.65        & 14.57 \cr  
-20.0 & 12.01 & 13.42 & 1.16 & 11 &  6.09        &  7.48 \cr  
-19.5 & 11.76 & 13.15 & 1.13 & 11 & 11.70        & 17.53 \cr  
-19.0 & 11.59 & 12.94 & 1.08 & 11 &  3.87        & 15.06 \cr  
-18.5 & 11.44 & 12.77 & 1.01 & 11 & 11.94        &  9.38 \cr  
-18.0 & 11.27 & 12.57 & 0.92 & 11 & 13.33        & 10.04 \cr  
\enddata
\tablenotetext{a}{HOD parameters listed here are for the parameterization 
with sharp cutoff in $\Navg$.}
\tablenotetext{b,c}{Mass is in unit of $\hMsun$.}
\tablenotetext{d}{This is the number of data points used in the fitting.}
\tablenotetext{e}{The value of $\chi^2$ quoted in parentheses is for the 
case of an exponential cutoff in $\Navg$.}
\end{deluxetable}

\begin{figure}[bp]
\plotone{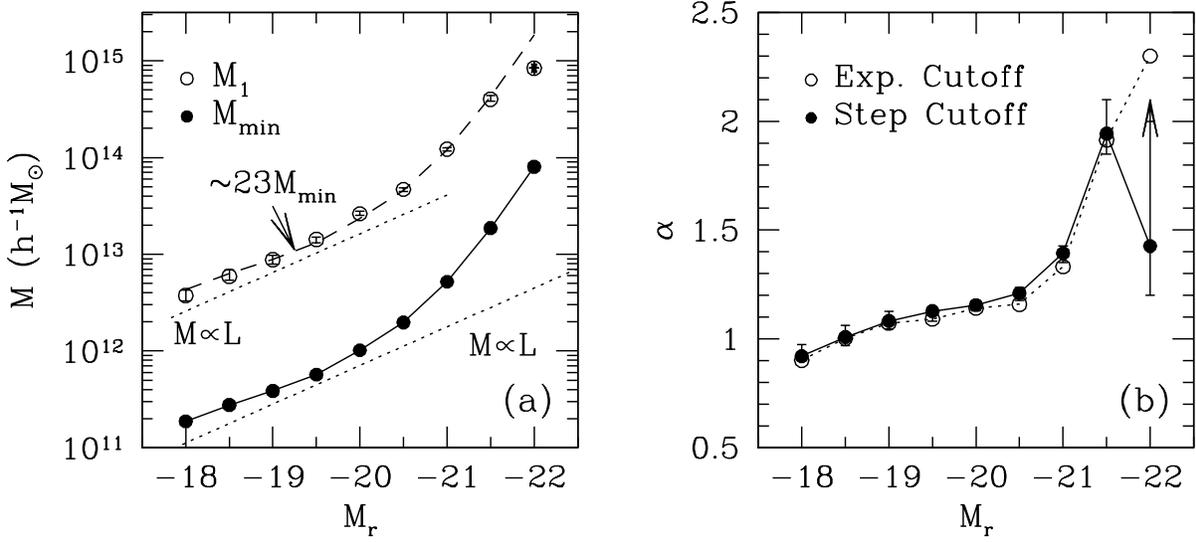}   
\caption[]{\label{fig:M_alpha_L}
HOD parameters as a function of the threshold luminosity. The left panel
shows $\Mmin$ (solid circles) and $M_1$ (empty circles). The dashed curve
is the $\Mmin$--$M_r$ (solid) curve scaled up by a factor of 23. 
The two dotted lines plot $M \propto L$
for comparison. The right panel shows the power-law slope $\alpha$ of 
$\Nsatavg$.  Adopting an exponential cutoff profile in 
$\Navg$ only has a small effect on the inferred $\alpha$
(empty circles versus solid circles).
}
\end{figure}

Figure~\ref{fig:M_alpha_L} plots the 
derived HOD parameters as a function of the threshold luminosity.  
The characteristic minimum mass $\Mmin$ of halos that can host a galaxy
increases as we go to high luminosity samples. For low luminosity samples 
($\Lthres<L^*$, with $\Mr^*\sim -20.5$), the minimum host halo mass $\Mmin$ is 
approximately proportional to the threshold luminosity. Halos near $\Mmin$
generally contain a single, central galaxy above the luminosity threshold,
and this linear relation suggests that the stellar light of this central
galaxy is approximately proportional to the halo mass in this low luminosity
regime. However, as we move to high luminosity galaxies ($\Lthres>L^*$), 
the minimum mass of hosting halos increases more steeply than a naive linear 
relation $\Mmin \propto L$. This departure is consistent with the well 
established fact that these luminous galaxies are found only in group or
cluster environments (see e.g., \citealt{loh03,blanton03e}). In these high 
mass halos, a larger fraction of baryon mass goes into satellites below the 
luminosity threshold and into a shock-heated intragroup medium, leaving 
less for the central galaxies. 
The steepening of the relation between $\Mmin$ and the threshold luminosity 
toward high luminosity is in good agreement with galaxy formation
models (see, e.g., \citealt{zheng04b}). Dynamical mass estimates
of galaxies from 
velocity dispersions of stars (e.g., \citealt{padmanabhan04}) or satellite 
galaxies (e.g., \citealt{prada03,mckay02}) and aperture mass 
measured from weak lensing (e.g., \citealt{mckay01,sheldon04,tasitsiomi04})
do not show as strong a dependence on galaxy luminosity, even after 
correcting to the halo virial mass. 
However, these results do not necessarily conflict with ours,
since $\Mmin$ represents the characteristic
minimum mass, not average mass, for galaxies above a given 
luminosity.  Scatter in the relation between galaxy luminosity and host
halo mass can substantially weaken the dependence of the average halo mass on 
galaxy luminosity, as shown by \citet{tasitsiomi04}.

Figure~\ref{fig:M_alpha_L}a also shows a small 
departure from the linear relation at the low luminosity end --- relatively 
larger halos are needed to host faint galaxies. This departure could be a 
hint of feedback processes suppressing the masses of galaxies in these low 
mass halos, but this subtle deviation from linearity is sensitive to our 
idealized assumption of a sharp $\Mmin$ threshold for central galaxies,
so with $\wrp$ data alone we cannot address this point reliably.
 
Open circles in Figure~\ref{fig:M_alpha_L}a show the mass scale $M_1$ of 
halos that on average host one satellite galaxy above the luminosity 
threshold (in addition to the central galaxy). For the samples analyzed
here, the derived $M_1$ and $\Mmin$ have an almost perfect scaling relation: 
$M_1\approx 23 \Mmin$ (dashed line). This striking result tells us that 
a halo hosting two galaxies above a luminosity threshold must be, on average,
at least 20 times as massive as a halo hosting only one galaxy above the
threshold. This result is consistent with the slowly rising plateau of 
$\Navg$ found for SPH and semi-analytic model galaxies by \citet{berlind03}
and for $N$-body subhalos by \citet{kravtsov04}. \citet{berlind03} show 
that in the regime where $1\leq \Navg \leq 2$, higher mass halos tend to 
host higher mass central galaxies rather than multiple galaxies of comparable
mass. The exact scaling factor depends on our assumption of the spatial 
distribution of galaxies inside halos.  If we reduce the concentration
parameter of the galaxy distribution at each halo mass by a factor of two,
we can still get reasonable fits to the data, but $M_1$ and $\alpha$
decrease to allow more galaxies in low mass, high concentration
halos, and the scaling factor drops to $M_1/\Mmin \approx 17$.
If we increase the concentrations by a factor of two, then the linear 
scaling relation becomes less  
accurate, and the factor is $M_1/\Mmin \approx 30$. 
The roughly constant factor of $\sim 20$ at all luminosities agrees 
qualitatively with predictions from $N$-body simulations, SPH simulations, 
and semi-analytic models (see \citealt{kravtsov04,zheng04b}), but establishing 
quantitative agreement over this large dynamic range in luminosity remains a 
challenge for further theoretical studies of galaxy formation.
The large value of this factor probably reflects the combination
of halo merger statistics and dynamical friction timescales; near-equal
mass mergers of halos are relatively rare, and they are followed fairly
quickly by mergers of their central galaxies.
 
The power-law slope $\alpha$ of the satellite mean occupation number 
$\Nsatavg$ rises slowly but steadily ($\alpha\simeq 0.9$ to $\alpha\simeq 1.2$)
with luminosity for thresholds $\Lthres < L^*$, then rises more steeply
for higher luminosity thresholds (Figure~\ref{fig:M_alpha_L}b). 
A straightforward interpretation of this trend would be that halos of higher 
mass have greater relative efficiency at producing multiple high luminosity
satellites. Studies on substructures in high resolution numerical simulations 
indicate that more massive halos tend to have relatively more substructures 
of higher masses (see, e.g., Figures~5 and 7 of \citealt{gao04} and Figure~1 
of \citealt{delucia04}), consistent with the trend we find in
$\alpha$. However, while the statistical errorbars on $\alpha$ (as 
indicated in the figure)  are small, 
systematic errors in $\Navg$ resulting from our restricted parameterization 
of the HOD could be more important. \citet{kravtsov04} typically find that 
$\alpha=1.00 \pm 0.05$ for samples of $N$-body subhalos selected based on 
maximum circular velocities, while their Figures 4--6 show that $\Nsatavg$ 
drops faster than a power-law at the low 
mass end. Motivated by this result, we tried changing our parameterization 
of the satellite mean occupation from $\Nsatavg=(M/M_1)^\alpha$ to 
$\Nsatavg=\exp[-M_{\rm cut,sat}/(M-\Mmin)] (M/M_1^\prime)$, with the same 
truncation at $M=\Mmin$. 
This formulation has the same number of free parameters, 
but it fixes $\alpha=1$ and changes the sharp cutoff at $M=\Mmin$ to an
adjustable exponential cutoff. Although the number of satellites is small
in the exponential cutoff region, the freedom afforded by 
$M_{\rm cut,sat}$ breaks the connection between the value of $\alpha$ and 
the normalization of $\Nsatavg$, so it has a significant impact on 
$w_p(r_p)$; making $\alpha>1$ or suppressing satellite numbers at low
halo mass both increase the one-halo contribution from higher mass halos.

\begin{figure}[bp]
\plotone{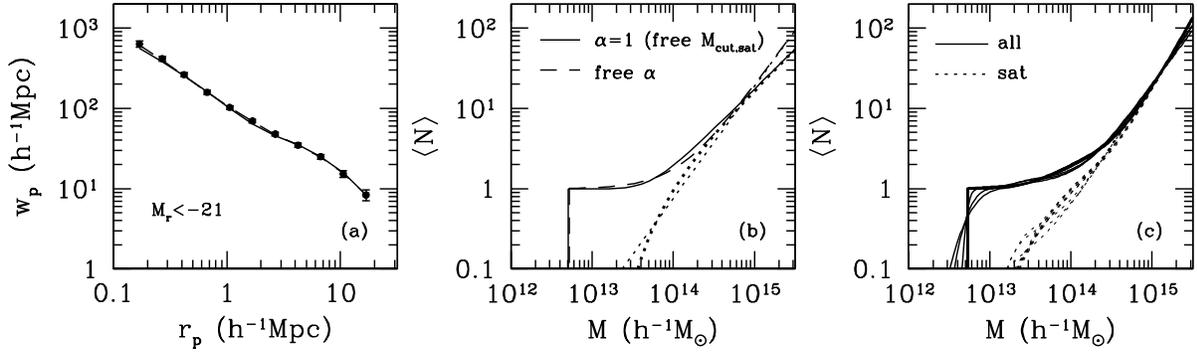}
\caption[]{\label{fig:satcut}
Fits to $\wrp$ of the $M_r<-21$ sample with different HOD parameterizations.
The left panel shows the measured and predicted values of $w_p(r_p)$ (with
two curves nearly superposed), and
the middle panel shows the corresponding $\Navg$. Dashed curves are results
of using our standard HOD parameterization ($\alpha$ is a free parameter),
and solid curves are from a parameterization that fixes $\alpha=1$ and 
adjusts the cutoff mass of $\Nsatavg$. In the middle panel, the 
thin and thick dotted curves are mean occupation numbers of satellite 
galaxies for the former and latter case, respectively. Fits from these 
two forms of HOD are very close to each other.
The right-hand panel shows $\Navg$ fits when using a more flexible HOD
parametrization (see text). Ten different models are shown, all
with $\Delta \chi^2 < 1$ relative to the best-fit model.
The solid lines correspond to $\Navg$ and
the dashed lines show $\Nsatavg$.
}
\end{figure}

We find that this parameterization yields $w_p(r_p)$ fits close to those 
of our sharp cutoff, variable $\alpha$ parameterization, as shown for the
$M_r<-21$ sample in Figure~\ref{fig:satcut}a.  However, the 
relation between $\Mmin$ and the mass scale $M_{\rm 1,sat}$ where 
$\Nsatavg=1$ remains very close to our original result of  
$M_{\rm 1,sat} \sim 23 \Mmin$, indicating that this scaling relation is 
robust.  
Another quantity that is robust to these changes is the central to satellite
galaxy ratio implied from the HOD model. 
For the $M_r<-21$ sample, $85\%$ of galaxies are central galaxies in the 
halos and only $15\%$ make up the satellite distribution. 
Central galaxies dominate over satellites for nearly all of our
samples, with an interesting exception that we discuss in 
\S\ref{subsec:hod_color} below.
Figure~\ref{fig:satcut}b shows that the mean occupations for the
two parameterizations are in fact very similar for $M < 10^{15}\hMsun$;
we cannot distinguish between the parameterizations because more massive 
halos are too rare to contribute significantly to $\wrp$.
Other complementary statistics, most notably the group multiplicity
function, are sensitive to $\Navg$ at high $M$. In the long run, we can use
the multiplicity function to pin down the high-$M$ regime and add greater 
flexibility to our parameterization of $\Navg$ in the regime of the
low mass cutoff and the plateau where $\Navg$ rises from one to several.
To illustrate the level of uncertainty in $\Navg$ with $\wrp$ alone, 
Figure~\ref{fig:satcut}c shows fits using a much more
flexible HOD parameterization (Z.\ Zheng \& D.\ Weinberg 2005, in preparation)
that allows a smooth cutoff in
$\Ncenavg$ and describes $\Nsatavg$ by a cubic spline connecting five
values specified at intervals in $\log_{10} M$, a total of seven
free parameters of which one is fixed by the mean galaxy density.
The ten models shown all have $\Delta\chi^2 \leq 1$ with respect
to the best-fit of the flexible HOD parameterization, which itself
has a $\chi^2$ that is 1.38 lower than that of our best-fit two-parameter 
model.  The central to satellite ratio is again quite robust to these 
changes, with a corresponding $1\sigma$ range of $14.5\% - 16.5\%$ for the 
satellite fraction.

Returning to our standard parameterization, Figure~\ref{fig:L_hod} shows 
that no choice of parameters yields a good fit for the brightest sample 
($M_r<-22$) --- the predicted $w_p(r_p)$ for the best fit model is 
$\sim$50\% too high at large $r_p$. 
Our HOD modeling is applied at $z\sim 0$ and does not incorporate 
evolution of the growth factor, while this bright sample extends to 
$z\sim 0.22$, where $\sigma_8$ is lower by $\sim$10\%. However, we find
that the fit does not improve substantially if we lower $\sigma_8$ to the
value at the sample's median redshift, $z_{\rm med}\sim 0.17$.  The 
masses of halos hosting these very luminous galaxies are above 
$\sim 10^{14} \hMsun$, and in this regime the analytic formula of 
\citet{sheth01} that we adopt for the halo bias factor may  
over-predict the halo bias (see their Figure~6).
A 10\% over-prediction of the halo bias factor would boost 
the large-scale correlation function by about 20\%. 
If we adopt the \cite{seljak04b} bias factors and keep our standard
treatment of halo exclusion and scale dependence of halo bias,
then we obtain a reasonably good fit to $\wrp$ for this sample, though
we now underpredict $\wrp$ for some fainter samples.

Another potential explanation for this discrepancy, and the most physically
interesting one, is that our assumption of a sharp threshold at $\Mmin$ is 
too idealized for this high luminosity sample. In general, the map from halo
mass to central galaxy luminosity is not one-to-one, so the transition from
no central galaxy to one central galaxy in luminosity-threshold galaxy 
samples should occur over some range of mass. To illustrate the effect, we 
apply an exponential cutoff profile $\exp(-\Mmin/M)$ to both $\Ncenavg$ and 
$\Nsatavg$. (The model shown previously in Figure~\ref{fig:satcut}b applied 
a smooth cutoff only to satellite galaxies.) This 
parameterization leads to a much better fit (dashed curve in the upper-left 
panel of Figure~\ref{fig:L_hod}), while the number of free parameters 
remains the same (still $M_1$ and $\alpha$, with $\Mmin$ fixed by the number
density constraint). A soft cutoff profile reduces the large-scale galaxy 
bias factor by allowing some galaxies in the sample to populate lower mass
halos with lower bias factors. We find that changing the HOD in this way 
yields slightly better fits for most other samples but that the derived 
parameters ($\Mmin$, $M_1$, and $\alpha$) are not very different from the
sharp threshold case, as demonstrated for $\alpha$ in  
Figure~\ref{fig:M_alpha_L}b (dotted line). Allowing a smooth cutoff
in $\Ncenavg$ has a much larger impact on the $M_r<-22$ sample than on the 
lower luminosity samples because the halo bias factor and halo space density 
change rapidly with mass for high mass halos. \citet{tasitsiomi04} also find
that, in their $N$-body model, 
a scatter in the luminosity-maximum velocity (mass) 
relation helps to reduce the predicted galaxy-mass correlation function of 
bright galaxies ($-22.2 \leq M_r\leq -21.7$) and thus reproduce that measured 
by \citet{sheldon04}. 
Definitive numerical results for the halo bias factor at high masses
would allow stronger conclusions on this interesting point,
since Figure~\ref{fig:L_hod} shows that the large scale amplitude of
$\wrp$ for bright samples should have an easily measurable dependence
on this scatter.

So far, we have concentrated on luminosity-threshold samples, for which the
HOD can be parameterized in a simple way.
Our step-function plus power-law parameterization is not appropriate for 
a luminosity-bin sample, since high mass halos have central galaxies
that fall out of the bin because they are too bright.
However, we can infer the HOD for a sample of galaxies in a luminosity
bin $L_{\rm thres,1}<L<L_{\rm thres,2}$ from the difference in the fitted 
HOD models for the two luminosity-threshold samples 
$L>L_{\rm thres,1}$ and $L>L_{\rm thres,2}$.
The separation of central and satellite galaxies in our parameterization 
simplifies this translation.  For a luminosity-bin 
sample, the mean occupation number of central galaxies is just the difference 
between two step functions, which becomes a square 
window, while that of satellite galaxies is the difference of two power 
law functions. 
We use the parameterization with $\alpha=1$ and
$\Nsatavg=\exp[-M_{\rm cut,sat}/(M-\Mmin)] (M/M_1^\prime)$ 
(see Fig.~\ref{fig:satcut}), 
so that small differences in $\alpha$ do not produce
anomalous behavior in the difference of occupation functions at high $M$.

\begin{figure}[bp]
\plotone{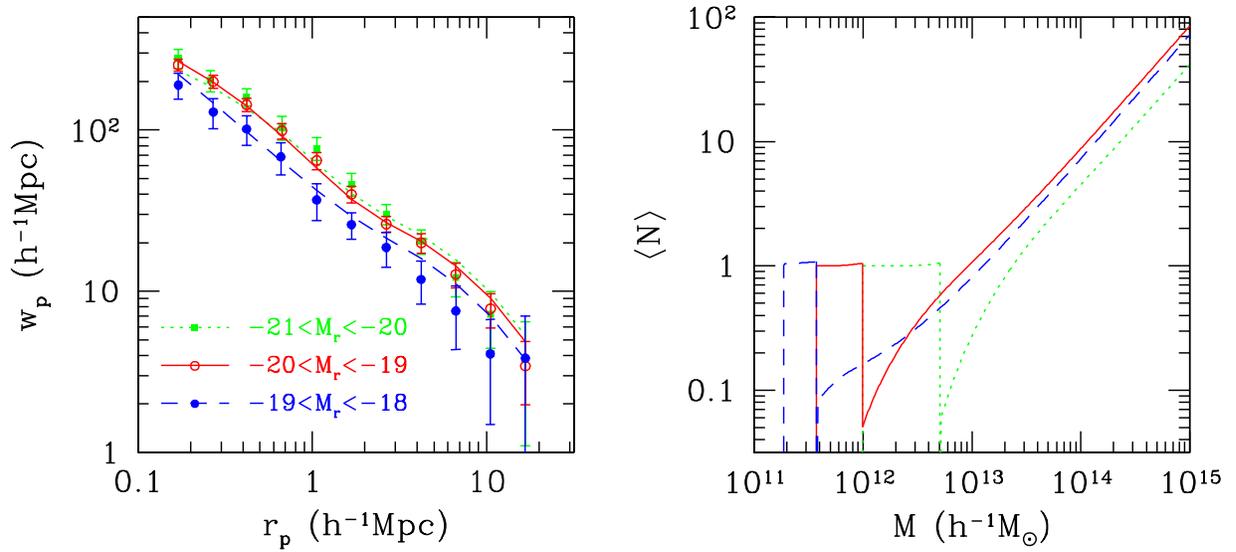}   
\caption[]{\label{fig:L_bin}
Predictions of $\wrp$ and HODs for luminosity-bin samples based on HOD 
parameters inferred from luminosity-threshold samples. The left panel 
compares measured (points) and predicted (curves) $\wrp$
for three luminosity-bin samples.  Right panels show
the $\Navg$ curves used for the predictions,
each of which is the difference of two luminosity-threshold $\Navg$ curves.
 }
\end{figure}

Figure~\ref{fig:L_bin}b shows the mean occupation functions for 
three luminosity bins derived in this way from our luminosity-threshold
results. The majority of galaxies in each luminosity bin are central
galaxies, with their relative fraction increasing with luminosity 
(about 55\%, 65\%, and 75\%, respectively, and 85\% for the $-22<\Mr<-21$ case
not shown in the plot). Our restricted HOD parameterization might lead to 
an underestimate of the fraction of central galaxies
for low luminosity samples; with the more flexible cubic spline
parameterization mentioned above we
find that the 1-$\sigma$ range of the central galaxy fraction is 69\%--78\%
for the  $-20<\Mr<-19$ sample and 75\%--81\% for the $-21<\Mr<-20$ sample, 
compared to 65\% and 75\% for the best-fit two-parameter model.
Nevertheless, the general trend of increased central galaxy fraction
with luminosity remains the same.
Figure~\ref{fig:L_bin}a compares the predicted $w_p(r_p)$ 
curves with the measured data points for each bin. 
While the threshold and bin correlation functions are obviously not
independent, it is nonetheless encouraging that 
predictions derived from the threshold samples match measurements for bin
samples fairly well, suggesting that our
adopted parameterization for luminosity-threshold samples is reasonable.
Our differencing of luminosity-threshold HODs yields a luminosity-bin
HOD parameterization similar to that adopted by \cite{guzik02} in their
models of SDSS galaxy-galaxy lensing.

\begin{figure}[bp]
\plotone{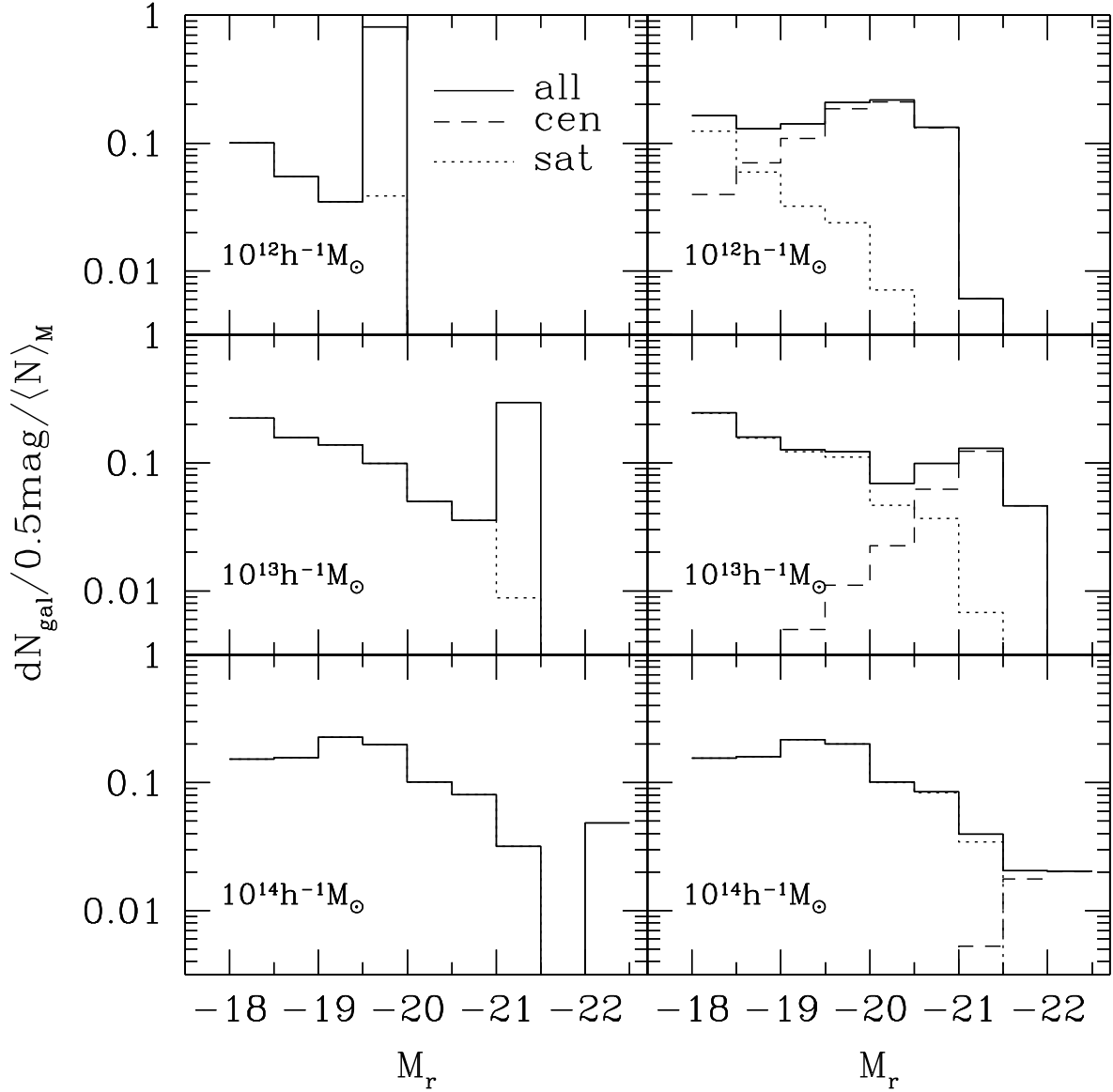}
\caption[]{\label{fig:clf}
Conditional luminosity functions of galaxies inferred from the best-fit HOD
parameters for the luminosity threshold samples, shown at three different
halo mass scales. Results from two variants of the HOD parameterizations are 
shown: a step-like cutoff in $\Navg$ (left-hand panels) and an exponential
cutoff profile (right-hand panels).  Dashed lines show the contribution
from central galaxies and dotted lines the contribution from 
satellites. In the step-like cutoff cases, the central galaxies
contribute only to the highest magnitude bin.
The CLF in each panel is arbitrarily normalized such that the total number 
of galaxies in each halo is one. 
}
\end{figure}

Knowing the mean occupation function of galaxies in each luminosity bin,
we can also easily predict the conditional luminosity function (CLF; 
\citealt{yang03}), defined as the average number of galaxies
per unit luminosity that reside in a halo of given mass. Through fitting 
luminosity functions and luminosity-dependent clustering simultaneously, 
the CLF offers an alternative approach to HOD modeling, and it has been 
used to model
observations from the 2dFGRS and the DEEP2 redshift survey and to construct 
mock galaxy catalogs (\citealt{yang03,bosch03a,bosch03b,mo03,yan03,
yan04}). These papers have parameterized the CLF as a \citet{schechter76} 
function with normalization, faint end slope, and characteristic luminosity 
depending on halo mass. Here, we take a different approach to the CLF --- 
instead of assuming an {\it a priori} functional form, we ask what the 
measured luminosity dependence of galaxy clustering can tell us about the 
shape of the CLF. The information for inferring the CLF at each halo mass 
is fully encoded in the best fit HOD parameters of different 
luminosity-threshold samples, and at a given halo mass $M$ 
one only needs to take differences of $\Navg$ for adjacent 
luminosity thresholds: 
$\Phi(L_1 < L < L_2 | M) \propto \langle N(>L_1) \rangle_M - 
                                 \langle N(>L_2) \rangle_M. 
$                                   
Figure~\ref{fig:clf} shows the inferred CLF at three 
halo masses for two forms of the HOD parameterization, a step-like cutoff 
in $\Navg$ in left-hand panels and an exponential cutoff in right-hand 
panels. Central galaxies produce 
a marked departure from a Schechter-like form, especially in low mass halos 
where they contain a larger fraction of the total luminosity. 
The prominence of the central
galaxy peak is much stronger for a step-function parameterization than for
an exponential cutoff form, but it is present in either case. 
An accurate empirical determination of the CLF via this route would require 
more complementary
clustering measurements so that the low mass cutoff of $\Ncenavg$ and 
$\Nsatavg$ can be well constrained. Semi-analytic galaxy formation models
and SPH simulations suggest that the CLF can be modeled as the sum of a 
truncated Schechter function representing satellite galaxies and a Gaussian 
function representing central galaxies (\citealt{zheng04b}), qualitatively 
consistent with the results in Figure~\ref{fig:clf}.

\subsection{Modeling the Color Dependence}
\label{subsec:hod_color}

As we saw in \S~\ref{subsec:xi_color},   
red galaxies are more strongly clustered than blue galaxies. 
A qualitative explanation is that red galaxies preferentially reside in
galaxy groups and clusters. Within the HOD framework, we can understand 
the color dependence in a quantitative way by inferring the relative 
distribution of blue and red galaxies as a function of halo mass from 
the clustering data.  We model the same sequence of luminosity
bins shown in Figure~\ref{fig:wrp_lumcol}, except that our
brightest sample consists of all galaxies with $\Mr<-21$ instead
of the $-22 < \Mr < -21$ magnitude bin.
We again use the luminosity-dependent color cut of Figure~\ref{fig:colormag}
to define red and blue subsamples, so that red galaxies correspond
roughly to the distinctive red sequence.
Red galaxies predominate in the most luminous sample, and blue
galaxies predominate in the two faintest samples.
Table~\ref{tab:hodfitcolor} lists the number of galaxies and number 
densities in each sample.

For each luminosity sample, we simultaneously fit the
the projected correlation functions of red, blue, and 
all (red+blue) galaxies to infer their HOD parameters.
We can obtain the mean occupation functions 
for blue and red galaxies from that of all galaxies by modeling the blue 
galaxy fraction, $f_b$, as a function of halo mass. Since our parameterization
distinguishes central and satellite galaxies, and the blue fractions for
these two populations could well be different, we separately parameterize 
$f_b$ for central galaxies and for satellites. We know that red galaxies are 
more common in high mass halos, so we adopt functional forms in which $f_b$ is 
a decreasing function of halo mass, such as a log-exponential,
\begin{equation}
\label{eqn:fb_exp}
f_b(M) = f_0\exp\left(-\frac{\log_{10} M-\log_{10} \Mmin}{\sigma_M}\right),
\end{equation}
or a log-normal 
\begin{equation}
\label{eqn:fb_gau}
f_b(M) = f_0\exp\left[-\frac{(\log_{10} M-\log_{10} \Mmin)^2}{2\sigma_M^2}\right].  
\end{equation}
We find that these two functions fit the data equally well. 
Motivated roughly by theoretical predictions (\citealt{zheng04b}), we adopt 
the log-normal form for the blue fraction $f_{b,{\rm cen}}$ in central 
galaxies and the log-exponential form for the blue fraction $f_{b,{\rm sat}}$ 
in satellite galaxies. There are two parameters in each function: $f_0$, 
the blue fraction in halos of $M=\Mmin$, and $\sigma_M$, a quantity 
characterizing how 
fast the blue fraction drops. Of the four new parameters, one (e.g., 
$f_{0,{\rm cen}}$) can be fixed by matching the global number density
of blue galaxies. We assume that red and blue satellite galaxies follow 
Poisson distributions with respect to their mean occupations $\langle \Nrsat
\rangle_M$ and $\langle \Nbsat \rangle_M$, just as in the full satellite 
sample. It is well known that there 
is color/morphology segregation within galaxy clusters (e.g., \citealt{oemler74,
melnick77,dressler80,adami98}) --- red galaxies are more centrally 
concentrated. With the $w_p(r_p)$ data alone, we have little power to 
constrain the relative concentration of red and blue galaxies, since the effect
shows up only on small scales and can be compensated by changing the relative 
satellite occupation numbers. We therefore do not 
consider the segregation effect here. We find that we can obtain good fits 
by assuming that both satellite populations follow the same NFW profile as 
the dark matter. Constraints on the profiles of red and blue satellites 
could be better obtained from direct analysis of identified groups, after 
which these profiles could be imposed in $w_p(r_p)$ fitting. 
Altogether, 
then, we have five free parameters ($M_1$, $\alpha$, $f_{0,{\rm sat}}$, 
$\sigma_{M,{\rm cen}}$, and $\sigma_{M,{\rm sat}}$) to simultaneously fit the
projected correlation functions of red, blue, and all galaxies, with
the parameters $\Mmin$ and $f_{0,{\rm cen}}$ fixed by number density 
constraints.

\begin{figure}[bp]
\plotone{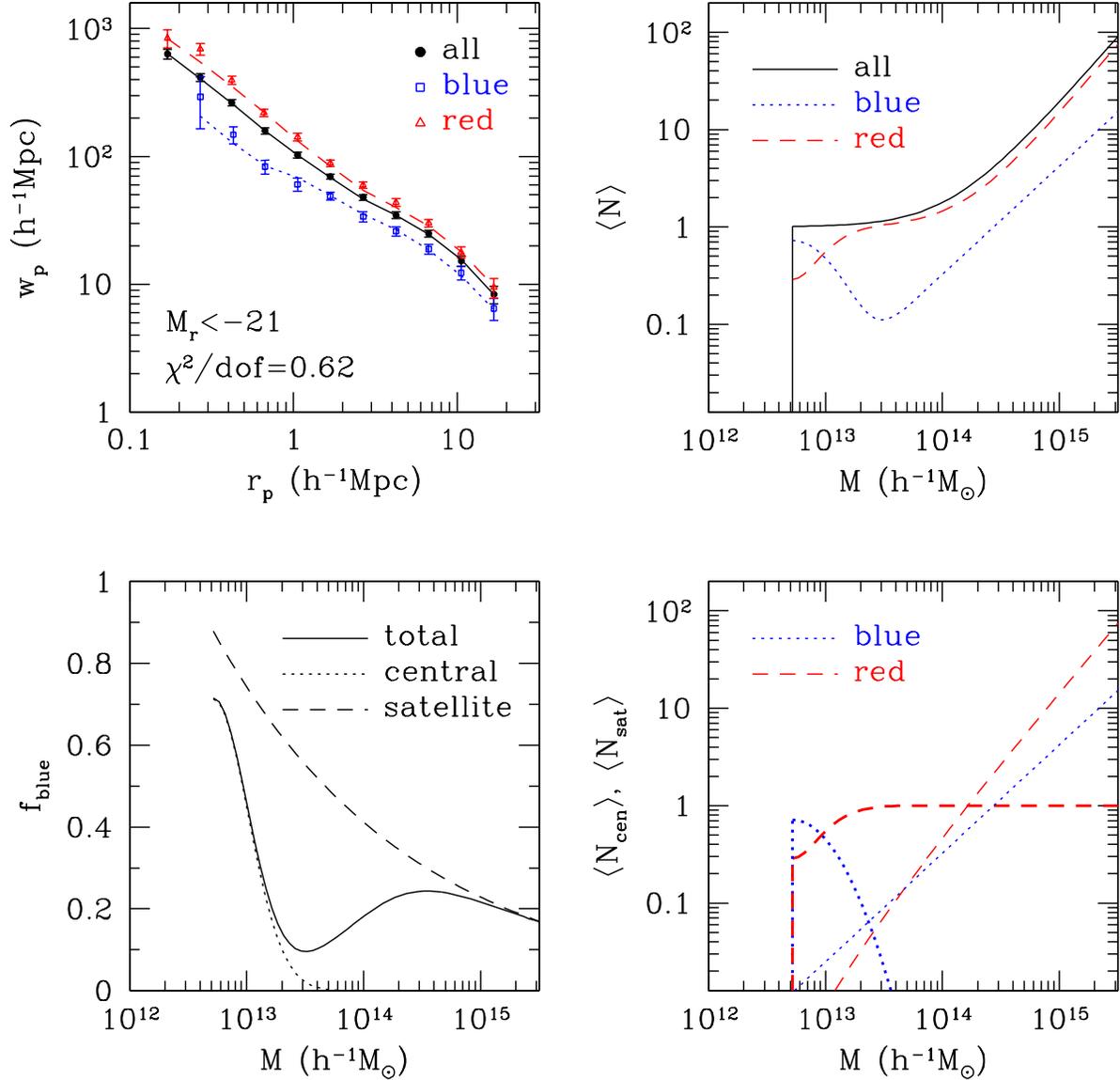}
\caption[]{\label{fig:color_hod_m21}
Color dependence of the HOD for the $M_r<-21$ sample. The upper-left
panel shows measurements of $\wrp$ and best HOD fits for red, blue, and all
galaxies in the sample. Mean occupation numbers of these three classes of
galaxies are plotted in the upper-right panel. The lower-right panel shows
mean occupation numbers of central (thick curves) and satellite (thin curves)
galaxies for red (dashed) and blue (dotted) galaxies. 
The lower-left panel plots the blue fraction for all galaxies
(solid curve), central galaxies (dotted curve), and satellite galaxies
(dashed curve).
}
\end{figure}

Figure~\ref{fig:color_hod_m21} shows fitting results for the luminous 
($M_r<-21$) sample. The best-fitting HOD parameters are listed in
Table~\ref{tab:hodfitcolor}. 
With the five-parameter model, we obtain an excellent fit to 32 data points,
with $\chi^2/{\rm d.o.f.}=0.62$ (upper-left panel)\footnote{We have 
estimated error covariance matrices separately for red, blue, and all 
galaxies and treated them as independent, because a jackknife
estimate of a $32\times 32$ covariance matrix would be too noisy to invert 
robustly. However, we may thereby underestimate error correlations.}, 
showing that the different spatial clustering of red and blue galaxies can 
be well explained by their different occupations of dark matter halos. 
In the fits, the mean occupation number of red galaxies 
rises continuously with halo mass, while $\Navg$ for blue galaxies shows a 
minimum near $3\times 10^{13}\hMsun$. As halo mass increases, the total blue 
{\it fraction} (lower-left panel) has a sharp drop, a small rise, then 
a gentle decline.  
The non-monotonic behavior is easily understood when we separate
the contributions of central and satellite galaxies.
In halos just above $\Mmin$, central galaxies are predominantly
blue, but above $\sim 2\Mmin$ they are predominantly red.
The minimum in the blue galaxy occupation occurs for halos that
are too massive to have a blue central galaxy but not massive enough to
have any satellite galaxies above our luminosity threshold.
This transition explains why the 
mean occupation number of blue galaxies can be approximated by a Gaussian 
bump (or a square window) plus a power-law, as used in some HOD models 
(e.g., \citealt{sheth01b,scranton03}); the two components represent blue 
central and satellite galaxies, respectively (see \citealt{guzik02}).
The blue satellite fraction declines slowly with halo mass in
the regime $M \ga 20\Mmin$ where satellite galaxies are common.

\begin{landscape}
\begin{deluxetable}{crrllcccccccc}
\tablewidth{0pt}
\tablecolumns{9}
\tablecaption{\label{tab:hodfitcolor} Volume-limited Correlation
Function Red/Blue Samples and Best-fit HOD Parameters}
\tablehead{
$M_r$ & $N_{\mathrm{gal,red}}$ & $N_{\mathrm{gal,blue}}$ &
${\bar n}_{\mathrm{red}}$ & ${\bar n}_{\mathrm{blue}}$ & $\log_{10}\Mmin$ &
$\log_{10}M_1$ & $\alpha$ & $f_{\mathrm{0,cen}}$ & $f_{\mathrm{0,sat}}$ &
$\sigma_{M,{\rm cen}}$ & $\sigma_{M,{\rm sat}}$ & $\frac{\chi^2}{d.o.f.}$
}
\tablecomments{The $M_r<-21$ sample use $14.5 < r \la 17.77$ and others use
$10.0 < r \la 17.5$. The mean number density (${\bar n}_{\mathrm{red}}$
or ${\bar n}_{\mathrm{blue}}$) is measured in units of $10^{-2}$ $h^{3}$
Mpc$^{-3}$. Mass is in unit of $\hMsun$.
}
\startdata

    $<$ -21 & 16,142 &  9,873 & 0.0726 & 0.0444 & 12.72 & 14.08 & 1.37 &
0.71 & 0.88 & 0.30 & 1.70 & 0.62 \cr

-21 $-$ -20 & 2,881 & 2,789 & 0.245  & 0.237  & 12.00 & 13.38 & 1.16 &
0.55 & 0.31 & 10.0\tablenotemark{a} & 20.0\tablenotemark{a} & 0.88 \cr

-20 $-$ -19 &  5,804 &  8,419 & 0.347  & 0.503  & 11.62 & 12.94 & 1.06 &
0.71 & 0.46 & 10.0\tablenotemark{a} & 7.99 & 0.72 \cr

-19 $-$ -18 &  1,195 &  3,350 & 0.267  & 0.747  & 11.38 & 12.58 & 0.95 &
0.86 & 4.00\tablenotemark{a,b} & 10.0\tablenotemark{a} & 0.69 & 1.48 \cr

\enddata
\tablenotetext{a}{The HOD fit is not sensitive to this value and only needs
it to be large, so the value is fixed at a large number for the HOD fit.}
\tablenotetext{b}{If the fraction of central or satellite galaxies becomes
greater than one, it is set to be one.}
\end{deluxetable}
\end{landscape}

Our detailed numbers for red and blue galaxy HODs should not be taken too
seriously because they depend to some degree on our particular choice of
parameterized model.  This model provides a good fit to the data, but we do
not claim that the fit is unique.  Nonetheless, when we initially investigated
the color dependence using a quite different parameterization that did not
distinguish central and satellite galaxies (but accounted for sub-Poisson
fluctuations at low halo masses), we were driven to a similar 
non-monotonic behavior of the blue galaxy $\Navg$. Furthermore, our results
agree qualitatively with the behavior predicted by SPH simulations and 
semi-analytic models of galaxy formation. This can be seen by comparing our 
Figure~\ref{fig:color_hod_m21} to Figure~13 of \citet{berlind03}, which 
divides the theoretical galaxy population based on mean stellar age (which
should correlate strongly with color). \citet{zheng04b} further analyze these 
predictions in terms of central and satellite galaxies, again showing good
qualitative agreement with our results (compare our 
Figure~\ref{fig:color_hod_m21} to their Figure~4). Semi-analytic and SPH 
calculations both predict that a halo's central galaxy is in general more
massive and older than its satellites and that the ages of both central and
satellite galaxies correlate with halo mass (\citealt{berlind03}, Figures~18
and 19). Physically, these trends reflect the earlier formation times and 
more active merger histories of central galaxies. Satellite galaxies have 
generally experienced most of their growth in lower mass halos that merged 
into their present day parent halos. These physical processes thus lead 
rather naturally to a trend in which low mass halos have blue central 
galaxies, higher mass halos have red central galaxies but a significant blue 
fraction in their satellite populations, and the highest mass halos have red 
central galaxies and predominantly red satellites. 
Figure~\ref{fig:color_hod_m21} shows the impact of these processes 
on the galaxy correlation function.

For galaxies of lower luminosities, we study the color division using 
luminosity-bin rather than luminosity-threshold samples, since the galaxy
color distribution depends strongly on luminosity, and we want to isolate
the color dependence of the HOD from luminosity dependence. 
Following our earlier procedure for luminosity-bin samples, we form the mean 
occupation function for galaxies with $L_{\rm thres,1}<L<L_{\rm thres,2}$ 
from the difference between two luminosity-threshold samples, 
$L>L_{\rm thres,1}$ and $L>L_{\rm thres,2}$, then predict $w_p(r_p)$. We 
determine HOD parameters for the full $L>L_{\rm thres,2}$ sample by fitting
$w_p(r_p)$ of that sample, and we parameterize blue fractions as for
the $M_r<-21$ sample discussed above, so that we still have five free 
parameters
in total to fit projected correlation functions for red, blue, and 
all galaxies. This approach of working our way down bin by bin is not
statistically optimal, since we do not fit all the data simultaneously,
and errors in the HOD fit at high luminosities will propagate into lower
luminosity bins. However, this procedure is straightforward, and it is 
adequate for our rather qualitative purposes here, where we seek to
understand general features of the HOD color dependence.

\begin{figure}[bp]
\epsscale{0.85}
\plotone{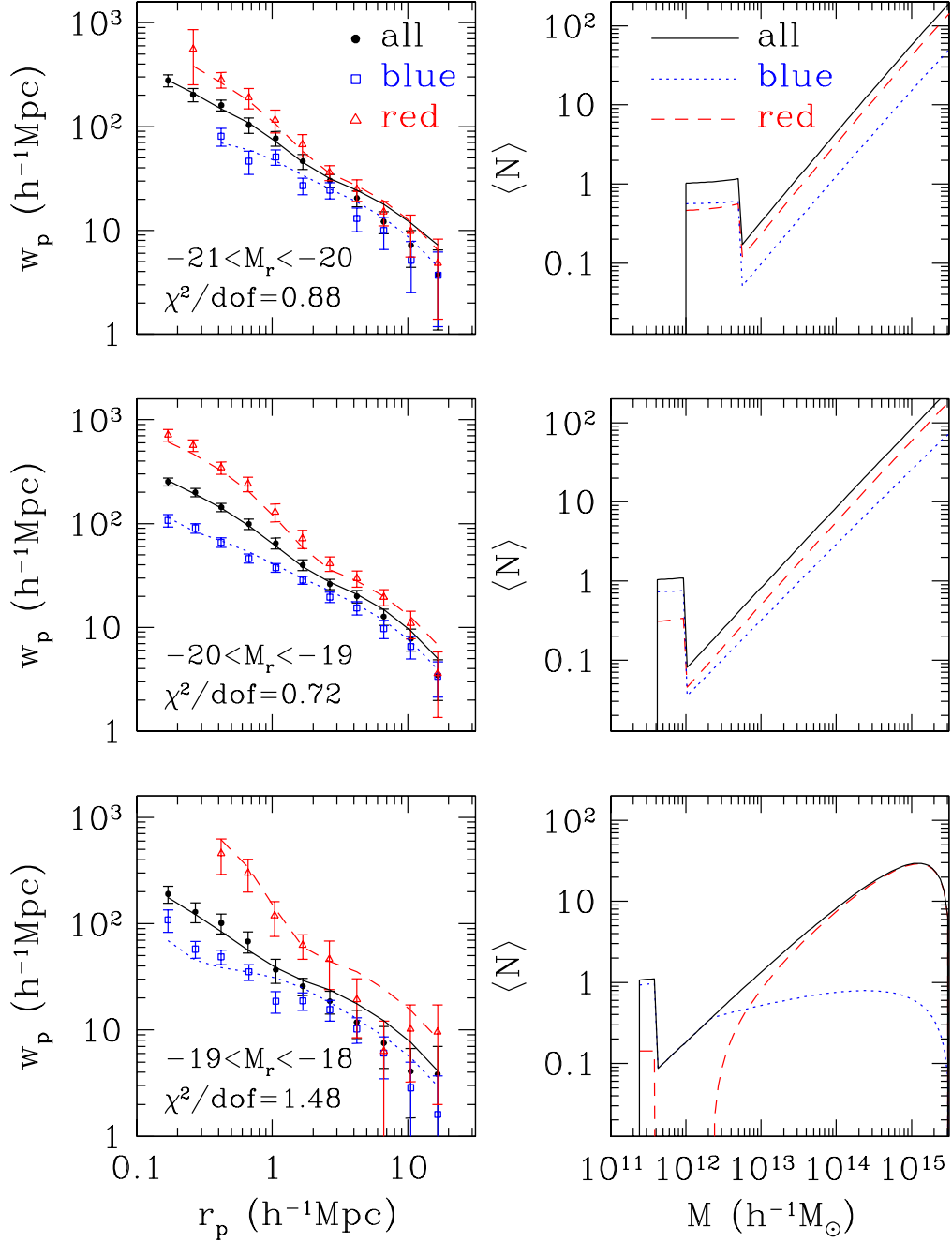}
\epsscale{1.0}
\caption[]{\label{fig:color_lumbin}
Color dependence of the HOD for the three luminosity-bin samples.
For each luminosity
bin, the left panel shows the $\wrp$ measurements and best HOD fits for
red, blue and all galaxies. The right panel shows the corresponding $\Navg$
for these three classes of galaxies.
}
\end{figure}

Figure~\ref{fig:color_lumbin} shows the results of these fits for the
three luminosity bin samples. The HOD parameters for all the samples
are specified in Table~\ref{tab:hodfitcolor}.
For $-21<M_r<-20$ (with $z_{\rm max}=0.06)$, the fit is formally 
acceptable ($\chi^2/{\rm d.o.f.}=0.88$), but the predicted correlations 
at large scales are
systematically too high, by $\sim 0.5-1\sigma$.
For $-20 < \Mr < -19$, the fit is excellent for all, red, and blue galaxies,
with the HOD model nicely explaining the strong inflection of the
red galaxy $\wrp$ near $2\hmpc$.
For $-19<M_r<-18$, the fit
overpredicts the correlation of galaxies on large scales, 
but the small volume of this faintest sample leaves it somewhat
susceptible to cosmic variance.
In every case our fits capture the qualitative difference between
red and blue galaxy clustering, and we do not know whether
the quantitative discrepancies reflect underestimates of observational
errorbars or inadequacy of this simple, five-parameter model.

If we take the fit results at face value, we see that as luminosity
decreases the central galaxies occupy lower and 
lower mass halos (as already seen in Figs.~\ref{fig:L_hod} 
and~\ref{fig:L_bin}) and become
more and more dominated by blue galaxies (56\%, 71\%, and 
85\%, respectively). For all three luminosity bins, 
the majority ($\sim 70-90\%$) of blue galaxies are central objects. 
The majority of red galaxies are also central in the brightest bins,
but the fraction of red galaxies that are satellites becomes larger
with decreasing luminosity (35\%, 54\%, and 72\% for the three samples).
Thus, faint red galaxies are predominantly
satellite systems in higher mass halos. 
To put the result in physical terms, low mass galaxies have late
star formation and blue colors if they reside in ``the field'' (making
them central objects of low mass halos) and become red only if they
enter dense environments that truncate their star formation,
a behavior predicted in SPH simulations \citep{berlind04}. Again, our detailed
numbers should be taken with a grain of salt because of the restrictions 
of the parameterization, but the fitting results show a sensible continuity
of behavior as we move from luminous galaxies to faint galaxies.

Since faint and bright red galaxies both reside in high mass halos,
as satellites and central objects, respectively, the {\it average}
host halo mass of red galaxies is actually lowest for intermediate 
luminosities.  From our fits, we find a mean host halo mass
$\sim 2-2.5\times 10^{14}\hMsun$ for red galaxies with $-18 > \Mr > -19$
and $M_r < -22$, compared to $\sim 10^{14}\hMsun$ for $-19 > \Mr > -22$.
This non-monotonic behavior explains why faint and bright red galaxies
have higher correlation amplitudes (Fig.~\ref{fig:wrp_lumcol} here
and \citealt{norberg02}) and denser local environments
\citep{hogg03} than intermediate luminosity red galaxies.

Dividing into blue and red subsamples allows another 
useful measurement, the cross correlation between blue and red galaxies.
At large scales, where the two-halo term dominates, there is little new 
information in the cross correlation because it is essentially guaranteed to
approach the geometric mean of the red and blue galaxy autocorrelations. 
However, in the one-halo regime the cross correlation encodes information 
about the halo-by-halo mixing of the red and blue populations. 
At one extreme, we could imagine that some halos contain only 
red galaxies and others only blue galaxies, and the ratio of the mean 
occupations represents the ratio of ``red halos" to ``blue halos".
In this case, the one-halo term of the cross-correlation function would be 
zero.  At the other 
extreme, the red and blue populations are fully mixed, and the typical 
red-to-blue ratio in each halo is just the ratio of mean occupations. In this
case, the one-halo contribution to the cross-correlation reaches its maximum 
amplitude. Our modeling of the autocorrelation functions has implicitly
assumed this fully mixed case, since we have taken the numbers of red and 
blue satellites within each halo to be Poisson distributed with respect to
their mean occupations. The cross-correlation measurement allows an 
independent test of this assumption. 

The one-halo term of the real space cross-correlation function of red and blue 
galaxies can be computed as
\begin{equation}
\label{eqn:br_1h}
1+\xi_{\rm rb,1h}(r)=\frac{1}{4\pi r^2\nravg\nbavg}
          \intdn \langle N_r N_b \rangle_M
          \frac{1}{2\Rvir(M)} F^\prime\left(\frac{r}{2\Rvir}; c_r, c_b\right),
\end{equation}
where $\nravg$ and $\nbavg$ are the mean number densities of red and blue 
galaxies, respectively, $\langle N_r N_b\rangle_M$ is the average number of 
red-blue galaxy pairs in a halo of mass $M$, $F(r/2\Rvir; c_r,c_b)$ is 
the cumulative radial distribution of red-blue galaxy pairs, and $c_r$ and 
$c_b$ are the concentration parameters of red and blue galaxies 
($c_r=c_b$ is assumed in this paper). This equation is similar to 
equation~(\ref{eqn:1halo}). With the separation of central and satellite 
galaxies, the pair distribution can be expressed as in 
equation~(\ref{eqn:dFdx}):
\begin{equation}
        \langle N_r N_b\rangle_M F^\prime(x; c_r,c_b) =    \newline
        \langle \Nrcen\Nbsat \rangle_M  F^\prime_{\rm cs}(x;c_b)
      + \langle \Nrsat\Nbcen \rangle_M  F^\prime_{\rm cs}(x;c_r)
      + \langle \Nrsat\Nbsat \rangle_M  F^\prime_{\rm ss}(x;c_r,c_b),
\label{eqn:br_dFdx}
\end{equation}
where $F^\prime_{\rm cs}$ has the same meaning as in equation~(\ref{eqn:dFdx}) 
and $F^\prime_{\rm ss}$ now represents the cross-convolution of two NFW 
profiles.
If red and blue galaxies are well mixed and
their occupation numbers are not correlated, then the average number of any 
kind of blue-red galaxy pairs on the right-hand side of 
equation~(\ref{eqn:br_dFdx}) 
can be replaced by the product of the mean occupation numbers of blue and red 
galaxies, e.g., 
$\langle \Nrcen\Nbsat \rangle_M= \langle \Nrcen \rangle_M \langle \Nbsat 
\rangle_M$. 
The extreme example of ``red halos'' and ``blue halos,'' on the other hand,
corresponds to 
$\langle \Nrcen\Nbsat \rangle_M=\langle \Nrsat\Nbcen \rangle_M=
\langle \Nrsat\Nbsat \rangle_M=0$. 

\begin{figure}[bp]
\plotone{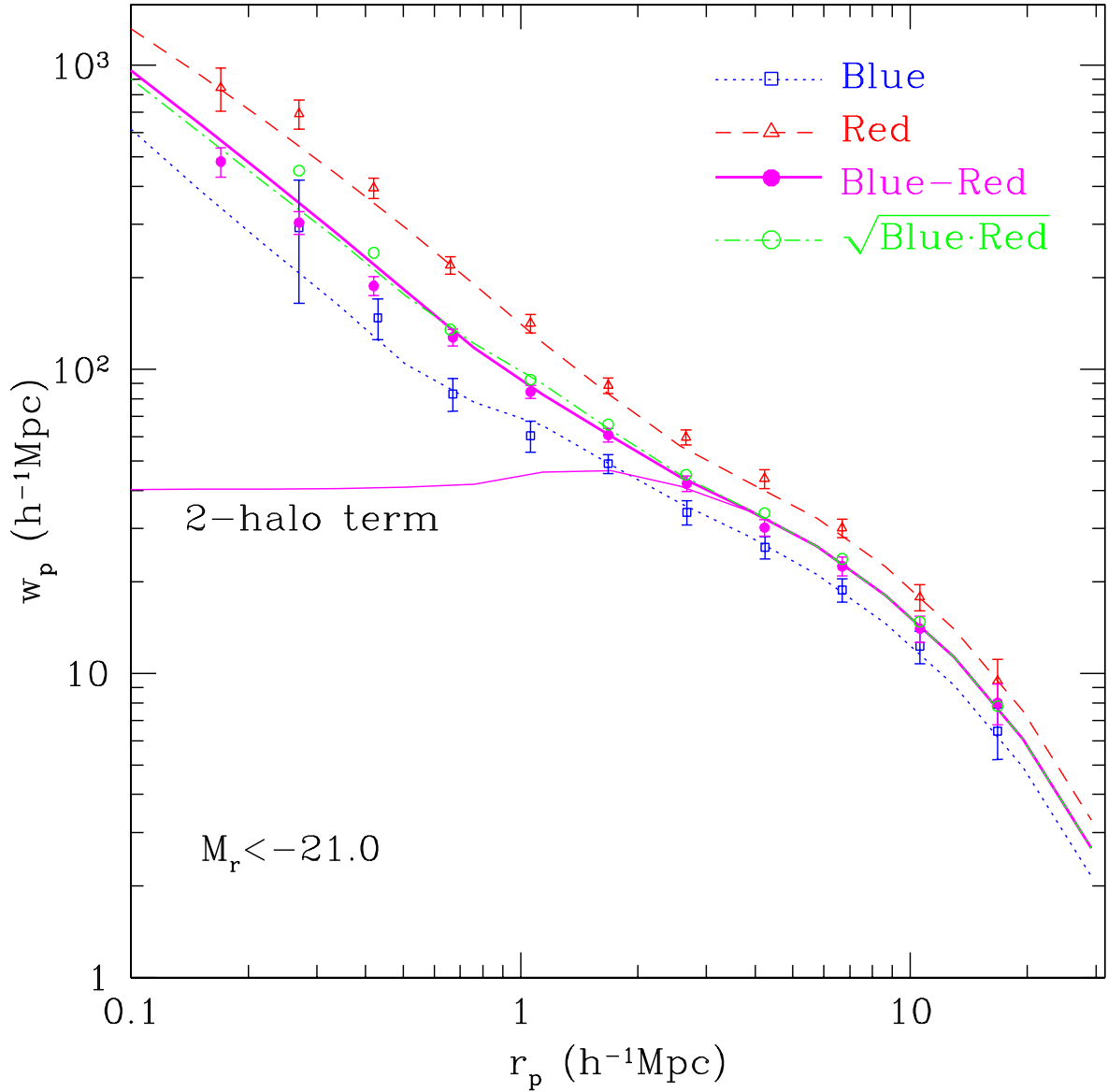}
\caption[]{\label{fig:color_cross}
Projected cross-correlation between red and blue galaxies in the $M_r<-21$ 
sample. Points are measurements. Dotted and dashed curves are HOD fits to 
auto-correlations of blue and red galaxies, respectively. The dot-dashed 
curve is the 
geometric mean of the two fits. HOD parameters of these fits are 
obtained by simultaneously fitting $w_p$ of red, blue, and all galaxies, 
i.e., the same as in Fig.~\ref{fig:color_hod_m21}. The thick solid curve is 
the predicted cross-correlation between red and blue galaxies based on these 
HOD parameters, with the assumption that blue and red galaxies are well mixed 
within halos. The thin solid curve is the two-halo term of the 
cross-correlation, representing the case in which blue and red galaxies avoid 
residing in the same halo (see the text). 
}
\end{figure}

We estimate the cross-correlation function from the data in an analogous 
way to the autocorrelations, using equation~(\ref{eq:LS}) with 
$DD$ replaced by $D_1 D_2$, $2DR$ by 
$D_1R_2+D_2R_1$, and $RR$ by $R_1 R_2$,
with the subscripts denoting the two subsamples. 
In Figure~\ref{fig:color_cross}, we compare the 
projected cross-correlation function for the $M_r<-21$ sample (filled circles) 
to the prediction assuming the fully mixed case (solid line). The other 
points and lines in the figure show the measured and predicted 
autocorrelations (from Figure~\ref{fig:color_hod_m21}) 
and their geometric means. The predicted cross-correlation is close to the 
measured one, though it is systematically higher at $r_p\lesssim 1 \hMpc$ 
(the $\chi^2$ for the 11 data points is about 15, most of which comes from 
small scales). At large scales, where the two-halo term dominates, the 
predicted and measured cross-correlation functions approach the geometric 
means as expected. For the extreme case of distinct red and blue halos,
there would be no one-halo contribution to the cross-correlation function, 
and the projected cross-correlation would flatten at $r_p \lesssim 2 \hMpc$
(see Figure~\ref{fig:color_cross}). This is clearly not the case in the
data. Our simple assumption of well mixed blue and red satellite populations 
with similar radial profiles appears to describe the one-halo regime of the
cross-correlation fairly accurately, though the fit is not perfect, and it 
is probably not unique. Direct measurement of the distributions of
red and blue galaxy numbers in identified groups can test our assumption.

\section{Conclusions}
\label{sec:disc}

We have exploited the large size and high quality imaging of the SDSS
galaxy redshift survey to study the detailed dependence of the galaxy
two-point correlation function on galaxy luminosity and color.
The amplitude of the projected correlation function $\wrp$ increases
monotonically with luminosity, slowly below $L_*$ and rapidly above.
The real space correlation functions of luminosity-bin or
luminosity-threshold samples are usually described to a first
approximation by a power-law $\xi(r)=(r/r_0)^{-1.8}$, with the 
exception of the brightest samples ($-23<\Mr<-22$ luminosity-bin
or $\Mr>-22$ luminosity-threshold), which
have a substantially steeper $\xi(r)$.
However, an inflection in $\wrp$ at $r_p \sim 1-3\hmpc$, which
is clearly identifiable for $\Mr<-21$ galaxies (Z04), is also
present to some degree in many of our other volume-limited samples,
and it is unmistakably present in the high-precision measurement from
the full, flux-limited data sample.
The projected correlation function of the $b_J$-selected, 2dFGRS survey
shows a similar feature (\citealt{hawkins03}, Figure~9).

Dividing galaxies by color, we find that red galaxies have stronger
clustering, steeper correlation functions at small scales, and much
stronger ``finger-of-God'' redshift-space distortions.
The blue galaxies in luminosity-bin samples usually have 
correlation functions close to an $r^{-1.7}$ power-law, with
amplitudes that increase with luminosity.  For red galaxies, the
situation is more complex.  Red galaxy correlation functions
are consistently steeper than $r^{-1.8}$, but the relative bias
as a function of luminosity is scale-dependent, with bright
red galaxies exhibiting the strongest clustering on large scales
and faint red galaxies exhibiting the strongest clustering
on small scales.   The luminosity and color dependence of $\wrp$
are not trivially separable, nor is one a simple consequence of the other.

Our conclusion that clustering increases with luminosity, most 
markedly above $L_*$, agrees with many previous investigations
\citep{hamilton88,park94,loveday95,benoist96,guzzo97}.
The relative bias $b/b_*$ as a function of $L/L_*$,
defined using the ratio of $\wrp$ values at $r_p=2.7\hmpc$, agrees
with that inferred (from essentially the same galaxy sample) using
the large scale power spectrum by \cite{tegmark04a}.
This measurement is also in good agreement with the fit of \cite{norberg01} 
derived from the 2dFGRS, over the (somewhat narrower) luminosity range of 
their measurements. 

Our finding of stronger clustering of ``early'' type
(which in our case means red) galaxies agrees with decades of results
obtained by a variety of techniques 
(e.g., \citealt{hubble36,zwicky68,davis76,dressler80,postman84,guzzo97,
willmer98,norberg02,zehavi02,madgwick03,goto03,hogg03}).
For our joint luminosity-color dependence analysis, the only
directly comparable study is that of \cite{norberg02}, who analyzed
a sample of similar size drawn from the 2dFGRS.  
In their case, ``early'' and ``late'' type galaxies are distinguished
by spectral class \citep{madgwick03} rather than broad-band color.
In many respects, our results agree well with theirs.
However, \cite{norberg02} found that early type galaxies have
correlation function slopes close to $-1.8$, while we find that
red galaxies have steeper correlation functions and exhibit scale-dependent
relative bias.  Our results are in good qualitative agreement with
those of \cite{hogg03}, who find that the mean local density around
SDSS galaxies increases with increasing luminosity and redder color,
with both luminous and faint red galaxies residing in high density
environments.

We have used HOD modeling assuming a $\Lambda$CDM cosmological model
to translate $\wrp$ measurements of pair counts into relations between
galaxies and dark matter halos.  We have adopted a simple, 
theoretically motivated HOD parameterization with a step-function
for central galaxies above a luminosity threshold and a power-law
mean occupation for satellite galaxies.  As in Z04, we find that HOD 
models naturally explain inflections in the observed $\wrp$ as
transitions from the one-halo to two-halo regime of $\xi(r)$.
With two parameters that can be adjusted to match $\wrp$, our
physically motivated HOD model fits the observations better than
a power-law for most but not all samples.
The large scale amplitude of $\wrp$ can be chosen at will in a power-law
fit, but it cannot be pushed arbitrarily low in an HOD model for a
given cosmology, and several of our fits overpredict this large
scale amplitude by modest amounts.  These discrepancies could 
reflect cosmic variance in the measurements that is not fully 
captured by jackknife error estimates, but future measurements 
from larger samples, together with more robust numerical results
for halo bias factors, could lead to
interesting conflicts with HOD predictions.

The luminosity dependence of $\wrp$ is well described by HOD models
in which the mean occupation $\Navg$ shifts ``horizontally'' with
increasing $L$, showing similar growth in the halo mass scale for
central and satellite galaxies.  For luminosity thresholds spanning
the range $\Mr=-18$ to $\Mr=-21.5$, we find that the mass at which
a halo hosts one satellite (on average) above the threshold is 23
times the minimum mass for hosting a central galaxy above the
threshold.  One consequence of this large factor is that most galaxies
at a given luminosity are central galaxies of their host dark matter
halos, not satellites in more massive systems.
Our derived satellite fraction for bright galaxies ($-22 < \Mr < -21$)
agrees well with that inferred by
\cite{seljak04c} from galaxy-galaxy lensing,
though we find higher satellite fractions at lower luminosities.
A second consequence is that the conditional luminosity function
at fixed halo mass has a spike or bump at the central galaxy 
luminosity and cannot be described by a \cite{schechter76} function.

By adding three adjustable parameters that describe the fraction of
blue central and satellite galaxies as a function of halo mass,
we are able to fit the 21 additional data points that represent
the blue and red galaxy $\wrp$ measurements for a given sample.
In a luminosity-threshold sample, central galaxies just above
$\Mmin$ are predominantly blue, while central galaxies in more massive
halos are predominantly red.  In the regime where satellites are
common, the blue galaxy fraction is a slowly declining function
of halo mass.  In luminosity-bin samples, the ratio of blue to 
red galaxies increases with decreasing luminosity.  The strong
small scale clustering of faint red galaxies reflects the fact
that nearly all such galaxies are satellite systems in massive
halos.  The cross correlation between red and blue galaxies 
supports the hypothesis that the two populations are well mixed 
at a given halo mass, rather than residing in distinct
``red galaxy halos'' and ``blue galaxy halos.''

Our derived trends of the HOD dependence on luminosity and color
are in qualitative agreement 
with the predictions of semi-analytic and numerical models of galaxy formation
\citep{berlind03,kravtsov04,zheng04b}.
They illustrate the power of HOD model fitting to extract 
physically informative insights from clustering measurements,
such as the scaling relation between central and satellite mass
thresholds, the dependence of central/satellite fractions on luminosity
and color, and the mixture of blue and red populations.
Our results for the luminosity and type dependence of the HOD
are in good qualitative agreement with those of
\citeauthor{bosch03a} (\citeyear{bosch03a}, figure~10), who fit
the conditional luminosity function of early and late type galaxies
in the 2dFGRS and use it to extract halo occupation functions.
In particular, both analyses show that blue (late-type) galaxies dominate
the low-$M$ end of $\Navg$ at low luminosities, that red galaxies
are prominent near $\Mmin$ for higher luminosities, and that faint red
(early-type) galaxies are predominantly satellites in massive halos.
Given the independent data sets and the very different analysis
methods and parameterizations, this qualitative agreement is reassuring.
\cite{magliocchetti03} find some similar results for the relative 
occupations of early and late type galaxies, though a comparison
is difficult because they model measurements from a flux-limited
sample rather than a well-defined class of galaxies.

The main limitation of the present analysis is that $\wrp$ alone
imposes limited constraints on the HOD \citep{berlind02}, forcing
us to adopt a restricted HOD parameterization and a fixed cosmological
model.  In a companion paper \citep{abazajian04b}, we bring in additional
constraints from CMB anisotropy measurements and show that these
data together with $\wrp$ measurements for the $\Mr<-21$ sample
impose tight constraints on cosmological parameters like $\Omega_m$, $h$,
and $\sigma_8$, similar to those obtained from the combination of
CMB data with the large scale SDSS galaxy power spectrum 
\citep{tegmark04b}.  The galaxy clustering measurements will become
much more powerful themselves as we bring in complementary information
from other clustering statistics (see \citealt{berlind02}).
For example, the group multiplicity function will pin down the high mass 
end of $\Navg$ (\citealt{peacock00,marinoni02,kochanek03}), allowing
us to explore more flexible parameterizations of the low occupancy
regime when fitting $\wrp$.  Void probability statistics
\citep{vogeley94,hoyle02} and the Tully-Fisher (\citeyear{tully77})
relation probe the single occupancy regime near $\Mmin$, while measured
profiles of groups and clusters can refine the assumption that
satellite populations trace halo dark matter profiles.
Three-point correlations probe the high mass regime of $P(N|M)$
\citep{ma00,scoccimarro01,takada03}, and they provide a diagnostic
for the amplitude of dark matter fluctuations \citep{fry94,feldman01,verde02}.
Finally, with real space clustering tightly constrained by complementary
statistics, dynamically sensitive measures such as redshift-space
distortions, weak lensing by galaxies and groups
\citep{sheldon01,sheldon04,seljak04c}, and virial masses of
groups and clusters become powerful tools for constraining
$\Omega_m$ and $\sigma_8$, even without auxiliary data from the CMB
or other observables (Z.\ Zheng \& D.\ Weinberg 2005, in preparation). 

Realizing this program will require considerable effort in the measurements
themselves, careful estimates of statistical and systematic uncertainties,
and development and testing of HOD calculational techniques accurate
enough to match the high precision afforded by the data.
By adopting theoretically motivated background assumptions, we
have learned an impressive amount from the projected correlation
function alone.  Further analysis of the growing
SDSS data set will allow us to test these background assumptions
and develop a thorough understanding of the relation between galaxies
and dark matter.

\acknowledgments

We thank Andrey Kravtsov and Daniel Eisenstein for useful discussions.
I.Z. was supported by NSF grants AST 00-98577 and AST 04-07200.
Z.Z. and D.H.W. were supported by NSF grants AST 00-98584 and AST 04-07125.
Z.Z. was also supported by a Presidential Fellowship from the Graduate School
of the Ohio State University.
Z. Z. acknowledges as well the support of NASA through Hubble Fellowship grant
HF-01181.01-A awarded by the Space Telescope Science Institute, which is
operated by the Association of Universities for Research in Astronomy, Inc.,
for NASA, under contract NAS 5-26555.

Funding for the creation and distribution of the SDSS Archive has been 
provided by the Alfred P. Sloan Foundation, the Participating Institutions, 
the National Aeronautics and Space Administration, the National Science 
Foundation, the U.S. Department of Energy, the Japanese Monbukagakusho, 
and the Max Planck Society. The SDSS Web site is http://www.sdss.org/.

The SDSS is managed by the Astrophysical Research Consortium (ARC) for the 
Participating Institutions. The Participating Institutions are The University 
of Chicago, Fermilab, the Institute for Advanced Study, the Japan Participation
Group, The Johns Hopkins University, the Korean Scientist Group, Los Alamos 
National Laboratory, the Max-Planck-Institute for Astronomy (MPIA), the 
Max-Planck-Institute for Astrophysics (MPA), New Mexico State University, 
University of Pittsburgh, University of Portsmouth, Princeton University, 
the United States Naval Observatory, and the University of Washington.


\begin{thebibliography}{}

\bibitem[Abazajian et al.(2004)]{abazajian04} 
Abazajian, K., et al. 2004, \aj, 128, 502

\bibitem[Abazajian et al.(2005)]{abazajian04b} 
Abazajian, K., Zheng, Z., Zehavi, I., Weinberg, D.\ W., Frieman, J.\ A., 
Berlind, A.\ A., Blanton, M.\ R., Bahcall, N.\ A., Brinkman, J., Schneider, 
D.\ P., \& Tegmark, M. 2005, \apj, 625, 613

\bibitem[Adami, Biviano, \& Mazure(1998)]{adami98}
Adami, C., Biviano, A., \& Mazure, A.\ 1998, \aap, 331, 439

\bibitem[Baldry et al.(2004)]{baldry04}
Baldry, I.\ K., Glazebrook, K., Brinkman, J., Ivezi\'{c}, \v{Z}., 
Lupton, R.\ H., Nichol, R.\ C., \& Szalay, A.\ S. 2004, \apj, 600, 681

\bibitem[Benoist et al.(1996)]{benoist96}
Benoist, C., Maurogordato, S., da Costa, L. N., Cappi, A., 
\& Schaeffer, R. 1996, \apj, 472, 452

\bibitem[Benson et al.(2000)]{benson00}
Benson, A. J., Cole, S., Frenk, C. S., Baugh, C. M., \& Lacey, C. G. 2000,
\mnras, 311, 793

\bibitem[Berlind \& Weinberg(2002)]{berlind02}
Berlind, A.\ A., \& Weinberg, D.\ H.\ 2002, \apj, 575, 587

\bibitem[Berlind et al.(2003)]{berlind03}
Berlind, A.\ A.,  Weinberg, D.\ H., Benson, A.\ J., Baugh, C.\ M., Cole, S.,
et al.\ 2003, \apj, 593, 1

\bibitem[Berlind et al.(2005)]{berlind04}
Berlind, A.\ A., Blanton, M.\ R., Hogg, D.\ W., Weinberg, D.\ H., 
Dav{\'e}, R., Eisenstein, D.\ J., \& Katz, N. 2005, \apj, 629, 625

\bibitem[Blanton et al.(2005a)]{blanton03e}
Blanton, M.\ R., Eisenstein, D.\ J., Hogg, D.\ W., Schlegel, D.\ J., \&
Brinkmann, J.\ 2005a, \apj, 629, 143
    
\bibitem[Blanton et al.(2003a)]{blanton03a}
Blanton, M.\ R., Lupton, R.\ H., Maley, F.\ M., Young, N.,
Zehavi, I., \& Loveday, J. 2003a, \aj, 125, 2276

\bibitem[Blanton et al.(2003b)]{blanton03b}
Blanton, M.\ R., et al. 2003b, \aj, 125, 2348 

\bibitem[Blanton et al.(2003c)]{blanton03c}
Blanton, M.\ R., et al.\ 2003c, \apj, 592, 819 

\bibitem[Blanton et al.(2005b)]{blanton05}
Blanton, M.\ R., et al.\ 2005b, \aj, 129, 2562

\bibitem[Brown, Webster \& Boyle(2000)]{brown00}
Brown, M.\ J.\ I., Webster, R. L., \& Boyle, B.\ J.\ 2000, \mnras, 317, 782

\bibitem[Bullock et al.(2001)]{bullock01}
Bullock, J.\ S., Kolatt, T.\ S., Sigad, Y., Somerville, R.\ S., Klypin, A.\ A.,
Primack, J.\ R., \& Dekel, A.\ 2001, \mnras, 321, 559

\bibitem[Bullock, Wechsler, \& Somerville(2002)]{bullock02}
Bullock, J.\ S., Wechsler, R.\ H., \& Somerville, Rachel, S.\ 2002, \mnras,
329, 246

\bibitem[Budavari et al.(2003)]{budavari03}
Budavari, T., et al., 2003, \apj, 595, 59

\bibitem[Colless et al.(2001)]{colless01}
Colless, M.~et al. 2001, \mnras, 328, 1039 

\bibitem[Davis \& Geller(1976)]{davis76}
Davis, M., \& Geller, M.\ J. 1976, \apj, 208, 13

\bibitem[Davis et al.(1988)]{davis88}
Davis, M., Meiksin, A., Strauss, M.\ A., da Costa, L.\ N., \& Yahil, A.
1988, \apj, 333, L9

\bibitem[Davis \& Peebles(1983)]{davis83}
Davis, M., \& Peebles, P.\ J.\ E. 1983, \apj, 267, 465

\bibitem[De Lucia et al.(2004)]{delucia04} 
De Lucia, G., Kauffmann, G., Springel, V., White, S.~D.~M., Lanzoni, B., 
Stoehr, F., Tormen, G., \& Yoshida, N.\ 2004, \mnras, 348, 333 

\bibitem[Dressler(1980)]{dressler80}
Dressler, A., 1980, \apj, 236, 351

\bibitem[Einasto(1991)]{einasto91}
Einasto, M.\ 1991, \mnras, 252, 261

\bibitem[Einasto, \& Einasto(1987)]{einasto87}
Einasto, M., \& Einasto, J.\ 1987, \mnras, 226, 543

\bibitem[Eisenstein et al.(2001)]{eisenstein01}
Eisenstein, D.\ J., et al. 2001, \aj, 122, 2267

\bibitem[Efstathiou, Bond, \& White(1992)]{efstathiou92}
Efstathiou, G., Bond, J.\ R., \& White, S.\ D.\ M. 1992, \mnras, 258, 1

\bibitem[Feldman et al.(2001)]{feldman01}
Feldman, H.\ A., Frieman, J.\ A., Fry, J.\ N., \& Scoccimarro, R. 2001,
\prl, 86, 1434

\bibitem[Fry(1994)]{fry94}
Fry, J.\ N. 1994, \prl, 73, 215

\bibitem[Fukugita et al.(1996)]{fukugita96}
Fukugita, M., Ichikawa, T., Gunn, J.\ E., Doi, M., Shimasaku, K., \&
Schneider, D.\ P. 1996, \aj, 111, 1748

\bibitem[Gao et al.(2004)]{gao04}
Gao, L., White, S.\ D.\ M., Jenkins, A., Stoehr, F., \& Springel, V. 2004,
\mnras, 355, 819

\bibitem[Gazta{\~ n}aga \& Juszkiewicz(2001)]{gaztanaga01}
Gazta{\~n}aga, E.~\& Juszkiewicz, R.\ 2001, \apjl, 558, L1

\bibitem[Gomez et al.(2003)]{gomez03}
Gomez, P. et al.\ 2003, \apj, 584, 210

\bibitem[Goto et al.(2003)]{goto03}
Goto, T., Yamauchi, C., Fujita, Y., Okamura, S., Sekiguchi, M., Smail, I.,
Bernardi, M., \& Gomez, P.\ L. 2003, \mnras 346, 601

\bibitem[Gott et al.(2005)]{gott03}
Gott, J.\ R., III, Juric, M., Schlegel, D.\ J., Hoyle, F., Vogeley, M.\ S., 
Tegmark, M., Bahcall, N.\ A., Brinkmann, J. 2005, ApJ, 624, 463

\bibitem[Gunn et al.(1998)]{gunn98}
Gunn, J.\ E., et al. 1998, \aj, 116, 3040

\bibitem[Guzik \& Seljak(2002)]{guzik02}
Guzik, J., \& Seljak, U. 2002, \mnras, 335, 311

\bibitem[Guzzo et al.(1997)]{guzzo97}
Guzzo, L., Strauss, M.\ 
A., Fisher, K.\ B., Giovanelli, R., \& Haynes, M.\ P.\ 1997, \apj, 489, 37

\bibitem[Hamilton(1988)]{hamilton88}
Hamilton, A. J. S. 1988, \apj, 331, L59

\bibitem[Hamilton(1992)]{hamilton92}
Hamilton, A.\ J.\ S. 1992, \apj, 385, L5

\bibitem[Hamilton(1993)]{hamilton93}
Hamilton, A. J. S. 1993, \apj, 417, 19

\bibitem[Hawkins et al.(2003)]{hawkins03}
Hawkins, E., et al.\ 2003, \mnras, 346, 78

\bibitem[Hogg et al.(2001)]{hogg01}
Hogg, D.\ W., Schlegel, D.\ J., Finkbeiner, D.\ P., \& Gunn, J.\ E. 2001, 
AJ, 122, 2129

\bibitem[Hogg et al.(2003)]{hogg03}
Hogg, D.\ W., et al. 2003, \apj, 585, L5 

\bibitem[Hoyle \& Vogeley(2002)]{hoyle02}
Hoyle, F., \& Vogeley, M.\ S. 2002, \apj, 566, 641

\bibitem[Hubble(1936)]{hubble36}
Hubble, E.P. 1936, The Realm of the Nebulae (Oxford University Press: Oxford),
79

\bibitem[Jenkins et al.(2001)]{jenkins01}
Jenkins, A., Frenk, C.\ S., White, S.\ D.\ M., Colberg, J.\ M., Cole, S.,
Evrard, A.\ E., Couchman, H.\ M.\ P., \& Yoshida, N.\ 2001, \mnras, 321, 372

\bibitem[Jing \& B\"{o}rner(1998)]{jing98b}
Jing, Y.\ P. \& B\"{o}rner, G.\ 1998, \apj, 503, 37

\bibitem[Jing, B\"{o}rner, \& Suto(2002)]{jing02}
Jing, Y.\ P., B\"{o}rner, G., \& Suto, Y.\ 2002, \apj, 564, 15

\bibitem[Jing, Mo, \& B\"{o}rner(1998)]{jing98a}
Jing, Y.\ P., Mo, H.\ J., \& B\"{o}rner, G.\ 1998, \apj, 494, 1

\bibitem[Kaiser(1984)]{kaiser84}
Kaiser, N. 1984, \apj, 294, L9

\bibitem[Kaiser(1987)]{kaiser87}
Kaiser, N. 1987, \mnras, 227, 1

\bibitem[Kauffmann, Nusser, \& Steinmetz(1997)]{kauffmann97}
Kauffmann, G., Nusser, A., \& Steinmetz, M. 1997, \mnras, 286, 795

\bibitem[Kauffman et al.(1999)]{kauffmann99}
Kauffmann, G., Colberg, J. M., Diaferio, A., \& White, S. D. M. 1999,
\mnras, 303, 188

\bibitem[Kauffmann \& Lemson(1999)]{kauffmann99a}
Kauffmann, G., \& Lemson, G. 1999, \mnras, 302, 111

\bibitem[Kauffmann et al.(2004)]{kauffmann04}
Kauffmann, G., White,
S.\ D.\ M., Heckman, T.\ M., Menard, B., Brinchmann, J., Charlot, S.,
Tremonti, C., \& Brinkmann, J.\ 2004, \mnras, 353, 713

\bibitem[Kayo et al.(2004)]{kayo04}
Kayo, I., et al. 2004, PASJ, 56, 415

\bibitem[Kochanek et al.(2003)]{kochanek03}
Kochanek, C.~S., White, M., Huchra, J., Macri, L., Jarrett, T. H.,
Schneider, S. E., \& Mader, J. 2003, \apj, 585, 161

\bibitem[Kravtsov et al.(2004)]{kravtsov04}
Kravtsov, A.\ V., Berlind, A.\ A., Wechsler, R.\ H., Klypin, A.\ A.,
Gottloeber, S., Allgood, B., \& Primack, J.\ R.\ 2004, \apj, 609, 35

\bibitem[Landy \& Szalay(1993)]{landy93}
Landy, S.\ D., \& Szalay, A.\ S. 1993, \apj, 412, 64

\bibitem[Lewis et al.(2002)]{lewis02}
Lewis, I.~et al.\ 2002, \mnras, 334, 673 

\bibitem[Loveday et al.(1995)]{loveday95}
Loveday, J., Maddox, S. J., Efstathiou, G., \& Peterson, B. A. 1995,
\apj, 442, 457

\bibitem[Loh(2003)]{loh03}
Loh, Y.\ S. 2003, Ph.D.\ thesis, Princeton University

\bibitem[Lupton et al.(2001)]{lupton01}
Lupton, R., Gunn, J. E., Ivezi\'{c}, \v{Z}., Knapp, G. R., Kent, S.,
\& Yasuda, N. 2001, in ASP Conf. Ser. 238, Astronomical Data Analysis
Software and Systems X, ed. F. R. Harnden, Jr., F. A. Primini, and
H. E. Payne (San Francisco: Astr. Soc. Pac.), p. 269 (astro-ph/0101420)

\bibitem[Ma \& Fry(2000)]{ma00}
Ma, C., \& Fry, J. N. 2000, \apj, 543, 503

\bibitem[Madgwick et al.(2003)]{madgwick03}
Madgwick, D.~S. et al. 2003, \mnras, 344, 847

\bibitem[Magliocchetti \& Porciani(2003)]{magliocchetti03}
Magliocchetti, M., \& Porciani, C.\ 2003, \mnras, 346, 186

\bibitem[Marinoni \& Hudson(2002)]{marinoni02}
Marinoni, C., \& Hudson, M.\ J.\ 2002, \apj, 569, 101

\bibitem[McKay et al.(2001)]{mckay01}
McKay T.\ A., et al.\ 2001, preprint (astro-ph/0108013)

\bibitem[McKay et al.(2002)]{mckay02}
McKay T.\ A., et al.\ 2002, \apj, 571, L85

\bibitem[Melnick \& Sargent(1977)]{melnick77}
Melnick, J., \& Sargent, W.\ L.\ W.\ 1977, \apj, 215, 401

\bibitem[Miller et al.(2003)]{miller03}
Miller, C. J., Nichol, R. C., Gomez, P. L., Hopkins, A. M., \&
Bernardi, M. 2003, \apj, 591, 142


\bibitem[Mo et al.(2004)]{mo03}
Mo, H.\ J., Yang, X.\ H., van den Bosch, F.\ C, \& Jing, Y.\ P.\ 2004, 
\mnras, 349, 205

\bibitem[Moustakas \& Somerville(2002)]{moustakas02}
Moustakas, L.\ A., \& Somerville, R.\ S.\ 2002, \apj, 577, 1

\bibitem[Navarro, Frenk, \& White(1995)]{NFW95}
Navarro, J.\ F., Frenk, C.\ S., \& White, S.\ D.\ M. 1995, \mnras, 275, 56

\bibitem[Navarro, Frenk, \& White(1996)]{NFW96}
Navarro, J.\ F., Frenk, C.\ S., \& White, S.\ D.\ M. 1996, \apj, 462, 563

\bibitem[Navarro, Frenk, \& White(1997)]{NFW97}
Navarro, J.\ F., Frenk, C.\ S., \& White, S.\ D.\ M. 1997, \apj, 490, 493

\bibitem[Norberg et al.(2001)]{norberg01}
Norberg, P., et al. 2001, \mnras, 328, 64

\bibitem[Norberg et al.(2002)]{norberg02}
Norberg, P., et al. 2002, \mnras, 332, 827

\bibitem[Oemler(1974)]{oemler74}
Oemler, A.\ 1974, \apj, 194, 1

\bibitem[Padmanabhan et al.(2004)]{padmanabhan04}
Padmanabhan, N., et al.\ 2004, New Astronomy, 9, 329

\bibitem[Park et al.(1994)]{park94}
Park, C., Vogeley, M. S., Geller, M. J., \& Huchra, J. P. 1994,
\apj, 431, 569

\bibitem[Peacock \& Smith(2000)]{peacock00}
Peacock, J.\ A., \& Smith, R.\ E.\ 2000, \mnras, 318, 1144

\bibitem[Percival et al.(2003)]{percival03}
Percival, W. et al.\ 2003, \mnras, 337, 1068

\bibitem[Pier et al.(2003)]{pier03}
Pier, J.\ R., Munn, J.\ A., Hindsley, R.\ B., Hennessy, G.\ S., Kent, S.\ M.,
Lupton, R.\ H., \& Ivezi\'{c}, \v{Z}. 2003, \aj, 125, 1559

\bibitem[Porciani, Magliocchetti, \& Norberg(2004)]{porciani04}
Porciani, C., Magliocchetti, M., \& Norberg, P.\ 2004, \mnras, 355, 1010

\bibitem[Postman \& Geller(1984)]{postman84}
Postman, M., \& Geller, M. J. 1984, \apj, 281, 95

\bibitem[Prada et al.(2003)]{prada03}
Prada, F., et al.\ 2003, \apj, 598, 260

\bibitem[Richards et al.(2002)]{richards02}
Richards, G.\ T., et al. 2002, \aj, 123, 2945

\bibitem[Sargent \& Turner(1977)]{sargent77}
Sargent, W.\ L.\ W., \& Turner, E.\ L. 1977, \apj, 212, L3

\bibitem[Saunders, Rowan-Robinson \& Lawrence(1992)]{saunders92}
Saunders, W., Rowan-Robinson, M., \& Lawrence, A. 1992, \mnras,
258, 134

\bibitem[Schechter(1976)]{schechter76}
Schechter, P.\ 1976, \apj, 203, 297

\bibitem[Schlegel et al.(1998)]{schlegel98}
Schlegel, D.\ J., Finkbeiner, D.\ P., \& Davis, M. 1998, \apj, 500, 525

\bibitem[Scoccimarro \& Sheth(2002)]{scoccimarro02}
Scoccimarro, R.,~\& Sheth, R.~K.\ 2002, \mnras, 329, 629

\bibitem[Scoccimarro et al.(2001)]{scoccimarro01}
Scoccimarro, R., Sheth, R.\ K., Hui, L., \& Jain, B.\ 2001, \apj, 546, 20

\bibitem[Scranton(2003)]{scranton03}
Scranton, R.\ 2003, \mnras, 339, 410

\bibitem[Seljak(2000)]{seljak00}
Seljak, U. 2000, \mnras, 318, 203

\bibitem[Seljak et al.(2005a)]{seljak04c}
Seljak, U., et al. 2005a, \prd, 71, 043511

\bibitem[Seljak et al.(2004b)]{seljak04a}
Seljak, U., et al. 2005b, \prd, 71, 103515
    
\bibitem[Seljak \& Warren(2004)]{seljak04b}
Seljak, U., \& Warren, M.\ S. 2004, \mnras, 355, 129 

\bibitem[Seljak \& Zaldarriaga(1996)]{seljak96}
Seljak, U., \& Zaldarriaga, M. 1996, \apj, 469, 437

\bibitem[Sheldon et al.(2001)]{sheldon01}
Sheldon, E.\ S., et al. 2001, \apj, 554, 881

\bibitem[Sheldon et al.(2004)]{sheldon04}
Sheldon, E.\ S., et al. 2004, \aj, 127, 2544

\bibitem[Sheth \& Diaferio(2001)]{sheth01b}
Sheth, R.\ K.\ \& Diaferio, A.\ 2001, \mnras, 322, 901

\bibitem[Sheth et al.(2001)]{sheth01a}
Sheth, R.\ K., Hui, L., Diaferio, A., \& Scoccimarro, R.\ 2001, \mnras, 325,
1288

\bibitem[Sheth, Mo, \& Tormen(2001)]{sheth01}
Sheth, R.\ K., Mo, H.\ J., \& Tormen, G.\ 2001, \mnras, 323, 1

\bibitem[Sheth \& Tormen(1999)]{sheth99}
Sheth, R.\ K.\ \& Tormen, G.\ 1999, \mnras, 308, 119

\bibitem[Sheth \& Tormen(2004)]{sheth04}
Sheth, R. K., \& Tormen, G. 2004, \mnras, 350, 1385

\bibitem[Smith et al.(2002)]{smith02}
Smith, J.\ A., et al.\ 2002, \aj, 123, 2121

\bibitem[Smith et al.(2003)]{smith03}
Smith, R.\ E., et al. 2003, \mnras, 341, 1311

\bibitem[Somerville et al.(2001)]{somerville01}
Somerville, R.\ S., Lemson, G., Sigad, Y., Dekel, A., Kauffmann, G.,
White, S.\ D.\ M. 2001, \mnras, 320, 289

\bibitem[Spergel et al.(2003)]{spergel03}
Spergel, D.\ N., et al. 2003, \apjs, 148, 175

\bibitem[Stoughton et al.(2002)]{stoughton02}
Stoughton, C., et al.\ 2002, \aj, 123, 485

\bibitem[Strateva et al.(2001)]{strateva01}
Strateva, I., et al. 2001, \aj, 122, 1861

\bibitem[Strauss et al.(2002)]{strauss02}
Strauss, M.A., et al.\ 2002, \aj, 124, 1810

\bibitem[Takada \& Jain(2003)]{takada03}
Takada, M., \& Bhuvnesh, J. 2003, \mnras, 340, 580

\bibitem[Tasitsiomi et al.(2004)]{tasitsiomi04}
Tasitsiomi, A., Kravtsov, A.\ V., Wechsler, R.\ H., \& Primack, J.\ R.\
2004, \apj, 614,533

\bibitem[Tegmark et al.(2004a)]{tegmark04a}
Tegmark, M., et al.\ 2004a, \apj, 606, 702 

\bibitem[Tegmark et al.(2004b)]{tegmark04b}
Tegmark, M., et al.\ 2004b, \prd, 69, 103501

\bibitem[Tully \& Fisher(1977)]{tully77}
Tully, R.\ B., \& Fisher, J.\ R. 1977, A\&A, 54, 661

\bibitem[van den Bosch, Yang, \& Mo(2003a)]{bosch03a}
van den Bosch, Frank, C., Yang, X.\ H., \& Mo, H.\ J.\ 2003, \mnras, 340, 771

\bibitem[van den Bosch, Mo, \& Yang(2003b)]{bosch03b}
van den Bosch, Frank, C., Mo, H.\ J., \& Yang, X.\ H.\ 2003, \mnras, 345, 923

\bibitem[Verde et al.(2002)]{verde02}
Verde, L., et al. 2002, \mnras, 335, 432

\bibitem[Vogeley et al.(1994)]{vogeley94}
Vogeley, M.\ S., Geller, M.\ J., Changbom, P., \& Huchra, J.\ P. 1994,
\aj, 108, 745

\bibitem[Weinberg(2002)]{weinberg02}
Weinberg, D.\ H.\ 2002, in A New Era in Cosmology, ASP Conference
Proceedings, Vol. 283, ed. Nigel Metcalfe \& Tom Shanks (San Francisco: 
Astronomical Society of the Pacific), 3 (astro-ph/0202184)

\bibitem[White(2002)]{white02}
White, M.\ 2002, \apjs, 143, 241

\bibitem[White, Hernquist, \& Springel(2001)]{white01}
White, M., Hernquist, L., \& Springel, V.\ 2001, \mnras, 550, 129

\bibitem[White, Tully, \& Davis(1988)]{white88}
White, S.\ D.\ M., Tully, R.\ B., \& Davis, M.\ 1988, \apjl, 333, L45

\bibitem[Whitmore, Gilmore, \& Jones(1993)]{whitmore93}
Whitmore, B. C., Gilmore, D. M., \& Jones, C. 1993, \apj, 407, 489

\bibitem[Willmer, da Costa \& Pellegrini(1998)]{willmer98}
Willmer, C. N. A., da Costa, L. N., \& Pellegrini, P. S. 1998, \aj, 115, 869

\bibitem[Yan, Madgwick, \& White(2003)]{yan03}
Yan, R., Madgwick, D.\ S., \& White, M.\ 2003, \apj, 598, 848

\bibitem[Yan, White \& Coil(2004)]{yan04}
Yan, R., White, M., \& Coil, A.\ L.\ 2004, \apj, 607, 739 

\bibitem[Yang, Mo, \& van den Bosch(2003)]{yang03}
Yang, X.\ H., Mo, H.\ J., \& van den Bosch, F.\ C.\ 2003, \mnras, 339, 1057

\bibitem[Yoshikawa et al.(2001)]{yoshikawa01}
Yoshikawa, K., Taruya, A., Jing, Y.\ P., \& Suto, Y.\ 2001, \apj, 558, 520

\bibitem[York et al.(2000)]{york00}
York, D. G. et al.\ 2000, \aj, 120, 1579

\bibitem[Zehavi et al.(2002)]{zehavi02}
Zehavi, I., Blanton, M.\ R., Frieman, J. A., Weinberg, D.\ H., Mo, H.\ J., 
et al.\ 2002, \apj, 571, 172, [Z02]

\bibitem[Zehavi et al.(2004)]{zehavi04}
Zehavi, I., Weinberg, D.\ H., Zheng, Z., Berlind, A.\ A., Frieman, J.\ A.,
et al.\ 2004, \apj, 608, 16, [Z04]
 
\bibitem[Zheng et al.(2002)]{zheng02}
Zheng, Z., Tinker, J.\ L., Weinberg, D.\ H., \& Berlind, A.\ A.\ 2002,
\apj, 575, 617

\bibitem[Zheng(2004)]{zheng04}
Zheng, Z.\ 2004, \apj, 610, 61

\bibitem[Zheng et al.(2004)]{zheng04b}
Zheng, Z., Berlind, A. A., Weinberg, D. H., Benson, A. J., 
Baugh, C. M., Cole, S., Dav\'e, R., Frenk, C. S., Katz, N., \& Lacey, C. G.
2004, \apj, submitted (astro-ph/0408564)

\bibitem[Zwicky et al.(1968)]{zwicky68}
Zwicky, F., Herzog, E., Wild, P., Karpowicz, M., \& Kowal, C.,
1961-1968, {\it Catalog of Galaxies and of Clusters of Galaxies},
Vols. 1-6, (Pasadena: California Institute of Technology)

\end{thebibliography}
\end{document}